\documentclass{amsart}  
\usepackage{verbatim,amssymb}
\usepackage{amscd}
\numberwithin{equation}{section}
\newtheorem{mthm}{Main Theorem}
\newtheorem{thm}{Theorem}[section]
\newtheorem{lem}[thm]{Lemma}
\newtheorem{prop}[thm]{Proposition}
\newtheorem{cor}[thm]{Corollary}
\newtheorem{defn}{Definition}[section]
\newtheorem{remark}{Remark}[section]

\newtheorem{claim}{Claim}[section]
\newtheorem{problem}[thm]{Problem} 
 
\newtheorem{comparison}{Comparison}[section]

\def\blev{{\mathfrak{B}}_{{\nogg},\mathrm{ev}}} 
\def\blod{{\mathfrak{B}}_{{\nogg},\mathrm{od}}} 
\def\G{\mathcal{G}}
\def\F{\mathcal{F}}
\def\sdet{\operatorname{sdet}}
\def\cbra(#1){\langle{#1}\rangle}

\def\unbt{\underline{t}}
\def\unbx{\underline{x}}
\def\unby{\underline{y}}
\def\unbq{\underline{q}}
\def\unbp{\underline{p}}
\def\unbxi{\underline{\xi}}
\def\unbtheta{\underline{\theta}}
\def\unbpi{\underline{\pi}}
\def\unbrho{\underline{\rho}}

\def\unbeta{\underline{\eta}}

\def\unbomega{\underline{\omega}}

\def\inidata{{\unbx},{\unbxi},{\unbtheta},{\unbpi}}

\def\inidatamodpara{{\unbx}+{\epsilon}{\unby},\unbxi+{\epsilon}{\unbeta},\unbtheta+{\epsilon}{\unbomega},\unbpi+{\epsilon}{\unbrho}}

\def\deg{{d}} 
\def\nogg{{L}} 

\def\dt{\frac{d}{dt}}

\def\where{\quad\text{where}\quad}
\def\when{\quad\text{when}\quad}
\def\with{\quad\text{with}\quad}
\def\for{\quad\text{for}\quad}
\def\forany{\quad\text{for any}\quad}

\def\et{\quad\text{and}\quad}

\def\proj{\operatorname{proj}}
\def\dist{\operatorname{dist}}
\def\dim{\operatorname{dim}} 

\def\euc{ {\mathbb{R}}}
\def\fR{ {\mathfrak{R}}}
\def\fC{ {\mathfrak{C}}}
\def\CSS{ {\mathcal{C}}_{S\!S}} 
\def\ccsl{{/\kern-0.5em {\mathcal{C}}}}

\def\superon{{ {\mathfrak{R}}{}^{0|n}}}
\def\supermn{{ {\mathfrak{R}}{}^{m|n}}}

\def\supermo{{ {\mathfrak{R}}{}^{m|0}}}

\def\cev{{\mathfrak{C}}_{\mathrm{ev}}}
\def\cod{{\mathfrak{C}}_{\mathrm{od}}}
\def\rev{{\mathfrak{R}}_{\mathrm{ev}}}
\def\rod{{\mathfrak{R}}_{\mathrm{od}}}

\begin{document}

\title[supersmooth functions]
{Definition and characterization\\of\\
supersmooth functions on superspace\\ based on Fr\'echet-Grassmann algebra}
\author[Atsushi Inoue]{Atsushi Inoue}
\address{Professor Emeritus, Department of Mathematics,
Tokyo Institute of Technology}
\curraddr{3-4-10, Kajiwara, Kamakura-city, Kanagawa-prefecture, 247-0063, Japan}
\dedicatory{Dedicated to the memory of late Professor N. Suita}
\email{atlom-inoue60@nifty.com}
\date {\today} 
\begin{abstract}
{Preparing the Fr\'echet-Grassmann (FG-)algebra ${\fR}$ composed with countably infinite Grassmann generators,
we introduce the superspace ${\fR}^{m|n}$.
After defining Grassmann continuation of smooth functions on ${\euc}^m$ to those on ${\fR}^{m|0}$, we introduce a class of functions on ${\fR}^{m|n}$ which are called supersmooth. 
In this paper, we characterize such supersmooth functions
in G\^ateaux (but not necessarily Fr\'echet) differentiable category 
on Fr\'echet but not on Banach space.
This type of arguments for $G^{\infty}$-functions is mainly done on the Banach-Grassmann (BG-)algebra,
but we find it rather natural to work within FG-algebra
when we treat systems of PDE such as Dirac, Weyl or Pauli equations.
In that application, we need to prove that the solution of the (super) Hamilton equation is supersmooth w.r.t. initial data.
Though we took this point of view in our previous works, but is managed rather insufficiently. 
Therefore, we re-treat this subject here to answer affirmatively.
We give also local or global inverse function theorems for supersmooth functions on ${\fR}^{m|n}$.}
\end{abstract}

\maketitle
\tableofcontents
\baselineskip=16pt

\section{Introduction}

In order to treat ``photon'' and ``electron''  on the same footing as is proposed in Berezin and Marinov~\cite{BM}, there are many trials to extend the fundamental fields ${\euc}$ or $\mathbb{C}$ to those such as 
Rogers' Banach-Grassmann(BG-)algebra ${\mathfrak{B}}_{\infty}$, De Witt algebra $\Lambda_{\infty}$ or
Fr\'echet-Grassmann(FG-)algebra (such as \cite{MK}, \cite{ChB} but  we use ${\fR}$ or ${\fC}$ 
explained below). 
On such extended ``field", we need to develop elementary and real analysis 
for treating what we have done over ${\euc}^m$ or ${\mathbb{C}}^m$.
By the way, in \cite{BM}, they doesn't mention clearly what type of the ground field they take, in other word, they doen't distinguish countable number of  Grassmann generators 
and finite number of odd variables.

Not only above mentioned reasons from mathematical physics but also to treat systems of PDE without diagonalizing matrix structures, we need, so-called, odd variables. For example, Feynman claimed the following
in p.355 of {Feynman and Hibbs~\cite{FH65}}, 
since `spin' has been the object outside Feynman's procedures at that time: 
(underlined by the author)
\begin{quote} 
$\cdots $ \underline{path integrals suffer grievously from a serious defect}.
They \underline{do not permit a discus-}
\underline{sion of spin operators or other such operators} 
in a simple and lucid way.
They find their greatest use in systems for which coordinates and their
conjugate momenta are adequate.
Nevertheless, spin is a simple and vital part of real quantum-mechanical
systems.
It is a serious limitation that the half-integral spin of the electron
does not find a simple and ready representation.
It can be handled if the \underline{amplitudes and quantities are considered}\\
\underline{as quarternions} instead of ordinary complex numbers,
but the \underline{lack of commutativity of }\\
\underline{such numbers is a serious complication}.
\end{quote}

On the other hand, a physicist Witten~\cite{witt82-1} explains the notion of supersymmetric quantum mechanics to mathematicians by re-interpreting Morse theory. 
That is, deforming the form Laplacian by Morse function and regarding it as the infinitesimal generator of heat flow type
corresponding to the Lagrangian represented by  ``odd variables",
he applies rather naively the asymptotic method to the path-integral representation of the heat flow
to get the Morse inequality. Though his representation is beautiful and persuading, but there exists no mathematical theory
to make rigorous his argument directly. Because there exists not only no rigorous
Feynman measure (i.e. roughly speaking,  no non-trivial Lebesgue-like measure in ${\infty}$-dimensional topological vector space), 
but also the lack of the consistent theory including even and odd variables on equal footing.

In \S 2, we explain our problem after constructing a countable Grassmann generators \`a la Rogers~\cite{rog80}.
After defining FG-algebra ${\fR}$ or ${\fC}$ and superspace ${\fR}^{m|n}$, 
we introduce Grassmann continuation of ordinary smooth functions on ${\euc}^m$ and define supersmooth functions (alias superfields by physicists).
By the way, it is well-known that the elementary differential calculus on Euclidean spaces
is extended straight forwardly to those on Banach spaces but not so on Fr\'echet spaces.
For example, though the dual of a Banach space is again a Banach space, but
the dual of a Fr\'echet space is not necessarily a Fr\'echet space. 
Therefore, to clarify the difference of Fr\'echet or G\^ateaux differentiability 
between Banach or Fr\'echet spaces, 
we enumerate definitions of those differentiable functions
on Fr\'echet spaces borrowing mainly  from Hamilton \cite{hamil}.
Preparing these notions, we introduce formally supersmooth functions on ${\fR}^{m|n}$ and ask how to characterize suitably these functions.

In \S 3, we recall the results when the number of Grassmann generators is finite.  It seems appropriate to mention here that 
though not only Lemma 2.2 of Vladimirov and Volovich~\cite{VV83} but also Lamma 1.7 of Boyer and Gitler~\cite{BG84} \underline{contain
unsatisfactory arguments, but their conclusions hold true in our case at}\\
\underline{hand}. This point is clarified with the aid by Kazuo Masuda.
Moreover, Rogers~\cite{rog80, rog07} does not remark 
the Cauchy-Riemann type  relation explicitly
which should be satisfied for her ${G}^{\infty}$ functions, 
though this point is not mentioned further here.

In \S 4, though the precise definitions such as superdomain, superdifferentiability, supersmoothness, etc. will be given later, we have

\begin{mthm}\label{mthm}
Let ${\mathfrak{U}}$ be a superdomain in ${\fR}^{m|n}$ and 
let a function $f:{\mathfrak{U}}\to {\fC}$ be given.
Following conditions are equivalent:\\
(a) $f$ is  super Fr\'echet (F-, in short) differentiable on ${\mathfrak{U}}$, i.e. $f\in {\F}_{S\!D}^{\infty}({\mathfrak{U}}:{\fC})$, \\
(b) $f$ is  super G\^ateaux (G-, in short) differentiable on ${\mathfrak{U}}$, i.e. $f\in {\G}_{S\!D}^{\infty}({\mathfrak{U}}:{\fC})$, \\
(c) $f$ is $\infty$-times G-differentiable 
and $f\in {\G}_{S\!D}^{1}({\mathfrak{U}}:{\mathfrak{C}})$,\\
(d) $f$ is $\infty$-times G-differentiable and its G-differential $df$ is ${\rev}$-linear,\\ 
(e) $f$ is $\infty$-times G-differentiable and its G-differential $df$ satisfies 
Cauchy-Riemann type  equations,\\
(f) $f$ is supersmooth, i.e. it has the following representation, called superfield expansion, 
such that
$$
f(x,\theta)=\sum_{|a|\le n}\theta^a \tilde{f}_a(x)\with
f_a(q)\in C^{\infty}(\pi_{\mathrm{B}}({\mathfrak{U}}))\et
 \tilde{f}_a(x)=\sum_{|\alpha|=0}^{\infty}\frac{1}{\alpha!}\frac{\partial ^{\alpha}f_a(q)}{\partial q^{\alpha}}\bigg|_{q=x_{\mathrm{B}}}
 x_{\mathrm{S}}^{\alpha}.
$$
\end{mthm}
\begin{remark}
In the above, (f) stands for the ``algebraic'' nature and (a) claims the ``analytic'' nature of ``superfields''.
Yagi~\cite{yag88} proves essentially the equivalence $(b)\Longleftrightarrow(e)\Longleftrightarrow(f)$.
\end{remark}

We give some applications of above characterization in appendices.
In Appendix A, we show that the Hamilton-flow corresponding to 
a certain super-Hamilton function is supersmooth w.r.t. the initial data. 
We also give the super version of Local or Global Inverse Function Theorem 
applying above characterization in Appendix B.

\begin{remark}
Concerning Feynman's problem mentioned above, as we need to define the Hamilton function and to solve Hamilton-Jacobi equation corresponding to the systems of PDE such as Dirac or Pauli equations.
Our solution of these problem is affirmative, see for example,  
Inoue~\cite{{ino98-1}, {ino98-2}, {In99-3}}.
This is based on the fact that any $2^d\times 2^d$-matrix is decomposed by matrices 
satisfying Clifford relations and the Clifford algebra has the differential operator representation 
on the Grassmann algebras. 
But this decomposition of matrices doesn't work directly for systems with sizes $3\times 3$, $5\times 5$ etc. Seemingly to treat those cases, we need new class of non-commutative numbers and analysis on it. See, for example as a candidate of this direction Campoamor-Stursberg and Rausch de Traubenberg~\cite{CS-RT}. 
\end{remark}

\section{Preliminaries and Problem}

\subsection{A construction of countable Grassmann generators \`a la Rogers}
In this subsection, we use the lexicographic representation for multiple indeces.
Denote by ${\mathcal M}_{\nogg}$ the set of integer sequences given by
\begin{equation}
{\mathcal M}_{\nogg}=\{\mu \;|\; \mu=(\mu_1,\mu_2,\cdots,\mu_k ),
\, 1\le \mu_1<\mu_2<\cdots<\mu_k \le {\nogg}\}
\et{\mathcal M}_\infty=\cup_{{\nogg}=1}^\infty {\mathcal M}_{\nogg}.
\label{EA1.1.0}
\end{equation}
 
For any $j\in{\mathbb{N}}$, we denote the sequence with just one element $j$ 
as $(j)\in {\mathcal M}_\infty$.
For each $r\in{\mathbb{N}}$, we may correspond 
a member $\mu \in {\mathcal M}_\infty$ by using
\begin{equation}
r={\frac12}(2^{\mu_1}+2^{\mu_2}+\cdots+2^{\mu_k})=r(\mu)
\longleftrightarrow
\mu=(\mu_1,{\cdots},\mu_k)=\mu(r).
\label{EA1.1.r}
\end{equation}

We put ${\mathfrak{e}}_{(r)}=({\overbrace{0,\cdots,0,1}^r},0,\cdots \ )\in \ell_{\infty}\cap \ell_1$.
Regarding $\emptyset\in {\mathcal M}_{\nogg}$, we put $e_{\emptyset}=1$ and
for each $\mu \in {\mathcal M}_\infty$, we define $e_\mu$ as
$e_\mu={\mathfrak{e}}_{(r(\mu))}$
where $r$ and $\mu$ are related through \eqref{EA1.1.r}.
Then, we identify
$$
\omega\ni w=(w_1,w_2,w_3,w_4,\cdots)
=\sum_{r=1}^{\infty}w_r{\mathfrak{e}}_{r}
\longleftrightarrow
(w_{(1)},w_{(2)},w_{(1,2)},w_{(3)},\cdots)=\sum_{\mu\in{\mathcal{M}}}w_\mu e_\mu.
$$

Now, we introduce the multiplication by
\begin{equation}
\left\{
\begin{aligned}
&e_{\mu} e_{\emptyset}=
e_{\emptyset}e_{\mu} =e_{\mu} \for \mu\in{\mathcal M}_\infty, \\
&e_{(i)} e_{(j)}=-e_{(j)} e_{(i)} \for i,j\in {\mathbb N},\\
&e_\mu=e_{(\mu_1)}e_{(\mu_2)}\cdots e_{(\mu_k)}\where \mu=(\mu_1,\mu_2,\cdots,\mu_k).
\end{aligned}
\right.
\end{equation}
Then,  putting $\sigma_{j}=e_{(j)}$ for $j\ge1$, we have a family of Grassmann generators 
$\{\sigma_j\}_{j=1}^{\infty}$.

\subsection{Superalgebra and Superspace}
Regarding the above constructed one as an example,
we prepare countable number of letters $\{\sigma_j\}_{j=1}^{\infty}$ 
with multiplication and addition satisfying
\begin{equation}
\sigma_i\sigma_j+\sigma_j\sigma_i=0\for i,j=1,2,{\cdots}.
\label{G-R}
\end{equation}
Denoting these letters as Grassmann generators, we put formally
$$
{\fC}=\{X=\sum_{{\bf{I}}\in{\mathcal{I}}} X_{\bf{I}}\sigma^{\bf{I}}\;\big|\; X_{\bf{I}}\in{\mathbb{C}}\}
$$
where
$$
\begin{gathered}
{\mathcal{I}}=\{{\bf{I}}=(i_k)\in \{0,1\}^{\mathbb{N}}\;\big|\;
|{\bf{I}}|=\sum_k i_k < \infty \},\\
\with \sigma^{\bf{I}}=\sigma_1^{i_1}\sigma_2^{i_2}\cdots,\quad
\sigma^{\tilde 0}=1, \quad {\bf{I}}=(i_1,i_2,\cdots ),\quad {\tilde 0}=(0,0,\cdots )\in{\mathcal{I}}.
\end{gathered}
$$
For the notational simplicity, we put also
$$
{\mathcal{I}}_{\mathrm{ev}}=\{{\bf{I}}\in{\mathcal{I}}\;|\; |{\bf{I}}|={\mathrm{ev}}\},\quad
{\mathcal{I}}_{\mathrm{od}}=\{{\bf{I}}\in{\mathcal{I}}\;|\; |{\bf{I}}|={\mathrm{od}}\}.
$$

Besides trivially defined linear operations of sums and scalar
multiplications, we have a product operation in ${\fC}$: 
For 
$$
X=\sum_{{\bf{J}}\in\mathcal{I}} X_{\bf{J}}\sigma^{\bf{J}}, \quad 
Y=\sum_{{\bf{K}}\in{\mathcal{I}}} Y_{\bf{K}}\sigma^{\bf{K}},
$$ 
we put
\begin{equation}
XY=\sum_{{\bf{I}}\in{\mathcal{I}}} (XY)_{\bf{I}}\sigma^{\bf{I}}\with
(XY)_{\bf{I}}=\sum_{{\bf{I}}={\bf{J}}+{\bf{K}}}
(-1)^{\tau({\bf{I}};{\bf{J}},{\bf{K}})}X_{\bf{J}}Y_{\bf{K}}.
\label{multiplic}
\end{equation}
Here, $\tau({\bf{I}};{\bf{J}},{\bf{K}})$ is an integer defined by
\begin{equation}
\sigma^{\bf{J}}\sigma^{\bf{K}} 
=(-1)^{\tau({\bf{I}};{\bf{J}},{\bf{K}})}\sigma^{\bf{I}}, \quad
{\bf{I}}={\bf{J}}+{\bf{K}},
\label{ncg}
\end{equation}
which is not necessary  to specify more concretely.

Identifying ${\fC}$ with the sequence space $\omega$ of K\"othe~\cite{kot}, we have
\begin{prop}
${\fC}$ forms an $\infty$-dimensional FG-algebra over ${\mathbb{C}}$,
that is, an associative, distributive and non-commutative ring with degree,
which is endowed with the Fr\'echet topology.
\end{prop}

\begin{remark}
For the proof, see, Inoue and Maeda~\cite{IM91}. By the way, 
though DeWitt himself wrote in p.3 of his book \cite{DW2} ``In the formal limit $N\to{\infty}$, they (i.e. $\Lambda_N={\mathfrak{B}}_N$ in \S 3  etc.) may continue to be regarded as vector spaces, but we shall not give them a norm or even a topology'',
but Pestov wrote in p.278 of \cite{pest91} 
``The DeWitt supernumber algebra $\Lambda_{\infty}$ was implicitly topologized, in fact, by DeWitt himself''.
By the way, we hesitate to use non-Hausdorff topology, in general, to construct the elementary analysis, that is the reason for our introduction of supernumbers above.
\end{remark} 

\begin{remark}
Degree in ${\fC}$ is defined by introducing subspaces
$$
{\fC}^{[j]}=\{X=\sum_{{\bf{I}}\in{\mathcal{I}},|{\bf{I}}|=j} X_{\bf{I}}\sigma^{\bf{I}}\} \for j=0,1,\cdots
$$
which satisfy
$$
{\fC}=\oplus_{j=0}^\infty {\mathfrak{C}}^{[j]},\quad
{\fC}^{[j]}\cdot {\fC}^{[k]} \subset {\fC}^{[j+k]}.
$$
\end{remark}

\begin{remark}
 Define 
$$
\proj_{\bf{I}}(X)=X_{\bf{I}} \for 
X=\sum_{{\bf{I}}\in{\mathcal{I}}}X_{\bf{I}}\sigma^{\bf{I}}\in{\fC}.
$$
The topology in ${\fC}$ is given by
$X\to 0$ in ${\fC}$ if and only if $\proj_{\bf{I}}(X)\to 0$ in ${\fC}$
for any ${\bf{I}}\in{\mathcal{I}}$.
\par
This topology is equivalent to the one introduced by the metric
$\dist(X,Y)=\dist(X-Y)$ where $\dist(X)$ is defined by
$$
\dist(X)=\sum_{{\bf{I}}\in{\mathcal{I}}}\frac 1{2^{r({\bf{I}})}} \frac
{|{\proj}_{\bf{I}}(X)|}{1+|{\proj}_{\bf{I}}(X)|} 
\with
r({\bf{I}})=1+\frac 12 \sum_{k=1}^\infty 2^k i_k \for {\bf{I}}\in{\mathcal{I}}.
$$
\par
To distinguish this topology, we show the following:
Even if $a_{\ell}\to{\infty}$ $({\ell}\to{\infty})$, we have
$X^{({\ell})}=a_{\ell}\sigma_1{\cdots}\sigma_{\ell}\to 0$ in ${\fC}$.
In fact, putting ${\bf{I}}^{({\ell})}=(\overbrace{1,{\cdots},1}^{\ell},0,{\cdots})\in{\mathcal{I}}$,
$r({\bf{I}}^{({\ell})})={2^{\ell}}$ and
$$
\dist(X^{({\ell})})=2^{-\ell}\frac{|a_{\ell}|}{1+|a_{\ell}|}\le 2^{-\ell}\;\;\mbox{or}\;\;
\mbox{for each ${\bf{I}}\in{\mathcal{I}}$, $\lim_{\ell\to{\infty}}{\proj}_{\bf{I}}(X^{({\ell})})=0$}.
$$
\end{remark}

\begin{remark}
We introduce parity in ${\fC}$ by setting
$$
p(X)=\begin{cases}
\bar{0} & \text{if $X=\sum_{{\bf{I}}\in{\mathcal{I}}_{\mathrm{ev}}}X_{\bf{I}}\sigma^{\bf{I}}$},\\
\bar{1} & \text{if $X=\sum_{{\bf{I}}\in{\mathcal{I}}_{\mathrm{od}}}X_{\bf{I}}\sigma^{\bf{I}}$},\\
\text{undefined} & \text{if otherwise}.
\end{cases}
$$
We put 
$$
\left\{
\begin{aligned}
&{\cev} = \oplus_{j=0}^\infty {\fC}^{[2j]}
=\{X\in{\fC}\,|\,p(X)={\mathrm{ev}}\},\\
&{\cod} = \oplus_{j=0}^\infty {\fC}^{[2j+1]}
=\{X\in{\fC}\,|\,p(X)={\mathrm{od}}\},\\
&{\fC} \cong {\cev}\oplus{\cod}\cong{\cev}\times{\cod}.
\end{aligned}
\right.
$$
We introduced the body (projection) map $\pi_{\mathrm{B}}$ by 
$$
\pi_{\mathrm{B}}X=\proj_{\tilde 0}(X)=X_{\tilde 0}=X^{[0]}=X_{\mathrm{B}}
\forany X\in\fC,
$$
and the soul part $X_{\mathrm{S}}$ of $X$ as
$$
X_{\mathrm{S}}=X-X_{\mathrm{B}}=\sum_{|{\bf{I}}|\ge 1}X_{\bf{I}}\sigma^{\bf{I}}.
$$
\end{remark}

Analogous to ${\fC}$, we define, as an alternative of ${\euc}$,
$$
\left\{
\begin{aligned}
&{\fR}=\{X\in{\fC}\,|\, \pi_{\mathrm{B}}X\in\mathbb{R}\},\;
{\fR}^{[j]}={\fR}\cap {\fC}^{[j]},\\
&{\rev}={\fR}\cap{\cev},\quad {\rod}={\fR}\cap{\cod}={\cod},\\
&{\fR}\cong {\rev}\oplus{\rod}\cong{\rev}\times{\rod}.
\end{aligned}
\right.
$$
We define the (real) superspace ${\fR}^{m|n}$ by
$$
{\fR}^{m|n}={{\fR}_{\mathrm{ev}}^m} \times {\rod^n}.
$$
The distance between $X,Y\in{\fR}^{m|n}$ is defined by, 
$$
\dist_{m|n}(X,Y)=\dist_{m|n}(X-Y)
$$ 
where
$$
\dist_{m|n}(X)=
\sum_{j=1}^m
\left(\sum_{{\bf{I}}\in{\mathcal{I}}_{\mathrm{ev}}} {\frac{1}{2^{r({\bf{I}})}}}
{\frac{|\proj_{\bf{I}}(x_j)|}{1+|\proj_{\bf{I}}(x_j)|}}
\right)
+\sum_{k=1}^n
\left(
\sum_{{\bf{I}}\in{\mathcal{I}}_{\mathrm{od}}}
{\frac {1}{2^{r({\bf{I}})}}}
{\frac{|\proj_{\bf{I}}(\theta_k)|}{1+|\proj_{\bf{I}}(\theta_k)|}}
\right). 
$$
We use the following notation:
$$
\begin{gathered}
X=(X_A)_{A=1}^{m+n}=(x,\theta)\in{\fR}^{m|n}  \with\\
 x=(X_A)_{A=1}^{m}=(x_j)_{j=1}^m\in\supermo,\quad
\theta=(X_A)_{A=m+1}^{m+n}=(\theta_k)_{k=1}^n\in\superon.
\end{gathered}
$$
We generalize the body map $\pi_{\mathrm{B}}$ from ${\fR}^{m|n}$ 
or $\supermo$ to ${\euc}$ by putting,
$$
{\fR}^{m|n} \ni X=(x,\theta)\longrightarrow
\pi_{\mathrm{B}}X=X_{\mathrm{B}}=(x_{\mathrm{B}},0)
\cong x_{\mathrm{B}}
=\pi_{\mathrm{B}}x
=(\pi_{\mathrm{B}}x_1,\cdots, \pi_{\mathrm{B}}x_m)\in{\euc}^m.
$$
We call $x_j\in\rev$ and
$\theta_k\in\rod$ as
even and odd (alias bosonic and fermionic) variable, respectively.

\subsection{G\^ateaux or Fr\'echet differentiable functions on Fr\'echet spaces}\label{GFonF}
\subsubsection{G\^ateaux-differentiability}
\begin{defn}[G\^ateaux-derivative, -differential and -differentiability]
(i) Let ${X}$, ${Y}$ be Fr\'echet spaces with countable seminorms $\{p_m\}$, $\{q_n\}$, respectively.
Let $U$ be an open subset of ${X}$. For a function $f:U{\to} {Y}$,
we say that $f$ is {1-time G\^ateaux (or $G$-)differentiable} at $x\in U$ in the direction $y{\in} X$ 
if there exists the following limit in ${Y}$:
$$
\lim_{t\to0}\frac{f(x+ty)-f(x)}{t}=\frac{df(x+ty)}{dt}\bigg|_{t=0}=d_Gf(x;y)=d_Gf(x)\{y\}=d_Gf(x)y=f'_G(x)y,
$$
i.e., for given $x\in U$ and $y\in {X}$  there exists an element $d_Gf(x;y)\in Y$ such that
for any $n\in{\mathbb{N}}$, we have
$$
q_n({f(x+ty)-f(x)}-{t}d_Gf(x;y))=o(t).
$$
We call this $d_Gf(x;y)$ the $G$-differential of $f$ at $x$ in the direction $y$ and denoted as above,
and $d_Gf(x)$ or $f_G'(x)$ are called  the {$G$-derivative}.
Moreover,  $f$ is said to be $G$-differentiable in $U$ and denoted by $f\in C_G^{1-}(U:{Y})$ if $f$ has
the $G$-differential $d_Gf(x;y)$ for every $x\in U$ and any direction $y\in {X}$. 
A map $f:U\to {Y}$ is said to be 1-time continuously $G$-differentiable on $U$, denoted by $f\in C_G^1(U:{Y})$, if $f$ has $G$-derivative in $U$ and
if $d_Gf: U\times {X}\ni(x,y)\to d_Gf(x;y)\in{Y}$ is jointly continuous.\\
(ii) If ${X}$, ${Y}$ are Banach spaces with norms $\Vert{\cdot}\Vert_X$, 
$\Vert{\cdot}\Vert_Y$, respectively,
then, $f$ has $G$-differential
$df(x;y)\in{Y}$ at $x\in U$ in the direction $y\in {X}$ {if and only if} 
$$
\Vert f(x+ty)-f(x)- t d_Gf(x;y)\Vert_Y=o(|t|)\;\;\mbox{as $t\to0$}.
$$
Moreover,  $f\in C_G^1(U:Y)$ {if and only if} $f$ is $G$-differentiable at $x$ 
and $d_Gf$ is continuous from $U\ni x$ to $d_Gf(x)\in{\bf{L}}(X:Y)$.
\end{defn}
\begin{prop}[pp.76-77 of \cite{hamil}]
Let ${X}$, ${Y}$ be Fr\'echet spaces and let $U$ be an open subset of ${X}$.
If $f\in C_G^1(U:{Y})$, then $d_Gf(x;y)$ is linear in $y$.
\end{prop}

\begin{remark}[see, p.70 of \cite{hamil}]
It should be remarked that even if $X, Y, Z$ are Banach spaces, 
there exists the difference between 
$$
``L:(X\supset U)\times Y\to Z\;\mbox{is continuous}"\et
``L: X\supset U\to {\bf{L}}(Y:Z)\;\mbox{is continuous}".
$$
\end{remark}
\begin{defn}[Higher order derivatives, p.80 of \cite{hamil}]
Let ${X}$, ${Y}$ be Fr\'echet spaces.\\
(i) If the following limit exists, we put
$$
d^2_Gf(x)\{y,z\}=d^2_Gf(x;y,z)=\lim_{t\to0}\frac{d_Gf(x+tz;y)-d_Gf(x;y)}{t}.
$$
Moreover, $f$ is said to be  $C_G^2(U:{Y})$ if $d_Gf$ is $C_G^1(U\times {X}:{Y})$, which happens {if and only if} $d^2_Gf$ exists and is continuous, that is,
$d^2_Gf$ is jointly continuous from $U\times{X}\times{X}\to {Y}$.\\
(ii) Analogously, we define
$$ 
d^n_Gf:U\times \overbrace{{X}\times{\cdots}\times{X}}^n\ni (x,y_1,{\cdots}, y_n)
\to d^n_Gf(x)\{y_1,{\cdots},y_n\}=d^n_Gf(x;y_1,{\cdots},y_n)\in {Y}.
$$
$f$ is  said to be $C_G^n(U:{Y})$ {if and only if} $d^n_Gf$ exists and is continuous. We put
$C_G^{\infty}(U:{Y})=\cap _{n=0}^{\infty}C_G^n(U:{Y})$.
\end{defn}
\begin{defn}[Many variables case]
(i) Let ${X}_1$, ${X}_2$, ${Y}$ be Fr\'echet spaces.
For  $x=(x_1,x_2)\in X_1\times X_2$ and $z=(z_1,z_2)\in X_1\times X_2$, we put
$$
\begin{gathered}
\partial_{x_1} f(x)\{z_1\}=f_{x_1}(x;z_1)=f_{x_1}(x)z_1=\lim_{t\to0}\frac{f(x_1+t z_1,x_2)-f(x_1,x_2)}{t},\\
\partial_{x_2} f(x)\{z_2\}=f_{x_2}(x;z_2)=f_{x_2}(x)z_2=\lim_{t\to0}\frac{f(x_1,x_2+t z_2)-f(x_1,x_2)}{t}.
\end{gathered}
$$
They are called partial derivatives.
We define the total $G$-derivative as
$$
d_Gf(x)\{z\}=f'_{x}(x;z)=\lim_{t\to0}\frac{f(x_1+tz_1,x_2+tz_2)-f(x_1,x_2)}{t}.
$$
For $f:X\to Y$ with $X=\prod_{i=1}^n X_i$, we define $\partial_{x_j}f(x)$ and $d_Gf(x)$ for $x=(x_1,{\cdots},x_n)$, analogously.\\
(ii) If ${X}_1$, ${X}_2$, ${Y}$ are Banach spaces, we may define analogously the above notion.
\end{defn}

\begin{prop}
Let $\{X_i\}_{i=1}^{n}$, $Y$ be Fr\'echet spaces and let $U$ be an open set in $X=\prod_{i=1}^{n}X_i$. \\
(a)  $f\in C_G^{1}(U:{Y})$, i.e. $d_Gf(x)\{y\}$ exists and is continuous, 
if and only if $\partial_{x_j}f(x)\{{\cdot}\}$ exist and are continuous, and we have
\begin{equation}
d_Gf(x;y)=d_Gf(x)\{y\}=\sum_{i=1}^{n}f_{x_i}(x;y_i)=\sum_{i=1}^{n}f_{x_i}(x)\{y_i\}\with x=(x_i)_{i=1}^{n},\; y=(y_i)_{i=1}^{n}\in X.
\label{total}
\end{equation}
(b)[Taylor's formula]  Moreover, if $f\in C_G^{p}(U:{Y})$, we have
\begin{equation}
f(x+y)=\sum_{k=0}^p\frac{1}{k!} d_G^kf(x)\{\overbrace{y,{\cdots},y}^k\}+R_pf(x,y)\with
\lim_{t\to0}t^{-p}R_pf(x,ty)=0 \for y\in{X}.
\label{tay-f}
\end{equation}
$$
R_{p}f(x,y)=\int_0^1\frac{(1-s)^{p-1}}{(p-1)!}\frac{d^{p}}{ds^{p}}f(x+sy) ds.
$$
\end{prop}
\par{\it Proof}.
\eqref{total} is proved in Theorem 3.4.3 of \cite{hamil} for $N=2$.
\eqref{tay-f} is given, for example, in p.101 of Keller \cite{keller}, et al.\qed

\subsubsection{Fr\'echet-differentiability}
\begin{defn}[Definition 1.8. of Schwartz \cite{schw69}]
(i) Let ${X}$, ${Y}$ be Fr\'echet spaces, and let $U$ be an open subset of ${X}$. 
A function $\varphi:U\to {Y}$ is said to be
{horizontal (or tangential) at $0$} {if and only if} for each neighbourhood $V$ of $0$ in
${Y}$ there exists a neighbourhood $U'$ of $0$ in ${X}$, and a
function $o(t):(-1,1)\to{\euc}$ such that
\begin{equation}
\varphi(tU')\subset o(t)V \with \lim_{t\to0}\frac{o(t)}{t}=0,
\label{FD11}
\end{equation}
i.e. for any seminorm $q_n$ on ${Y}$ and $\epsilon>0$, there exists a seminorm $p_m$ on $E$ and $\delta>0$ such that 
\begin{equation}
q_n(\varphi(tx))\le \epsilon t \for
p_m(x)<1,\;\;|t|\le \delta
\label{FD12}
\end{equation}
From \eqref{FD12}, putting
$V=\{z\in {Y}\;|\; q_n(z)<1\},\; U'=\{x\in {X}\;|\; p_m(x)<1\}$,
we may recover \eqref{FD11}.\\
(ii) For given Banach spaces $(X,\Vert{\cdot}\Vert_X)$ 
and $(Y,\Vert{\cdot}\Vert_Y)$, ``horizontal" implies
$$
\Vert\varphi(x)\Vert_Y\le\Vert x\Vert_X\psi(x)\with \psi:X\to\mathbb{R},\;\; \lim_{x\to0}\psi(x)=0
\quad \mbox{i.e. $\Vert\varphi(x)\Vert_Y=o(\Vert x\Vert_X)$ as $\Vert x\Vert_X\to0$.}
$$
\end{defn}
\begin{defn}[Fr\'echet differentiability] 
(i)(Definition 1.9. of \cite{schw69}) Let ${X}$, ${Y}$ be Fr\'echet spaces with $U$ being an open subset of ${X}$. 
We say that $f$ has a Fr\'echet (or is $F$-)derivative (or $f$ is $F$-differentiable) at $x\in U$,  
{if} there exists a continuous
linear map $A=A_x:{X}\to {Y}$ such that $\varphi(x;y)$ is horizontal w.r.t $y$ at $0$, where $\varphi(x;y)$ is defined by
$$
\varphi(x;y)=f(x+y)-f(x)-A_xy.
$$
We call $A=A_x$ the $F$-derivative of $f$ at $x$, and we denote
$A_xy$ as $d_Ff(x;y)$. 
Moreover, we denote $f\in C_F^1(U:Y)$ if $f$ is $F$-differentiable and 
$d_F f : U{\times} X{\ni} (x,y){\to} d_F f(x;y){\in}Y$ is jointly continuous. \\
(ii) For Banach spaces, $f$ is $F$-differentiable at $x$ {if} there exists a continuous
linear map $A=A_x:{X}\to {Y}$ satisfying
$$
\Vert f(x+y)-f(x)-Ay\Vert_Y=o(\Vert y\Vert_X)\;\;\mbox{as $\Vert y\Vert_X\to0$}.
$$
Moreover, $f\in C_F^1(U:Y)$ if $f$ is $F$-differentiable and $X\ni x\to A_x\in {\bf{L}}(X:Y)$ is continuous.
\end{defn}
\begin{remark}
If $f$ is $F$-differentiable, then it is also $G$-differentiable. Moreover, 
$$
f'_G(x;y)=d_Gf(x;y)=d_Ff(x;y)=f'_F(x;y).
$$
\end{remark}
\begin{defn}[Higher order derivatives]
(i) Let ${X}$, ${Y}$ be Fr\'echet spaces with $U$ being an open subset of ${X}$. 
A $F$-differentiable function $f:U\to Y$ is twice $F$-differentiable at $x\in U$ 
if  $d_Ff:U\times X\ni (x,y)\to d_Ff(x;y)\in Y$ is $F$-differentiable at $x\in X$. That is, the function
$$
\psi(x;y,z)=d_Ff(x+z;y)-d_Ff(x;y)-d_F^2f(x)\{y,z\},
$$
is horizontal w.r.t. $z$ at $0$.\\
(ii) (p.72 of \cite{berg77}) Let $X$, $Y$ be Banach spaces.
A $F$-differentiable function $f:U\to Y$ is twice $F$-differentiable at $x\in U$ 
if $f_F':X\to {\bf{L}}(X:Y)$ is $F$-differentiable at $x\in X$ and
$f_F''(x)$, the derivative of $f_F'(x)$, belongs to ${\bf{L}}(X: {\bf{L}}(X:Y))= {\bf{L}}(X\times X:Y)$.
$f\in C^2(U:Y)$ if (a) $f$ is twicely $F$-differentiable for each $x\in U$ 
and (b) $f_F''(x):U\to  {\bf{L}}(X\times X:Y)$ is continuous.\\
(iii) Analogously  $N$-times $F$-differentiability is defined.
\end{defn}
\begin{defn}[Many variables case]
(i) Let $U=\prod_{i=1}^N U_i$ with each $U_i$ being an open subset of Fr\'echet spaces $X_i$. 
For $x=(x_1,{\cdots}, x_N)\in U$ with $x_i\in U_i$ and $h_i\in X_i$ s.t. $x_i+h_i\in U_i$, if there exists $F_i(x;h_i)\in Y$ such that
$$
\varphi_i(x;h_i)=f(x_1,{\cdots}, x_{i-1},x_i+h,x_{i+1},{\cdots} x_N)-f(x_1,{\cdots}, x_{i-1},x_i,x_{i+1},{\cdots} x_N)-F_i(x;h_i)
$$
is horizontal w.r.t. $h_i$. We denote $F_i(x;h_i)$ as $\partial_{x_i}f(x)h_i$, the partial derivative of $f$ w.r.t. $x_i$.\\
(ii)(p.69 of \cite{berg77})
 In case $X_i$ are Banach spaces,
the partial derivative of $f$ w.r.t. $x_i$, $\partial_{x_i}f(x)$, is defined by
$$
f(x_1,{\cdots}, x_{i-1},x_i+h_i,x_{i+1},{\cdots} x_N)-f(x_1,{\cdots}, x_{i-1},x_i,x_{i+1},{\cdots} x_N)=\partial_{x_i}f(x)h_i+o(\Vert h_i\Vert).
$$
More generally, for each $x$ if there exists a continuous linear map $d_F f:X\ni h\to d_Ff(x;h)\in Y$ such that
$$
\Vert f(x+h)-f(x)-d_Ff(x;h)\Vert_Y=o(\Vert h\Vert_X) \for h\in X.
$$
We denote also $d_Ff(x;h)=f'_F(x)\{h\}$ with $f'_F(x)\in  {\bf{L}}(X:Y)$.
Moreover, there exist operators $\partial_{x_i}f(x)\in  {\bf{L}}(X_i:Y)$ such that
\begin{equation}
f'_F(x;h)=f'_F(x)\{h\}=\sum_{i=1}^N\partial_{x_i}f(x)\{h_i\}=\sum_{i=1}^N\partial_{x_i}f(x;h_i)\with h=(h_1,{\cdots}, h_N).
\label{FD11-1}
\end{equation}
\end{defn}

\subsection{A definition of supersmooth functions}
\begin{lem}[Proposition 2.2 of \cite{IM91}]\label{GC-lemma}
(i) For $f\in C^\infty({\euc}:{\mathbb{C}})$, we may define 
\begin{equation}
\tilde{f}(x)=\sum_{|\alpha|=0}^\infty{\frac 1{\alpha!}}
\partial_q^\alpha f(q)\big|_{q=x_{\mathrm{B}}} x_{\mathrm{S}}^\alpha
\quad\text{for \; $x=x_{\mathrm{B}}+x_{\mathrm{S}}\in{\fR}^{m|0}$}
\label{GC1}
\end{equation}
which is called the Grassmann continuation of $f$, denoted by $\tilde{f}\in \ccsl_{S\!S}({\fR}^{m|0})$.\\
(ii) If $f\in C^\infty({\euc}:{\fC})$, i.e. $f(q)=\sum f_{\bf{I}}(q)\sigma^{\bf{I}}$ with $f_{\tilde0}\in C^\infty({\euc}:{\euc})$ and
$f_{\bf{I}}\in C^\infty({\euc}:{\mathbb{C}})$, 
we may define analogously
$\tilde{f}(x)=\sum \tilde{f}_{\bf{I}}(x)\sigma^{\bf{I}}$, which is denoted by $\tilde{f} \in \CSS({\fR}^{m|0})$.
\end{lem}

\begin{remark}
(i) To fix our idea, we re-prove this lemma in \S 4, where the weak topology in ${\fC}$ is crucial.\\
(ii) Because of this definition, not only functions in $C^{\infty}({\euc}:{\mathbb{C}})$ but also those in 
${\mathcal{D}}'({\euc}:{\mathbb{C}})$, we may define the Grassmann continuation 
by using the duality between the test sequence space $c_0$ and the space $\omega$.
This generalization is necessary to introduce Sobolev spaces on superspace, but this point is not discussed here.
\end{remark}

\begin{defn} [Supersmooth functions on FG-algebra]
We define a set of functions 
$$
\ccsl_{S\!S}({\fR}^{m|n})=\big\{u(X)=u(x,\theta)=\sum_{|a|\le n}\theta^a \tilde{u}_a(x)\;|\; 
u_a(q)=\partial_\theta^a u(q,0)\in C^{\infty}({\euc}^m:{\mathbb{C}})\big\}.
$$
An element $u\in\ccsl_{S\!S}({\fR}^{m|n})$ is called a supersmooth function on ${\fR}^{m|n}$.
\par
Analogously, we put
$$
\CSS({\fR}^{m|n}:{\fC})=
\big\{u(X)=\sum_{|a|\le n}\theta^a \tilde{u}_a(x) \big|\,
u_a(q)=\partial_\theta^a u(q,0)\in  C^\infty({\euc}^m:{\fC})
\;\;\mbox{for any}\;\;
a\big\}.
$$
\end{defn}
\begin{remark}
Since we prefer to use the left derivative w.r.t. odd variables, we use the representation $\theta^a \tilde{u}_a(x)$ but not $\tilde{u}_a(x)\theta^a $. 
\end{remark}

\subsection{How to characterize supersmooth functions?}
Though we introduce  supersmooth functions as a polynomial of odd variables with a  special class of coefficient functions, 
we have
\begin{problem}
How do we characterize a supersmooth function 
$u(X)\in\ccsl_{S\!S}({\fR}^{m|n})$ defined above?\\
(1) Is it possible to say that G\^ateaux infinitely differentiability with super $F$-differentiability is necessary and sufficient 
for supersmoothness? Here, a function $u:{\fR}^{m|n}\to{\fC}$ is said to be super $F$-differentiable,
if for any $X=(X_A)=(x,\theta), Y=(Y_A)=(y,\omega)\in {\fR}^{m|n}$, 
there exist $(\gamma_A(X))=(\gamma_j(X), \gamma_{m+k}(X))\in{\fC}^{m+n}$ such that
$$
\Phi(X;Y)=u(X+Y)-u(X)-\sum_{j=1}^m y_j\gamma_j(X)-\sum_{k=1}^n\omega_k\gamma_{m+k}(X)
$$
is ``horizontal'' w.r.t. $Y$.\\
(2) How does the Cauchy-Riemann type equation relate to super $F$- or $G$-differentiability?
\end{problem}

\begin{remark}
This problem itself is posed implicitly by Inahama, one of author's colleague at T.I.T. at that time. Especially when we construct a parametrix of Pauli equation in superanalysis category~\cite{IM02}, we solve the Hamilton equation corresponding to the Hamilton function derived from Pauli equation. There, we don't mention clearly the solution obtained by what explained is supersmooth. In Appendix A, we prove that for given supersmooth initial data, the solutions obtained there stay in supersmooth category.
\end{remark}

\begin{remark}
(i) For example,  Jadczyk  and Pilch~\cite{JaPi81} prove that
the Fr\'echet infinitely differentiability with
its differential being ``$Q_{\mathrm{ev}}$''- linear is necessary and sufficient for $G^{\infty}$-superdifferentiability of Rogers, in BG-algebra category
satisfying their conditions, called self-duality.\\
(ii) In order to make clear our point, we recall what is well-known for analytic functions:
\end{remark}
{\begin{comparison}
Let a function $f(z)$ from ${\mathbb{C}}$ to ${\mathbb{C}}$ be given, which is decomposed as
$$
f(z)=u(x,y)+iv(x,y),\quad u(x,y)=\Re{f(z)}\in{\mathbb R},\; v(x,y)=\Im{f(z)}\in{\mathbb R},
$$
where $z=x+iy$, $|z|=\sqrt{x^2+y^2}$ with $z_0=x_0+iy_0$.
\par
$\bullet$ $f$ is said to be F-differentiable in ${\mathbb{C}}$ at $z=z_0$ if the following limit exists in ${\mathbb{C}}$;
\begin{equation}
\lim_{w\to0}\frac{f(z_0+w)-f(z_0)}{w}=\gamma\in{\mathbb{C}}.
\end{equation}
In other word, there exists a number $\gamma\in{\mathbb{C}}\equiv {\bf{L}}({\mathbb{C}}:{\mathbb{C}})$
such that
\begin{equation}
|f(z_0+w)-f(z_0)-\gamma w|=o(|w|)\qquad (|w|\to0).
\label{c-bibun}
\end{equation}
\par
$\bullet$ $f(z)$ is called analytic in $D=\{z\in{\mathbb{C}}\;|\; |z-z_0|<R\}$ if one of the following conditions is satisfied:
\par (a) For any $z\in D$, $f(z)$ is differentiable in the above sense \eqref{c-bibun}.
\par (b) Identifying $D\subset{\mathbb{C}}$ with $\tilde{D}=\{(x,y)\in {\euc}^2\;|\; |x-x_0|^2+|y-y_0|^2<R^2\}$, we have\\
\hspace{8mm}
(b-1) $u,v\in C^1(\tilde{D}:\mathbb{R}^2)$ and $df(x,y)$ is not only linear w.r.t. ${\euc}$ but also linear w.r.t. ${\mathbb{C}}$,\\
\hspace{8mm}
or\\
\hspace{8mm}
(b-2) $u,v\in C^1(\tilde{D}:{\euc}^2)$ and $u,v$ satisfies Cauchy-Riemann type  equation.
\par (c) $f(z)$ has the convergent power series expansion $f(z)=\sum_{n=0}^{\infty} a_n(z-z_0)^n$ for $z\in D$.\\
\end{comparison}}
\begin{remark}[The meaning of (b-1) and (b-2)]
For $f:{\mathbb{C}}\to {\mathbb{C}}$ given above, we define a map
$$
\Phi:{\euc}^2\ni \binom{x}{y}\to\binom{u(x,y)}{v(x,y)}\in{\euc}^2.
$$
Since $u,v\in C^1(\tilde{D}:{\euc}^2)$,
$\Phi$ is F-differentiable at $({x_0},{y_0})$, that is,  there exists 
$\Phi'_F({x_0},{y_0})\in {\bf{L}}({\euc}^2:{\euc}^2)$ such that
$$
\Big\Vert\Phi({x_0+h},{y_0+k})-\Phi({x_0},{y_0})-\Phi'_F({x_0},{y_0})\binom{h}{k}\Big\Vert=o(\Big\Vert\binom{h}{k}\Big\Vert).
$$
Here, we have
\begin{equation}
\Phi'_F(x,y)=\begin{pmatrix}
u_x(x,y)&u_y(x,y)\\
v_x(x,y)&v_y(x,y)
\end{pmatrix},
\label{mult}
\end{equation}
and if we require 
$\Phi'_F({x_0},{y_0})\in {\bf{L}}({\euc}^2:{\euc}^2)$ is not only ${\euc}$-linear but also ${\mathbb{C}}$-linear, 
that is, for any $a, b\in{\euc}$, especially for $b\neq0$,
$$
\Phi'_F({x_0},{y_0})
\bigg(\begin{pmatrix}
a&-b\\
b&a
\end{pmatrix}\binom{h}{k}\bigg)
=
\begin{pmatrix}
a&-b\\
b&a\end{pmatrix}
\Phi'_F({x_0},{y_0})
\binom{h}{k}\forany
\binom{h}{k}\in{\euc}^2,
$$
holds, we need $u_x(x,y)=v_y(x,y)$ and $u_y(x,y)=-v_x(x,y)$.
\end{remark}

\begin{remark}
In order to prove the equivalence to (a) to (c), one uses the Cauchy's integral representation, in general.
But to prove the equivalence of (b) and (c) without integral representation, it seems useful to recall the notion of Pringsheim regularity as follows, see Pestov~\cite{pest93}: 
\begin{quotation}
A function $f$ is said to be Pringsheim regular in $U$ if
the Taylor series of $f$ converges in a neighborhood of every point of $x\in U$ (though not necessarily to the function $f$ itself).
\end{quotation}
\end{remark}

\section{The case with finite Grassmann generators}

To clarify the problem, we first gather results when the number ${\nogg}$ of Grassmann generators is finite,
i.e. the special case of Banach-Grassmann algebra,
which are mainly treated by Vladimirov and Volovich \cite{VV83} and 
Boyer and Gitler~\cite{BG84},
though the terminology is slightly modified from them.

\subsection{The Grassmann algebras with finite Grassmann generators}
Preparing Grassmann generators $\{\sigma_j\}_{j=1}^{\nogg}$ for finite ${\nogg}$, we put
$$
\begin{aligned}
&{\mathfrak{B}}_{\nogg}
=\{X=\sum_{{\bf{I}}\in{\mathcal{I}}_{\nogg}}X_{\bf{I}}\sigma^{\bf{I}}\;|\; X_{\bf{I}}\in{\euc}\et \Vert X\Vert=\sum_{{\bf{I}}\in{\mathcal{I}}_{\nogg}}|X_{\bf{I}}|<{\infty}\},\\
&{\blev}
=\{X=\sum_{|{\bf{I}}|=\mathrm{ev}, {\bf{I}}\in{\mathcal{I}}_{\nogg}}X_{\bf{I}}\sigma^{\bf{I}}\;|\; X_{\bf{I}}\in{\euc}\et \Vert X\Vert=\sum_{|{\bf{I}}|=\mathrm{ev},{\bf{I}}\in{\mathcal{I}}_{\nogg}}|X_{\bf{I}}|<{\infty}\},\\
&{\blod}
=\{X=\sum_{|{\bf{I}}|=\mathrm{od}, {\bf{I}}\in{\mathcal{I}}_{\nogg}}X_{\bf{I}}\sigma^{\bf{I}}\;|\; X_{\bf{I}}\in{\euc}\et \Vert X\Vert=\sum_{|{\bf{I}}|=\mathrm{od},{\bf{I}}\in{\mathcal{I}}_{\nogg}}|X_{\bf{I}}|<{\infty}\},\\
&\quad \where
{\mathcal{I}}_{\nogg}=\{{\bf{I}}=(i_1,{\cdots},i_{\nogg})\in \{0,1\}^{L}\},\quad {\mathcal{I}}_{\nogg}\ni {\bf{I}}=(i_1,{\cdots},i_{\nogg})\hookrightarrow (i_1,{\cdots},i_{\nogg},0,{\cdots})\in{\mathcal{I}}.
\end{aligned}
$$

Regarding ${\mathfrak{B}}_{\nogg}$ as a vector space ${\euc}^{2^{\nogg}}$, we introduce its topology as Euclidean space, 
or $\Vert X\Vert=\sum_{{\bf{I}}\in{\mathcal{I}}_{\nogg}} |X_{\bf{I}}|$ for $X\in {\mathfrak{B}}_{\nogg}$.
We define a superspace as
$$
{\mathfrak{B}}_{\nogg}^{m|n}=\blev^m\times \blod^n, 
$$
identified with ${\euc}^{2^{({\nogg}-1)}(m+n)}$ as vector space.

\begin{prop}[Proposition 2.4 of \cite{rog80}]\label{prop2.4rog} 
If $X=(x,\theta)\in {\mathfrak{B}}_{\nogg}^{m|n}$ satisfies
$$
\begin{gathered}
\langle Y|X\rangle=\langle y|x\rangle+\langle \omega|\theta\rangle=0\quad \forany Y=(y,\omega)\in {\mathfrak{B}}_{\nogg}^{m|n}\\
\with
x=(x_1,{\cdots},x_m),\, y=(y_1,{\cdots},y_m)\in\blev^m, \,\theta=(\theta_1,{\cdots},\theta_n), \,\omega=(\omega_1,{\cdots},\omega_n)\in \blod^n,\\
\langle y|x\rangle=\sum_{j=1}^my_jx_j,\quad \langle \omega|\theta\rangle=\sum_{k=1}^n\omega_k\theta_k,
\end{gathered}
$$
then,
$$
\left\{
\begin{aligned}
&\;\;\; (i)\; x_j=0 \qquad\qquad\qquad\qquad\qquad\qquad\qquad\, \mbox{for $j=1,{\cdots},m$},\\
&{\begin{cases}
(ii{\mathrm{-1}})\; \theta_k=\lambda_k \sigma_1{\cdots}\sigma_{\nogg}\;\mbox{with some $\lambda_k\in{\euc}$}
&\mbox{if ${\nogg}$ is finite, for $k=1,{\cdots},n$},\\
(ii{\mathrm{-2}})\;\theta_k=0 &\mbox{if ${\nogg}={\infty}$, for $k=1,{\cdots},n$}.
\end{cases}}
\end{aligned}
\right.
$$
\end{prop}

Though the following property is important and mentioned in \cite{VV83} and also in Khrennikov~\cite{khr99}, 
but their proofs are \underline{not necessarily correct}.
Therefore, we re-prove the following statement:
\begin{prop}[Lemma 2.2 of \cite{VV83}]\label{VVlem2.2}
Suppose that there exist elements $\{A_i\}_{i=1}^{\nogg}\subset {\mathfrak{B}}_{\nogg} $ satisfying
\begin{equation}
\sigma_j A_i+ \sigma_i A_j=0 \forany i,j=1,{\cdots},{\nogg}.
\label{VV2.19}
\end{equation}
Then there exists an element $F\in {\mathfrak{B}}_{\nogg}$ such that $A_i=\sigma_i F$ for $i=1,{\cdots},{\nogg}$.
\end{prop}
 
To prove this, we prepare the following lemma, modified version of Lemma 1.7 in \cite{BG84}:
\begin{lem}\label{bglem} 
If $A$ is any algebra and $I_1,{\cdots},I_n$ are ideals in $A$ satisfying
\begin{equation}\tag{KM}
\cap_{j=1}^{k-1} (I_j+I_k)=\cap_{j=1}^{k-1}I_j+I_k \forany k,
\label{ka-masuda}
\end{equation}
then there exists an exact sequence
\begin{equation}
\cap_{k}I_k\overset{\iota}{\rightarrow}
A\overset{\alpha}{\rightarrow}
\oplus_{k} A/I_k\overset{\beta}{\rightarrow}\oplus_{(k,j)}A/(I_k+I_j).
\label{ka-BG}
\end{equation}
Here $\iota$ is the injection, $\alpha$ is the diagonal followed by the natural projection,
denoting $[a_k]\in A/I_k$, and $[a_k-a_j]\in A/(I_k+I_j)$ where
$(k,j)$ runs over all pairs $n\ge k>j\ge 1$ and we put
$$
\beta([a_1],{\cdots},[a_n])=(\overbrace{[a_2-a_1],[a_3-a_1],[a_3-a_2],{\cdots}\,{\cdots},[a_n-a_1],{\cdots},[a_n-a_{n-1}]}^{n(n-1)/2}).
$$
\end{lem}
\par{\it Proof}.
(i) By the definition of maps, it is obvious that $\mbox{Image of $\alpha$}\subset\mbox{Kernel of $\beta$}$.\\
(ii) We prove the following claim which clearly implies $\mbox{Image of $\alpha$}\supset\mbox{Kernel of $\beta$}$.
\begin{claim}
Let ${\ell}$ be a positive finite integer, then we  may prove the following;
\begin{equation}
a_i=a_j  \!\!\mod (I_i+I_j) \forany 1\le i, j\le {\ell}
\Longrightarrow \exists b\;\;\mbox{such that $b=a_i \!\!\mod I_i$ for $1\le i\le {\ell}$}.
\label{KM-ind}
\end{equation}
\end{claim}
{\it Proof of Claim}: We prove this by induction.\\
(i) In case $(E_2)$,
from $a_1=a_2 \!\!\mod (I_1+I_2)$, there exist  $x\in I_1$ and $y\in I_2$ such that $a_1-a_2=x+y$.
Putting $b=a_1-x=a_2+y$, we have $b=a_i \mod I_i$ for $1\le i\le 2$.\\
(ii) To prove $(E_{k})$, we assume not only $(E_{k-1})$ holds, i.e.
$$
(E_{k-1}): a_i=a_j  \!\!\mod I_i+I_j\;\mbox{for any $1\le i, j\le k-1$}
\Longrightarrow \exists c\;\;\mbox{such that $c=a_i\! \mod I_i$ for $1\le i\le k-1$},
$$
but also  the left hand side of $(E_{k})$ holds. Then, we have, for $1\le i\le k-1$,
$$
c-a_k\overset{\!\!\!\mod I_i}{=}a_i-a_k\overset{\!\!\!\mod  (I_i+I_k)}{=}0\Longrightarrow
c-a_k\in \cap_{i=1}^{k-1} (I_i+I_k).
$$
This implies by \eqref{ka-masuda} that  $c-a_k={\exists}y+{\exists}z$ with $y\in\cap_{i=1}^{k-1} I_i$ and $z\in I_k$.
Putting $b=c-y=a_k+z$, we have
$b=c-y\overset{\!\!\!\mod I_i}{=}a_i-0$ for $1\le i\le k-1$ and $b=a_k+z\overset{\!\!\mod\, I_k}{=}a_k+0$.
This $b$ guarantees that $(E_k)$ holds. Therefore, we may take $k={\ell}$.  \qed

\par{\it Proof of Lemma \ref{VVlem2.2}}.
Putting $I_i=\langle\sigma_i\rangle=\{\sigma_i X\;|\; X\in {\mathfrak{B}}_{\nogg}\}$, we have
$I_i\cap I_j=\langle \sigma_i \sigma_j\rangle$, etc. Then it is clear that these $\{I_i\}$ satisfy \eqref{ka-masuda}.
Taking $i=j$ in \eqref{VV2.19}, we have $\sigma_i A_i=0$ which implies $\exists B_i\in{\mathfrak{B}}_{\nogg}$ such that
$A_i=\sigma_i B_i$. Putting this into \eqref{VV2.19}, we have $\sigma_i\sigma_j(B_j-B_i)=0$, i.e., 
$B_i-B_j\in I_i+I_j=$Kernel of multiplication $\sigma_i\sigma_j$.
Therefore, $(B_1,{\cdots},B_{\nogg})\in$Kernel of $\beta$. 
This implies, by Lemma \ref{bglem}, there exists $F$ such that $F=B_i+a_i$ $(a_i\in I_i)$,
then $\sigma_iF=\sigma_i B_i=A_i$.  \qed

\begin{remark}
(i) The above condition \eqref{ka-masuda}, the proofs and the following facts are due to Kazuo Masuda.\\
(ii) Without the condition \eqref{ka-masuda}, there exists a counter-example for Lemma 1.7 of \cite{BG84}: 
\begin{quote}
Adding to ${\euc}^2$ the trivial multiplication, $a{\cdot}b=0$ for any $a, b\in{\euc}^2$, we take this as $A$.
Since $I_1={\euc}+0, I_2=0+{\euc}, I_3=\{(x,x)\in{\euc}^2\;|\; x\in{\euc}\}$ are ideals satisfying
$I_1+I_2=I_1+I_3=I_2+I_3=A$, we have
$\cap_{j=1}^3I_j=0$, $A={\euc}^2$, $\oplus_{j=1}^3A/I_j={\euc}^3$, $\oplus_{(i,j)}A/(I_i+I_j)=0$.
Since the sequence $0\to{\euc}^2\to{\euc}^3\to0$ is never exact, we don't have \eqref{ka-BG} in this case.
\end{quote}
(iii) We want to claim that the inductive argument proposed in \cite{VV83} \underline{seems not correct}.
To explain this, we change slightly the notation in \cite{VV83} as follows;
$$
\begin{cases}
\;\;\;\;\;{e_j}&{\longrightarrow \;\;\sigma_j},\\
\mbox{right-multiplication}&{\longrightarrow} \;\;\mbox{left-multiplication},\\
\;\;\;{G(q)}&{\longrightarrow \;\;{\mathfrak{B}}_{\nogg}}.
\end{cases}
$$
\par
As same as above proof of Lemma \ref{VVlem2.2}, they have $\sigma_i\sigma_j(B_j-B_i)=0$ for $i,j=1,{\cdots},L$. 
From this, they claim that there exists an element $F\in{\mathfrak{B}}_{\nogg}$ such that 
\begin{equation}\tag{VV2.22}
\sigma_iB_i=\sigma_iF\for i=1,{\cdots},L.
\label{VV2.22}
\end{equation}
They try to prove this claim by induction w.r.t. $L$ as follows:\\
(i) $L=2$:  As
$\sigma_i A_i=0$ for each $i=1,2$, $A_i\in\langle\sigma_i\rangle$, that is,
there exist elements $B_j\in[\sigma_1,\sigma_2]$=the algebra generated by $\{\sigma_1,\sigma_2\}$ such that
$$
A_1=\sigma_1 B_1,\; A_2=\sigma_2 B_2.
$$
From $\sigma_1\sigma_2(B_2-B_1)=\sigma_1A_2+\sigma_2 A_1=0$, we should have
$$
B_2-B_1=\exists{x}+\exists{y}\in\langle\sigma_1\rangle+\langle\sigma_2\rangle.
$$
Therefore, putting
$$
F=B_2-y=B_1+x,
$$
we have $\sigma_iB_i=\sigma_i F$ for $i=1,2$.\\
(ii) $L=3$: 
By the same talk as above,
there exist $B_1, B_2, B_3\in[\sigma_1,\sigma_2,\sigma_3]$ such that $A_k=\sigma_k B_k\in\langle\sigma_k\rangle$,
By the inductive hypothesis for $L=2$, there exist elements 
$F_3\in[\sigma_1,\sigma_2],\;F_2\in[\sigma_3,\sigma_1], \;F_1\in[\sigma_2,\sigma_3]$ such that
$$
\begin{cases}
A_i=\sigma_iB_i=\sigma_iF_3\with F_3=a_0+a_1\sigma_1+a_2\sigma_2+a_3\sigma_1\sigma_2 \for i=1,2,\\
A_i=\sigma_iB_i=\sigma_iF_2\with F_2=b_0+b_1\sigma_3+b_2\sigma_1+b_3\sigma_3\sigma_1\for i=3,1,\\
A_i=\sigma_iB_i=\sigma_iF_1\with F_1=c_0+c_1\sigma_2+c_2\sigma_3+c_3\sigma_2\sigma_3\for i=2,3.
\end{cases}
$$
From these, we have
$$
\begin{cases}
A_1=\sigma_1F_3=\sigma_1F_2\Longrightarrow \sigma_1(F_2-F_3)=0,\\
A_2=\sigma_2F_3=\sigma_2F_1\Longrightarrow \sigma_2(F_3-F_1)=0,\\
A_3=\sigma_3F_2=\sigma_3F_1\Longrightarrow \sigma_3(F_1-F_2)=0.
\end{cases}
\Longrightarrow 
\begin{cases}
a_0=b_0,\; a_2=0,\; b_1=0.\\
b_0=c_0,\;a_1=0, \;c_2=0,\\
c_0=a_0,\; c_1=0,\; b_2=0.
\end{cases}
$$
That is, we have 
$$
F_3=a_0+a_3\sigma_1\sigma_2,\; F_2=a_0+b_3\sigma_2\sigma_3,\; F_1=a_0+c_3\sigma_3\sigma_1.
$$

Therefore, without using \eqref{VV2.19}, we may take $F=a_0$ which satisfies $A_j=\sigma_ja_0$.
On the other hand, since a superdifferentiable function satisfies
the second equation in \eqref{II-3-12-1} below, there occurs some \underline{discrepancy}.
This stems from their argument that they assume as if $F_j\in[\sigma_1,\sigma_2,\sigma_3]$ instead of $F_3\in[\sigma_1,\sigma_2]$ etc.
\end{remark}

\subsection{Super $F$- or $G$-differentiable functions on ${\mathfrak{B}}_{\nogg}^{m|n}$}
\subsubsection{Super $F$- or $G$-differentiability}
\begin{defn}[Definition 2.5 of \cite{rog80}] 
Let ${\mathfrak{U}}_{\nogg}$ be an open set in ${\mathfrak{B}}_{\nogg}^{m|n}$ and $f: {\mathfrak{U}}_{\nogg}\to{\mathfrak{B}}_{\nogg}$.
Then,\\
(a) $f$ is said to be in ${\F}_{R}^0$ on ${\mathfrak{U}}_{\nogg}$, denoted by $f\in {\F}_{R}^0({\mathfrak{U}}_{\nogg}:{\mathfrak{B}}_{\nogg})$,  
if $f$ is continuous on ${\mathfrak{U}}_{\nogg}$,\\
(b) $f$ is said to be 1-time super $F$-differentiable on ${\mathfrak{U}}_{\nogg}$, denoted by $f\in {\F}_{R}^1({\mathfrak{U}}_{\nogg}:{\mathfrak{B}}_{\nogg})$, or $f\in {\F}_{R}^1$,   if there exist $m+n$ {continuous} functions $F_A:{\mathfrak{U}}_{\nogg}\to{\mathfrak{B}}_{\nogg}$, $(A=1,{\cdots},m+n)$
and a function $\rho:{\mathfrak{B}}_{\nogg}^{m|n}\to{\mathfrak{B}}_{\nogg}$ such that, if $X=(x,\theta)$ and $Y=(y,\omega)$ are in ${\mathfrak{U}}_{\nogg}$ with $X+Y\in {\mathfrak{U}}_{\nogg}$, 
\begin{equation}
f(X+Y)=f(X)+\sum_{j=1}^A Y_A{\,} F_A(X)+\Vert Y\Vert {\,} \rho(Y;X) 
\label{G-ss}
\end{equation}
with $\rho(Y;X)\to0$ when $\Vert Y\Vert\to 0$. $F_A(X)$ is denoted by $f_{X_A}(X)$. \\
(c) For any positive integer $p$, 
if $f$ is in ${\F}_{R}^1({\mathfrak{U}}_{\nogg}:{\mathfrak{B}}_{\nogg})$ and $F_k\in {\F}_{R}^{p-1}({\mathfrak{U}}_{\nogg}:{\mathfrak{B}}_{\nogg})$, we denote it 
$f\in {\F}_{R}^p({\mathfrak{U}}_{\nogg}:{\mathfrak{B}}_{\nogg})$ or simply $f\in {\F}_{R}^p$.  
We remark here, $F_{m+k}$ $(k=1,{\cdots},n)$ is not necessarily unique, but unique up to $\langle\sigma_1{\cdots}\sigma_L\rangle=\{\sigma_1{\cdots}\sigma_LX\;|\; X\in{\mathfrak{B}}_{\nogg}\}$.\\
(d) $f\in {\F}_{R}^{\infty}({\mathfrak{U}}_{\nogg}:{\mathfrak{B}}_{\nogg})$ or $f$ is said to be ${\F}_{R}^{\infty}$ (or $\infty$-times super $F$-differentiable) on ${\mathfrak{U}}_{\nogg}$, if $f$ is ${\F}_{R}^p$ on ${\mathfrak{U}}_{\nogg}$ for any positive integer $p$.\\
(e) $f\in {\F}_{R}^{\omega}({\mathfrak{U}}_{\nogg}:{\mathfrak{B}}_{\nogg})$, or $f$ is said to be in ${\F}_{R}^{\omega}$ (or superanalytic) on ${\mathfrak{U}}_{\nogg}$,  if $f$ is expanded as absolutely convergent power series in Banach topology;
$$
\begin{gathered}
f(X+Y)=\sum_{|\mathfrak{a}|=0}^{\infty}Y^{\mathfrak{a}}f_{\mathfrak{a}}
 \with {\mathfrak{a}}=(\alpha,a)\in{\mathbb{N}}^m\times\{0,1\}^n,\; |{\mathfrak{a}}|=|\alpha|+|a|,\\
Y^{\mathfrak{a}}=y^{\alpha}\theta^{a},\; y^{\alpha}=y_1^{\alpha_1}{\cdots}y_m^{\alpha_m},\;
\theta^a=\theta_1^{a_1}{\cdots}\theta_n^{a_n},\;
 f_{\mathfrak{a}}= f_{\alpha,a}\in {\mathfrak{B}}_{\nogg}.
 \end{gathered}
$$
(f) Let $g:{\mathfrak{U}}_{\nogg}\to {\mathfrak{B}}_{\nogg}^{r|s}$.
Then, $g$ is said to be in ${\F}_{R}^{\infty}$ (or ${\F}_{R}^{\omega}$) on ${\mathfrak{U}}_{\nogg}$ if each of the $r+s$ components of $g$ is ${\F}_{R}^{\infty}$(or ${\F}_{R}^{\omega}$).
\end{defn}

\begin{remark}
(i) Rogers calls above functions as $G$-differentiable, but to distinguish it from G\^ateaux (or $G$-) differentiability, we prefer to denote them as
${\F}_{R}$, $\F$ stands for Fr\'echet and ${}_{R}$ comes from Rogers.
But remark her definitions are applicable also to functions on ${\mathfrak{B}}_{\infty}^{m|n}$.\\
(ii) Though Rogers nor Volvich and Vladimirov don't introduce $G$-differentiability, but we may define ${\mathcal{G}}_{S\!D}^{\infty}({\mathfrak{U}}_{\nogg}:{\mathfrak{B}}_{\nogg})$ analogously as above.
\end{remark}

We have the following Taylor's formula:
\begin{prop}[Theorem 4.2.1 of \cite{rog07}] 
Let ${\mathfrak{U}}_{\nogg}$ be an open set in ${\mathfrak{B}}_{\nogg}^{m|n}$. 
Assume $f\in {\F}_{R}^{\infty}({\mathfrak{U}}_{\nogg}:{\mathfrak{B}}_{\nogg})$.
Take any $X=(x,\theta), Y=(y,\omega)\in{\mathfrak{B}}_{\nogg}^{m|n}$ such that $X+tY\in {\mathfrak{U}}_{\nogg}$ for all $t\in[0,1]$.
Then, for any positive integer $N$, we have
\begin{equation}
f(X+Y)=\sum_{p=0}^{N}\sum_{\scriptstyle{|\alpha|+|a|=p}\atop\scriptstyle{|a|\le n}} 
\frac{1}{\alpha!}y^{\alpha}\omega^a\partial_x^{\alpha}\partial_\theta^a f(X)
+\sum_{\scriptstyle{|\alpha|+|{a}|=N+1}\atop\scriptstyle{|a|\le n}}
\frac{1}{\alpha!}y^\alpha\omega^a\int_0^1dt \, \partial_x^{\alpha}\partial_{\theta}^a f(x+ty,\theta+t\omega)
\label{taylor-exp-1}
\end{equation}
with
$$
\omega^a=\omega_1^{a_1}{\cdots}\omega_n^{a_n},\;
\partial_{\theta}^a f=\partial_{\theta_n}^{a_n}(\partial_{\theta_{n-1}}^{a_{n-1}}{\cdots}(\partial_{\theta_1}^{a_1} f)).
$$
\end{prop}
{\it Proof}.
For any $N$ and $u\in C^{N+1}([0,1])$, we have
$$
u(1)=\sum_{\ell=0}^{N}\frac{1}{\ell!}u^{(\ell)}(0)+\int_0^1 dt\frac{(1-t)^N}{N!}u^{(N+1)}(t).
$$
Take $u(t)=f(X+tY)$ and remarking 
$$
(x_1+{\cdots}+x_k)^{\ell}=\sum_{\scriptstyle{\alpha\in{\mathbb{N}}^k}\atop\scriptstyle{|\alpha|={\ell}}}\binom{{\ell}}{\alpha}x^{\alpha}
\with
\binom{{\ell}}{\alpha}=\frac{{\ell}!}{{\alpha_1}!{\cdots}{\alpha_k}!},
$$
$a!=1$ for $a\in\{0,1\}^n$ and components in $\sum_{j=1}^m y_j\partial_{x_j}$ and $\sum_{k=1}^n\omega_k\partial_{\theta_k}$ are commutative,
we have
$$
\begin{aligned}
\frac{1}{\ell!}\big(\dt\big)^{\ell}&f(X+tY)\bigg|_{t=0}
=\sum_{\scriptstyle{|\alpha|+|{a}|=\ell}\atop\scriptstyle{|a|\le n}}\frac{1}{{|\alpha|}!|a|!}
\big(\sum_{j=1}^m y_j\partial_{x_j}\big)^{|\alpha|}
\big(\sum_{k=1}^n\omega_k\partial_{\theta_k}\big)^{|a|}f(x,\theta)\\
&=
\sum_{\scriptstyle{|\alpha|+|{a}|=\ell}\atop\scriptstyle{|a|\le n}}
\sum_{\scriptstyle{|\beta|=|\alpha|}\atop\scriptstyle{|b|=|a|}}\frac{1}{{|\alpha|}!|a|!}
\binom{|\alpha|}{\beta}y^{\beta}\partial_x^{\beta}\binom{|a|}{b}{\omega}^b\partial_{\theta}^b f(x,\theta)
=\sum_{\scriptstyle{|\alpha|+|{a}|=\ell}\atop\scriptstyle{|a|\le n}}\frac{1}{\alpha!}y^{\alpha}\omega^a\partial_x^{\alpha}\partial_{\theta}^af(x,\theta).
\qed
\end{aligned}
$$

Inspired by the above expansion, we put
\begin{defn}[Definition 4.2.2 of  \cite{rog07}]
Let $V$ be an open set of ${\euc}$ and let ${\mathfrak{V}}_{\nogg}$ be an open set in ${\mathfrak{B}}_{\nogg}^{m|0}$ such that $V=\pi_{\mathrm{B}}({\mathfrak{V}}_{\nogg})$.
For any $f\in C^{\infty}(V:{\mathfrak{B}}_{\nogg})$, we define the Grassmann continuation of $f$ on $\euc$ to ${\mathfrak{V}}_{\nogg}$ as
\begin{equation}
\tilde{f}(x)=\sum_{|\alpha|=0}^{\infty}\frac{1}{\alpha!}
{\partial_q^{\alpha}f(q)}\big|_{q=x_{\mathrm{B}}}x_{\mathrm{S}}^{\alpha}
\where
x=x_{\mathrm{B}}+x_{\mathrm{S}}.
\label{gc-finite}
\end{equation}
\end{defn}
\begin{remark}
Since $x_{\mathrm{S}}^{\alpha}=0$ if $|\alpha|\ge \log_2 ({\nogg}+1)$, the summation above is finite for fixed ${\nogg}<{\infty}$.
Though Rogers calls this expression \eqref{gc-finite} as Grassmann analytic continuation, but we prefer to abbriviate ``analytic'' 
because in case ${\nogg}={\infty}$, we don't claim the ``absolute convergence'' of the series obtained.
\end{remark}

We have readily
\begin{lem}[Theorem 4.2.3 of \cite{rog07}] \label{geff}
Let ${\mathfrak{U}}\subset{\mathfrak{B}}_{\nogg}^{m|0}$ be open.
For $f, g\in C^{\infty}({\mathfrak{U}}_{\mathrm{B}}:{\mathfrak{B}}_{\nogg})$, we have
$$
(a)\; \tilde{f}\in{\mathcal{F}}_R^{\infty},\quad
(b)\; \partial_{x_j}\tilde{f}(x)=\widetilde{\partial_{q_j}f}(x),\quad
(c)\; \widetilde{f{\cdot}g}(x)=\tilde{f}(x)\tilde{g}(x).
$$
\end{lem}
\par{\it Proof}.
(c) stems from the fact that the classical Taylor expansion of the sum or the product of functions is the sum or product of the Taylor expansion of the functions, respectively. \qed

\begin{prop}[Theorem 4.2.4 of \cite{rog07}]\label{SS-SD}
Let ${\mathfrak{U}}_{\nogg}$ be an open set in ${\mathfrak{B}}_{\nogg}^{m|n}$. 
Putting
$$
{\mathcal{C}}_{\mathrm{SS}}({\mathfrak{U}}_{\nogg}:{\mathfrak{B}}_{\nogg})
=\{f:{\mathfrak{U}}_{\nogg}\to{\mathfrak{B}}_{\nogg}\;|\; f(x,\theta)=\sum_{|a|\le n}\theta^a\tilde{f}_a(x)
\with f_a(x_{\mathrm{B}})\in C^{\infty}(\pi_{\mathrm{B}}({\mathfrak{U}}_{\nogg}):{\mathfrak{B}}_{\nogg})\},
$$
we have
\begin{equation}
{\F}_{R}^{\infty}({\mathfrak{U}}_{\nogg}:{\mathfrak{B}}_{\nogg})={\mathcal{C}}_{\mathrm{SS}}({\mathfrak{U}}_{\nogg}:{\mathfrak{B}}_{\nogg}).
\label{ss=sd}
\end{equation}
\end{prop}

{\it Proof}. 
Let $f\in{\F}_{R}^{\infty}({\mathfrak{U}}_{\nogg}:{\mathfrak{B}}_{\nogg})$.
Putting $u(t)=f(x_{\mathrm{B}}+tx_{\mathrm{S}},t\theta)$ in \eqref{taylor-exp-1}, 
we have
$$
f(x_{\mathrm{B}}+x_{\mathrm{S}},\theta)
=\sum_{p=0}^{N}\sum_{\scriptstyle{|\alpha|+|a|=p}\atop\scriptstyle{|a|\le n}} 
\frac{1}{\alpha!}x_{\mathrm{S}}^{\alpha}\theta^a\partial_x^{\alpha}\partial_\theta^a f(x_{\mathrm{B}},0)
+\sum_{\scriptstyle{|\alpha|+|{a}|=N+1}\atop\scriptstyle{|a|\le n}}
\frac{1}{\alpha!}x_{\mathrm{S}}^\alpha\theta^a\int_0^1dt \, \partial_x^{\alpha}\partial_{\theta}^a f(x_{\mathrm{B}}+tx_{\mathrm{S}},t\theta).
$$
Since $x_{\mathrm{S}}^{\alpha}=0$ if $2|\alpha|>{\nogg}$, for sufficiently large $N$, the reminder term is $0$.
Therefore, rearranging the finite sum in the first term above, we have
$$
\tilde{f}_a(x)=\sum_{|\alpha|=0}^{N}\frac{1}{\alpha!}x_{{\mathrm{S}}}^{\alpha}\partial_x^{\alpha}\partial_{\theta}^af(x_{\mathrm{B}},0)=\partial_{\theta}^a f(x,0)
\with f_a(q)=\partial_{\theta}^af(x,0)\big|_{x=q}.
$$
Conversely, if $f(x,\theta)=\sum_{|a|\le n}\theta^a\tilde{f}_a(x)$ with $f_a(q)\in C^{\infty}(\pi_{\mathrm{B}}({\mathfrak{U}}_{\nogg}):{\mathfrak{B}}_{\nogg})$, then $f$ is ${\F}_{R}^{\infty}({\mathfrak{U}}_{\nogg}:{\mathfrak{B}}_{\nogg})$ by using Lemma \ref{geff}. \qed 

\begin{remark}
An element in ${\mathcal{C}}_{\mathrm{SS}}$ is called supersmooth here, but is also called $Z$-expansion by Rogers or superfield expansion by physicist.
\end{remark}

\subsubsection{Characterization of functions in ${\F}_{R}^1({\mathfrak{B}}_{\nogg}^{m|n}:{\mathfrak{B}}_{\nogg})$}
We denote 
$$
{\bf{e}}_A=(\underbrace{{\overbrace{0,{\cdots},0,1}^{A}},0,{\cdots},0}_{m+n})\in{\euc}^{m+n}\et
H=(H_A)_{A=1}^{m+n}=\sum_{A=1}^{m+n}H_A{\bf{e}}_A\in{\mathfrak{B}}_{\nogg}^{m|n}.
$$
By definition of G-differentiability in Banach space $ {\mathfrak{B}}_{\nogg}^{m|n}$, using \eqref{total}, we have
\begin{equation}
d_Gf(X;H)=\dt f(X+tH)\big|_{t=0}=\sum_{A=1}^{m+n} f'_{X_A}(X;H_A{\bf{e}}_A)\with H=(H_A)_{A=1}^{m+n}\in{\mathfrak{B}}_{\nogg}^{m|n}.
\label{total-bis}
\end{equation}
\begin{prop}[Lemma 2.1 of  \cite{VV83}] \label{L2.1VV}
Let ${\mathfrak{U}}_{\nogg}$ be an open subset of ${\mathfrak{B}}_{\nogg}^{m|n}$.
Then, $f\in {\F}_{R}^1({\mathfrak{U}}_{\nogg}:{\mathfrak{B}}_{\nogg})$ {if and only if} $f$ is $F$-differentiable and there exist continuous functions $F_A(X)$ defined on ${\mathfrak{U}}_{\nogg}$ such that
\begin{equation}
f'_{X_A}(X;H_A{\bf{e}}_A)=H_A{\,} F_A(X),\quad A=1,{\cdots},m+n \forany H\in{\mathfrak{B}}_{\nogg}^{m|n}.
\label{vv2.8}
\end{equation}
\end{prop}
\par{\it Proof}.
$\Longrightarrow$) $f\in {\F}_{R}^1({\mathfrak{U}}_{\nogg}:{\mathfrak{B}}_{\nogg})$ implies $F$- and also $G$-differentiability and using \eqref{total-bis},  we have readily
$$
\sum_{A=1}^{m+n}f'_{X_A}(X;H_A{\bf{e}}_A)=d_Gf(X;H)=d_Ff(X;H)=\sum_{A=1}^{m+n}H_A{\cdot}F_A(X).
$$
Therefore, we have \eqref{vv2.8} by taking $H_A$ appropriately.\\
$\Longleftarrow$) Clearly, $F$-differentiability with \eqref{total-bis} and \eqref{vv2.8} implies $f\in {\F}_{R}^1({\mathfrak{U}}_{\nogg}:{\mathfrak{B}}_{\nogg})$. \qed

Following characterization of superdifferentiability is announced as Theorem 2.1 of \cite{VV83}:
\begin{prop}[Theorem 2.1 of \cite{VV83}]\label{Thm2.1VV} 
Let ${\mathfrak{U}}_{\nogg}$ be an open subset of ${\mathfrak{B}}_{\nogg}^{m|n}$.
Then, $f\in {\F}_{R}^1({\mathfrak{U}}_{\nogg}:{\mathfrak{B}}_{\nogg})$ {if and only if} $f$ is $F$-differentiable 
and its derivatives satisfy the following equations for $A=1,{\cdots},m+n$,
\begin{equation}
\left\{
\begin{aligned}
&G_A f_{X_A}'(X;H_A{\bf{e}}_A)-(-1)^{p(H_A)p(G_A)}H_Af_{X_A}'(X;G_A{\bf{e}}_A)=0,\\ &f_{X_A}'(X;H_AG_A{\bf{e}}_A)=H_A f_{X_A}'(X;G_A{\bf{e}}_A) \for p(H_A)=0,\; p(X_A)=p(G_A).
\end{aligned}
\right.
\label{vv2.10-1}
\end{equation}
\end{prop}
\par
{\it Proof}. $\Longrightarrow)$ Multiplying $G_A$ to \eqref{vv2.8} from the left and changing the role of $G_A$ and $H_A$ in the obtained equality, we have the first equation of \eqref{vv2.10-1}.
The second equation in \eqref{vv2.10-1} is derived from \eqref{vv2.8} by
$$
f_{X_A}'(X;H_AG_A{\bf{e}}_A)=H_AG_AF_A(X)=H_Af_{X_A}'(X;G_A{\bf{e}}_A) \for p(H_A)=0,\; p(X_A)=p(G_A).
$$
$\Longleftarrow)$ Putting $G_A=1$ in \eqref{vv2.10-1} and defining $F_A(X)= f_{X_A}'(X;1)$ which yields
\eqref{vv2.8} for $A=1,{\cdots},m$.
The case for $A=m+1,{\cdots}, m+n$, i.e. $p(X_A)=1$, \underline{will be given at the end of Proof of Proposition \ref{propII-3.5} in the}\\
\underline{next subsection}. \qed
\begin{remark}
(a) The second equation in \eqref{vv2.10-1} is said to be that the differential is ${\mathfrak{B}}_{\nogg,{\mathrm{ev}}}$-linear.\\
(b) Concerning the construction of $F_A(X)$ for  $A=m+1,{\cdots}, m+n$ from \eqref{vv2.10-1},
there seems  some flaw in Lemma \ref{VVlem2.2} of \cite{VV83}  or 
in Lemma  (1.7) of \cite{BG84}. In spite of these, thanks to Masuda's reasoning, 
we justify the key statement in Lemma \ref{VVlem2.2} of \cite{VV83}.
\end{remark}

\paragraph{\bf{Cauchy-Riemann relation}}
To understand the meaning of supersmoothness, we need to give  more precisely the dependence w.r.t  ``coordinates". 

Let ${\mathfrak{U}}_{\nogg}$ be an open set in ${\mathfrak{B}}_{\nogg}^{m|n}$ 
and let a function $f:{\mathfrak{U}}_{\nogg}\ni X\to f(X)=\sum_{\bf{I}} f_{\bf{I}}(X)\sigma^{\bf{I}}\in{\mathfrak{B}}_{\nogg}$ be given such that
$f_{\bf{I}}(X+tY)\in C^\infty([0,1]:{\mathbb{R}})$ for each fixed $X,Y\in {\mathfrak{U}}_{\nogg}$.
Let $X=(X_{A})=(x_j,\theta_k)$ be represented by 
$X_{A}=\sum_{\bf{I}} X_{{A},{\bf{I}}}\sigma^{\bf{I}}$ with $ X_{{A},{\bf{I}}}\in {\euc}$ where ${A}=1,\cdots, m+n$.

Now, we put 
\begin{equation}
\left\{
\begin{aligned}
&\frac{\partial f(X)}{\partial  X_{{A},{\bf{I}}}}=
\left.{\frac d{dt}}f(X+tE_{{A},{\bf{I}}})\right|_{t=0}\;\mbox{with}\;
E_{{A},{\bf{I}}}=\sigma^{\bf{I}} {\bf{e}}_A, 
|{\bf{I}}|=\begin{cases}
\mbox{ev}&\mbox{for}\; {1\le{A}\le m},\\
\mbox{od} &\mbox{for}\; {m+1\le{A}\le m+n},
\end{cases}\\ 
&{\frac{\partial f(X)}{\partial  X_{{A},(k)}}}=
\left.{\frac d{dt}}f(X+tE_{{A},(k)})\right|_{t=0}\with
E_{{A},(k)}=\sigma_k {\bf{e}}_A,\;
\mbox{for}\;  {m+1\le{A}\le m+n},
\end{aligned}
\right.
\label{II-3-13}
\end{equation}
Here $1\le k\le L$.
As $H_A{\bf{e}}_A=\sum_{{\bf{I}}\in{\mathcal{I}}_{\nogg}}H_{A,{\bf{I}}}E_{{A},{\bf{I}}}$, we have
\begin{equation}
f'_{X_A}(X;H_A{\bf{e}}_A)=\dt f(X+t\sum_{{\bf{I}}\in{\mathcal{I}}_{\nogg}}H_{A,{\bf{I}}}E_{{A},{\bf{I}}})\big|_{t=0}=\sum_{{\bf{I}}\in{\mathcal{I}}_{\nogg}}H_{A,{\bf{I}}}\frac{\partial f(X)}{\partial  X_{{A},{\bf{I}}}} .
\label{vv2.15}
\end{equation}

Using above coordinate $\{X_{{A},{\bf{I}}}\}$, we rewrite Proposition \ref{L2.1VV} as
\begin{prop}[Proposition 1.5 of \cite{BG84}] \label{P1.5BG}
$f\in {\F}_{R}^1({\mathfrak{U}}_{\nogg}:{\mathfrak{B}}_{\nogg})$ {if and only if} $f\in C_F^1({\mathfrak{U}}_{\nogg}:{\mathfrak{B}}_{\nogg})$ and there exists continuous functions
$$
g_A(f):{\mathfrak{U}}_{\nogg}\to {\mathfrak{B}}_{\nogg}, \quad 1\le A\le m+n,
$$
such that
\begin{equation}
\frac{\partial f(X)}{\partial X_{A,{\bf{I}}}}=\sigma^{\bf{I}} {\cdot} g_A(f)(X)\quad \mbox{for}\quad|{\bf{I}}|=\begin{cases}
\mbox{ev}&\mbox{for}\;  1\le A\le m,\\
\mbox{od}&\mbox{for}\;  m+1\le A\le m+n.
\end{cases}
\label{BG1.5}
\end{equation}
\end{prop}
\par
{\it Proof}.
$\Longrightarrow)$ Since $f\in {\F}_{R}^1({\mathfrak{U}}_{\nogg})$, it is G-differentiable and \eqref{vv2.8} holds, therefore
$$
\frac{\partial f(X)}{\partial X_{A,{\bf{I}}}}=\frac{d}{dt}f(X+tE_{A,{\bf{I}}})\big|_{t=0}=\sigma^{\bf{I}}{\cdot} F_A(X)
$$
with $F_A(X)$ being continuous w.r.t. $X$. Put $g_A(f)(X)=F_A(X)$, then the assertion holds.\\
$\Longleftarrow)$ Multiplying $H_{A,{\bf{I}}}\in\mathbb{R}$ to both side of \eqref{BG1.5} and adding w.r.t. ${\bf{I}}$, we have
$$
\begin{aligned}
f'_G(X;H_A{\bf{e}}_A)&=\frac{d}{dt}f(X+tH_A{\bf{e}}_A)\big|_{t=0}\\
&=\sum_{{\bf{I}}\in{\mathcal{I}}_{\nogg}}H_{A,{\bf{I}}}\frac{\partial f(X)}{\partial X_{A,{\bf{I}}}}=\sum_{{\bf{I}}\in{\mathcal{I}}_{\nogg}}H_{A,{\bf{I}}}\sigma^{\bf{I}}{\cdot} g_A(f)(X)=H_A{\cdot} g_A(f)(X)=f'_F(X;H_A{\bf{e}}_A).
\end{aligned}
$$
Therefore, by Proposition \ref{L2.1VV}, we get the result. \qed  

\begin{prop}[Theorem 2.2 of \cite{VV83} and Theorem 1.6 of \cite{BG84}] \label{propII-3.5}
Let ${\mathfrak{U}}_{\nogg}\subset {\mathfrak{B}}_{\nogg}^{m|n}$ be open and let $f\in C_G^1({\mathfrak{U}}_{\nogg}:{\mathfrak{B}}_{\nogg})$
be considered as a function 
of $2^{(L-1)}{(m+n)}$ variables $\{X_{{A},{\bf{I}}}\}$ with values in ${\mathfrak{B}}_{\nogg}$.
$f(X)$ is ${\F}_{R}^1({\mathfrak{U}}_{\nogg})$-differentiable {if and only if}
$f(X)$ satisfies the following (Cauchy-Riemann type) equations.
\begin{equation}
\left\{
{\begin{aligned}
&{\frac{\partial f(X)}{\partial  X_{{A},{\bf{I}}}}}=
(-1)^{\tau({\bf{I}};{\bf{I}}-{\bf{J}},{\bf{J}})}\sigma^{{\bf{I}}-{\bf{J}}} {\frac{\partial f(X)}{\partial X_{{A},{\bf{J}}}}}
\;\;\mbox{for}\;\; |{\bf{I}}-{\bf{J}}|={\mbox{ev}},\;\;1\le{A}\le m+n,\\
&\sigma_i{\frac{\partial f(X)}{\partial X_{{A},(j)}}}
+\sigma_j{\frac{\partial f(X)}{\partial X_{{A},(i)}}}=0\;\;\mbox{for}\;\;i,j=1,{\cdots},L, \;\; m+1\le{A}\le m+n.
\end{aligned}}
\right.
\label{II-3-12-1}
\end{equation}
Here, integer $\tau({\bf{I}};{\bf{I}}-{\bf{J}},{\bf{J}})$ is defined in \eqref{ncg}.
\end{prop}

{\it Proof due to \cite{BG84}}. 
$\Longrightarrow)$ 
From \eqref{BG1.5}, if ${\bf{J}}\subset{\bf{I}}$ and $|{\bf{I}}-{\bf{J}}|=$even, then
$$
{\frac{\partial f(X)}{\partial  X_{{A},{\bf{I}}}}}=\sigma^{\bf{I}} {\cdot} g_A(f)(X)=(-1)^{\tau({\bf{I}};{\bf{I}}-{\bf{J}},{\bf{J}})}\sigma^{{\bf{I}}-{\bf{J}}}\sigma^{\bf{J}} {\cdot} g_A(f)(X)=(-1)^{\tau({\bf{I}};{\bf{I}}-{\bf{J}},{\bf{J}})}{\frac{\partial f(X)}{\partial  X_{{A},{\bf{J}}}}}.
$$
Also from \eqref{BG1.5} with $i,j=1,{\cdots},L$ and ${\bf{I}}=1$, 
$$
\sigma_i{\frac{\partial f(X)}{\partial X_{{A},(j)}}}
+\sigma_j{\frac{\partial f(X)}{\partial X_{{A},(i)}}}=\sigma_i\sigma_jg_A(f)(X)+\sigma_j\sigma_ig_A(f)(X)=0.
$$
\par
$\Longleftarrow)$ We have to construct functions 
$F_A(X)$($1\le A\le m+n$) such that
$$
{\dt}f(X+tH)\big|_{t=0}=\sum_{A=1}^{m+n} H_A{\,}  F_A(X) \for X\in {\mathfrak{U}}_{\nogg}\et H\in{\mathfrak{B}}_{\nogg}^{m|n}.
$$

Putting ${\bf{J}}={\tilde{0}}=(\overbrace{0,{\cdots},0}^L)$, $|{\bf{I}}|$=even and multiplying $H_{A,{\bf{I}}}$ to both sides of the first equation of \eqref{II-3-12-1}, we have
$$
f'_G(X;H_A{\bf{e}}_A)=\frac{d}{dt}f(X+tH_A)\big|_{t=0}
=\sum_{{\bf{I}}\in{\mathcal{I}}_{\nogg}}H_{A,{\bf{I}}}{\frac{\partial f(X)}{\partial X_{{A},{\bf{I}}}}}
=\sum_{{\bf{I}}\in{\mathcal{I}}_{\nogg}}H_{A,{\bf{I}}}\sigma^{\bf{I}}\frac{\partial f(X)}{\partial X_{{A},\tilde 0}}
=H_A \frac{\partial f(X)}{\partial X_{{A},\tilde 0}}.
$$
Therefore, for $A=1,{\cdots},m$, we define
$$
F_{A}(X)=\frac{\partial f(X)}{\partial X_{{A},\tilde 0}}
\for  1\le{A}\le m,\;\; X\in U.
$$

To define $F_{A}(X)$ for $m+1\le {A}\le m+n$,
we need to use the second equation of \eqref{II-3-12-1}.
From that, applying Proposition \ref{VVlem2.2}, we know there exists an element $F_A(X)$ satisfying
$$
{\frac{\partial f(X)}{\partial X_{{A},(i)}}}=\sigma_i F_A(X) \for m+1\le {A}\le m+n.
$$
Applying the first equation of  \eqref{II-3-12-1} with $|{\bf{I}}|$=odd, we have, when $i_1=1$,
$$
{\frac{\partial f(X)}{\partial X_{{A},{\bf{I}}}}}=\sigma^{\check{\bf{I}}}{\frac{\partial f(X)}{\partial X_{{A},(i_1)}}}
=\sigma^{\check{\bf{I}}}\sigma_{i_1}F_A(X)=\sigma^{\bf{I}} F_A(X)\with
\check{\bf{I}}=(0,i_2,i_3,{\cdots},i_L)\in\{0,1\}^{L}. 
$$
Therefore,
$$
f'_{X_A}(X;H_A{\bf{e}}_A)=\sum_{{\bf{I}}\in{\mathcal{I}}_{\nogg}}
H_{A,{\bf{I}}}{\frac{\partial f(X)}{\partial X_{{A},{\bf{I}}}}}
=\sum_{\bf{I}\in{\mathcal{I}}_{\nogg}}H_{A,{\bf{I}}}\sigma^{\bf{I}} F_A(X)
=H_AF_A(X)\for m+1\le {A}\le m+n.
$$
This implies the superdifferentiability of $f$ and also finishes Proof of Proposition \ref{Thm2.1VV}. \qed

\subsubsection{Characterization of superdifferentiability on ${\mathfrak{B}}_{\nogg}^{m|n}$}
\begin{lem}\label{F2Finfty}
$$
{\F}_{R}^1({\mathfrak{U}}_{\nogg}:{\mathfrak{B}}_{\nogg})\cap C_G^{\infty}({\mathfrak{U}}_{\nogg}:{\mathfrak{B}}_{\nogg})\Longrightarrow {\F}_{R}^{\infty}({\mathfrak{U}}_{\nogg}:{\mathfrak{B}}_{\nogg}).
$$
\end{lem}
\par{\it Proof}. Since we prove this analogously as will be given in Theorem \ref{yag-thm3}, we omit it here.
But as ${\mathfrak{B}}_{\nogg}^{m|n}\cong{\euc}^{2^{(L-1)}{(m+n)}}$, it is unnecessary to introduce ``admissibility'' and to distinguish the Fr\'echet and G\^ateaux differentiability in $C^{\infty}$-category.\qed

\begin{thm}\label{finiteCRBG}
Let ${\mathfrak{U}}_{\nogg}$ be a open set in ${\mathfrak{B}}_{\nogg}^{m|n}$ and let a function $f:{\mathfrak{U}}_{\nogg}\to {\mathfrak{B}}_{\nogg}$ be given.
Following conditions are equivalent:\\
(a) $f$ is super F-differentiable on ${\mathfrak{U}}_{\nogg}$, i.e. $f\in {\F}_{R}^{\infty}({\mathfrak{U}}_{\nogg})$,\\
(b) $f\in C_G^{\infty}({\mathfrak{U}}_{\nogg})\cap {\F}_{R}^{1}({\mathfrak{U}}_{\nogg})$,\\
(c) $f\in C_G^{\infty}({\mathfrak{U}}_{\nogg})$ and its (G--) differential $df$ is ${\blev}$-linear,\\
(d) $f\in C_G^{\infty}({\mathfrak{U}}_{\nogg})$ and its (G--) differential $df$ satisfies Cauchy-Riemann equations,\\
(e) $f$ is supersmooth, i.e. it has the following representation, called superfield expansion, such that
$$
f(x,\theta)=\sum_{|a|\le n}\theta^a \tilde{f}_a(x)\with
f_a(q)\in C^{\infty}(\pi_{\mathrm{B}}({\mathfrak{U}}_{\nogg}))\et
 \tilde{f}_a(x)=\sum_{|\alpha|=0}^{\infty}\frac{1}{\alpha!}\frac{\partial ^{\alpha}f_a(q)}{\partial q^{\alpha}}\bigg|_{q=x_{\mathrm{B}}}
 x_{\mathrm{S}}^{\alpha}.
$$
\end{thm}
\par{\it Proof}.
(a)$\Longrightarrow$(b) trivially,
(b)$\Longrightarrow$(c) by Proposition \ref{Thm2.1VV},
(c)$\Longrightarrow$(d) by Proposition \ref{propII-3.5},
(d)$\Longrightarrow$(e) by Lemma \ref{F2Finfty},
(e)$\Longrightarrow$(a) by Proposition \ref{SS-SD}. \qed.

\section{The definition and characterization of supersmooth functions on FG-algebra}

\subsection{Remarks on FG-algebras}
Though we introduced FG-algebras in \S 2 by using the sequence space $\omega$, 
we prepare another definition using projective limits in order to clarify the relation to \S 3.
 We define index sets as
$$
\begin{gathered}
{\mathcal{I}}=\{{\bf{I}}=(i_1,i_2,{\cdots})\in\{0,1\}^{\mathbb{N}}\;|\; |{\bf{I}}|=\sum_{j=1}^{\infty} i_j<{\infty}\}
=\cup_{{\deg}=0}^{\infty} {\mathcal{I}}^{(\deg)}=\oplus_{{\deg}=0}^{\infty}{\mathcal{I}}^{[\deg]},\\
\where {\mathcal{I}}^{(\deg)}=\{{\bf{I}}=(i_1,i_2,{\cdots})\in {\mathcal{I}}\;|\; |{\bf{I}}|\le {\deg}\},\;\;
{\mathcal{I}}^{[\deg]}=\{{\bf{I}}=(i_1,i_2,{\cdots})\in {\mathcal{I}}\;|\; |{\bf{I}}|= {\deg}\},\\
{\mathcal{I}}_{\nogg}=\{{\bf{I}}=(i_1,i_2,{\cdots},i_{\nogg})\in\{0,1\}^{L}\}=\cup_{{\deg}=0}^{L} {\mathcal{I}}_{\nogg}^{(\deg)}=\oplus_{{\deg}=0}^{L}{\mathcal{I}}_{\nogg}^{[\deg]},\\
\where {\mathcal{I}}_{\nogg}^{(\deg)}=\{{\bf{I}}=(i_1,i_2,{\cdots},i_{\nogg})\;|\; |{\bf{I}}|\le {\deg}\},\;\;
{\mathcal{I}}_{\nogg}^{[\deg]}=\{{\bf{I}}=(i_1,i_2,{\cdots},i_{\nogg})\;|\; |{\bf{I}}|= {\deg}\}.
\end{gathered}
$$
Clearly, we have
$$
{\mathcal{I}}_{\nogg}^{(\deg)}\to {\mathcal{I}}^{(\deg)}\et
{\mathcal{I}}_{\nogg}\to {\mathcal{I}}\;\;\mbox{in $\omega$ when $L\to{\infty}$}.
$$

Besides ${\mathfrak{C}}$, for any ${\nogg}$ and ${\deg}\le {\nogg}$, we put
$$
\begin{aligned}
{\fC}_{\nogg}&=\{X=\sum_{{\bf{I}}\in{\mathcal{I}}_{\nogg}}X_{\bf{I}}\sigma^{\bf{I}}\;|\; X_{\bf{I}}\in\mathbb{C}\}\\
&\cong  \bigwedge_{\mathbb{C}}({\euc}^{\nogg})\mbox{=the exterior algebra of forms 
on ${\fC}^{\nogg}$ with coefficients in $\mathbb{C}$}
\cong {\mathbb{C}}^{2^{\nogg}},\\ 
{\fC}_{\nogg}^{({\deg})}
&=\bigg\{
X=\sum_{{\bf{I}}\in{\mathcal{I}}_{\nogg}^{(\deg)}} X_{\bf{I}}\sigma^{\bf{I}} \;|\; X_{\bf{I}}\in{\mathbb{C}}
\bigg\}
\et
{\fC}_{\nogg}^{[{\deg}]}
=\bigg\{
X=\sum_{{\bf{I}}\in{\mathcal{I}}_{\nogg}^{[{\deg}]}} X_{\bf{I}}\sigma^{\bf{I}} \;|\; X_{\bf{I}}\in{\mathbb{C}}
\bigg\}.
\end{aligned}
$$
${\fC}_{\nogg}$ is called ${\nogg}$-skelton of ${\mathfrak{C}}$, etc.
Since the family $\{{\fC}_{\nogg}\}_{{\nogg}\ge0}$ and the natural projections $\{\psi_{{\nogg}{K}}\}$ for ${K}>{\nogg}$,
defined by $\psi_{{\nogg}{K}}:{\fC}_{{K}}\to {\fC}_{\nogg}$ 
with 
$\psi_{{\nogg}{K}}(\sum_{{\bf{I}}\in{\mathcal{I}}_{K}}X_{\bf{I}}\sigma^{\bf{I}})=\sum_{{\bf{I}}\in{\mathcal{I}}_{\nogg}}X_{\bf{I}}\sigma^{\bf{I}}$,
we have the set $({\fC}_{\nogg},\psi_{{\nogg}{K}})$ which forms a projective system
and yields a projective limit ${\fC}_{\infty}$. 
More precisely, the topology of ${\fC}_{\infty}$ is defined as follows:
Elements $X^{(n)}$ converges to $X$ in ${\fC}_{\infty}$ if and only if
for any $\epsilon >0$ and ${\bf{I}}$, 
there exists an integer $n_0=n_0(\epsilon,{\bf{I}})$ such that
$|X^{(n)}_{\bf{I}}-X_{\bf{I}}|<\epsilon $ when $ n>n_0$.

\begin{claim}
When ${\nogg}\to{\infty}$, we have the projective limits 
{\allowdisplaybreaks
\begin{gather*}
{\fC}_{\nogg}\to {\mathfrak{C}}_{\infty}={\mathfrak{C}},\quad
{\fC}_{{\nogg},{\mathrm{ev}}}=\{X=\sum_{{\bf{I}}\in{\mathcal{I}}_{\nogg}, |{\bf{I}}|={\mathrm{ev}}}X_{\bf{I}}\sigma^{\bf{I}}\;|\; X_{\tilde0}\in\mathbb{C},\;X_{\bf{I}}\in\mathbb{C}\}\to\cev, \\
{\fC}_{{\nogg},{\mathrm{od}}}=\{X=\sum_{{\bf{I}}\in{\mathcal{I}}_{\nogg}, |{\bf{I}}|={\mathrm{od}}}X_{\bf{I}}\sigma^{\bf{I}}\;|\; X_{\bf{I}}\in\mathbb{C}\}\to\cod\et
{\fC}_{\nogg}^{m|n}={\mathfrak{C}}_{{\nogg},{\mathrm{ev}}}^m\times {\fC}_{{\nogg},{\mathrm{od}}}^n\to {\fC}^{m|n}.\\
{\fC}_{\nogg}^{(\deg)}
=\bigg\{
X=\sum_{{\bf{I}}\in{\mathcal{I}}_{\nogg}^{({\deg})}} X_{\bf{I}}\sigma^{\bf{I}} \;|\; X_{\bf{I}}\in{\mathbb{C}}\bigg\}\to{\fC}^{({\deg})},\;\;
{\fC}_{\nogg}^{[{\deg}]}
=\bigg\{
X=\sum_{{\bf{I}}\in{\mathcal{I}}_{\nogg}^{[{\deg}]}} X_{\bf{I}}\sigma^{\bf{I}} \;|\;X_{\bf{I}}\in{\mathbb{C}}
\bigg\}
\to {\fC}^{[{\deg}]}.
\end{gather*}
}
\end{claim}

\begin{claim}
We denote the natural projection from ${\fC}$ to ${\fC}_{\nogg}$ as $p_{\nogg}$ defined by
$$
p_{\nogg}:{\fC} \ni X=\sum_{{\bf{I}}\in{\mathcal{I}}}X_{\bf{I}}\sigma^{\bf{I}} \to X_{\nogg}=p_{\nogg}(X)=\sum_{{\bf{I}}\in{\mathcal{I}}_{\nogg}}X_{\bf{I}}\sigma^{\bf{I}}.
$$
The projection $p_{\nogg}$ from ${\fC}_{\infty}$ onto ${\fC}_{\nogg}$ is continuous and open for any ${\nogg}\ge0$.
\end{claim}

\begin{remark}
Same holds for ${\fR}$, ${\fR}_{\nogg}$, ${\fR}_{{\nogg},{\mathrm{ev}}}$, ${\fR}_{{\nogg},{\mathrm{od}}}$, ${\fR}_{\nogg}^{(\deg)}$ and ${\fR}_{\nogg}^{[\deg]}$.
We remark also
$$
\dim_{\mathbb{C}}{\fC}_{\nogg}=2^{\nogg},\; \dim_{\mathbb{R}}{\fR}_{\nogg}=2^{{\nogg}+1}-1.
$$
For supersmooth functions on ${\mathfrak{U}}_{\nogg}\subset {\fR}_{\nogg}^{m|n}$, we have the same conclusion as Theorem \ref{finiteCRBG}.
\end{remark}

\begin{lem}\label{yg-vv}
Suppose that there exist elements $\{A_i\}_{i=1}^{\infty}\subset {\rod}$ satisfying
\begin{equation}
\sigma_j A_i+ \sigma_i A_j=0 \forany i,j\in{\mathbb{N}}
\label{VV2.19-1}
\end{equation}
Then there exists a unique element $F\in {\fR}$ such that $A_i=\sigma_i F$ for $i=1,{\cdots},{\infty}$.
\end{lem}
\par{\it Proof}. We follow the argument in Lemma 4.4 of \cite{yag88}. Since $A_i$ is represented by $A_i=\sum_{{\bf{J}}\in{\mathcal{I}}}a_{\bf{J}}^i\sigma^{\bf{J}}$ with $a_{\bf{J}}^i\in{\mathbb{C}}$
and $\sigma_iA_i=0$, we have $\sum_{\{{\bf{J}}\,|\,j_i=0\}}a_{\bf{J}}^i\sigma^{\bf{J}}=0$.
Therefore, each $A_i$ can be written uniquely as
$A_i=(\sum_{\{{\bf{J}}\,|\,j_i=0\}}b_{\bf{J}}^i\sigma^{\bf{J}})\sigma_i$ for some $b_{\bf{J}}^i\in\mathbb{C}$.
From the condition \eqref{VV2.19-1}, we have $b_{\bf{J}}^i=b_{\bf{J}}^j$ for ${\bf{J}}$ with $j_i=j_j=0$.
Letting $b_{\bf{J}}=b_{\bf{J}}^i$ for ${\{{\bf{J}}\,|\,j_i=0\}}$, we put 
$$
F=\sum_{{\bf{J}}\in{\mathcal{I}}}b_{\bf{J}}\sigma^{\bf{J}}=\sum_{i=1}^{\infty}(\sum_{\{{\bf{J}}\,|\,j_i=0\}\in{\mathcal{I}}}b_{\bf{J}}^i\sigma^{\bf{J}})
$$
which is well-defined and further more $A_i=\sigma_ i F$ holds for each $i$. 
Since we may change the order of summation freely in $\fR$, we have
$$
F=\sum_{\{{\bf{J}}\,|\,j_i=0\}}b_{\bf{J}}\sigma^{\bf{J}}+\sum_{\{{\bf{J}}\,|\,j_i\neq0\}}b_{\bf{J}}\sigma^{\bf{J}}
=\sum_{\{{\bf{J}}\,|\,j_j=0\}}b_{\bf{J}}\sigma^{\bf{J}}+\sum_{\{{\bf{J}}\,|\,j_j\neq0\}}b_{\bf{J}}\sigma^{\bf{J}}. \qed
$$

Repeating above argument, we have
\begin{cor}[Lemma 4.4 of \cite{yag88}]
Let $\{A_{\bf{J}}\in{\mathfrak{R}}\;|\; |{\bf{J}}|={\mathrm{od}}\}$ satisfy 
$$
\sigma^{\bf{K}}A_{\bf{J}}+\sigma^{\bf{J}}A_{\bf{K}}=0\for {\bf{J}},{\bf{K}}\in{\mathcal{I}}_{\mathrm{od}}.
$$
Then there exists a unique element $F\in {\fR}$ such that $A_{\bf{J}}=\sigma^{\bf{J}} F$ for ${\bf{J}}\in{\mathcal{I}}_{\mathrm{od}}$.
\end{cor}

\begin{defn}
We denote the set of maps $f:{\rod}\to{\fR}$ which are 
continuous and ${\fR}_{\mathrm{ev}}$-linear
(i.e. $f(\lambda X)=\lambda f(X)$ for $\lambda\in {\rev}$, $X\in {\rod}$)
by $f\in {\bf{L}}_{\rev}({\rod}: {\fR})$.
\end{defn}
\begin{cor}[The \underline{self-duality} of ${\fR}$]\label{sd-FA}
For $f\in {\bf{L}}_{\rev}({\rod}: {\fR})$, 
there exists an element $u_f\in{\fR}$ satisfying
$$
f(X)=X\!{\cdot} u_f\for X\in {\rod}.
$$
\end{cor}
\par{\it Proof}. 
Since $f:{\rod}\to {\fR}$ is ${\rev}$-linear, we have $f(XYZ)=XYf(Z)=-XZf(Y)$ for any $X, Y, Z\in {\rod}$.
By putting $X=\sigma_k, Y=\sigma_j, Z=\sigma_i$ and $f_i=f(\sigma_i)\in{\fR}$ for $i=1,{\cdots},{\infty}$,
we have $\sigma_k(\sigma_jf_i+\sigma_if_j)=0$ for any $k$. 
Therefore, $\sigma_jf_i+\sigma_if_j=0$, and by Lemma above,
there exists $u_f\in {\mathfrak{R}}$ such that $f_i=\sigma_i u_f$ for $i=1,{\cdots},{\infty}$.
\par
For ${\bf{I}}=(i_1,{\cdots})\in \{0,1\}^{\mathbb{N}}$ with $|{\bf{I}}|={\mathrm{odd}}$ and $i_k=1$ for some $k$, 
rewriting ${\bf{I}}=(-1)^{i_1+{\cdots}+i_{k-1}}\sigma^{\check{{\bf{I}}}_k}\sigma_k$ 
with ${\check{{\bf{I}}}_k}=({i_1,{\cdots},i_{k-1},0},i_{k+1},{\cdots})$,
by  ${\rev}$-linearity of $f$, we have $f(\sigma^{\bf{I}})=(-1)^{i_1+{\cdots}+i_{k-1}}\sigma^{\check{{\bf{I}}}_k}f(\sigma_k)$. 
Then, this map is well-defined because of $\sigma_jf_i+\sigma_if_j=0$, that is, it doesn't depend on other decomposition of ${\bf{I}}$. 
In fact, for ${\bf{I}}=({{i_1,{\cdots},i_{j-1},1},i_{j+1},{\cdots},i_{k-1},1},i_{k+1},{\cdots})$, we put
$\tilde{\bf{I}}=({{i_1,{\cdots},i_{j-1},0},i_{j+1},{\cdots},i_{k-1},0},i_{k+1},{\cdots})$.
Then, for ${\ell}=i_1+{\cdots}+i_{j-1}+i_{j+1}+{\cdots}+i_{k-1}\in{\mathbb{N}}$, 
remarking $|\tilde{\bf{I}}|=$odd, we have
$$
{\sigma^{\bf{I}}=\begin{cases}
(-1)^{\ell}\sigma_k\sigma_j\sigma^{\tilde{\bf{I}}}\\
-(-1)^{\ell}\sigma_j\sigma_k\sigma^{\tilde{\bf{I}}},
\end{cases}}
\Longrightarrow
{\begin{cases}
f((-1)^{\ell}\sigma_k\sigma_j\sigma^{\tilde{\bf{I}}})=(-1)^{\ell}\sigma_j\sigma^{\tilde{\bf{I}}}f(\sigma_k),\\
f(-(-1)^{\ell}\sigma_j\sigma_k\sigma^{\tilde{\bf{I}}})=-(-1)^{\ell}\sigma_k\sigma^{\tilde{\bf{I}}}f(\sigma_j).
\end{cases}}
$$
By $\sigma_jf_k+\sigma_kf_j=0$, we have 
$$
f((-1)^{\ell}\sigma_k\sigma_j\sigma^{\tilde{\bf{I}}})-f(-(-1)^{\ell}\sigma_j\sigma_k\sigma^{\tilde{\bf{I}}})
=-(-1)^{\ell}\sigma^{\tilde{\bf{I}}}[\sigma_jf_k+\sigma_kf_j]=0.
$$

We extend $\tilde{f}$ as
$\tilde{f}(X)=\sum_{{\bf{I}}\in{\mathcal{I}}} X_{\bf{I}}{f}(\sigma^{\bf{I}})$ for $X=\sum_{{\bf{I}}\in{\mathcal{I}}} X_{\bf{I}}\sigma^{\bf{I}}\in {\rod}$.
Then, since $X_{\bf{I}}\in\mathbb{C}$,
$\tilde{f}(X)=\sum_{{\bf{I}}\in{\mathcal{I}}} X_{\bf{I}} \sigma^{\bf{I}} u_f=X\!{\cdot} u_f$.
In fact, if ${\bf{I}}$ with $|{\bf{I}}|$=odd with $i_k\neq0$, then,
by $f_k=f(\sigma_k)=\sigma u_f$ and $\rev$-linearity,
$$
\tilde{f}(\sigma^{\bf{I}})={f}(\sigma^{\bf{I}})=(-1)^{i_1+{\cdots}+i_{k-1}}\sigma^{{\check{{\bf{I}}}_k}}f(\sigma_k)
=(-1)^{i_1+{\cdots}+i_{k-1}}\sigma^{{\check{\bf{I}}_k}}\sigma_k u_f=\sigma^{\bf{I}} u_f.
$$
Clearly $\tilde{f}(X)=f(X)$. \qed

\subsubsection{${\mathfrak{B}}_{\nogg}$ is not self-dual}
K. Masuda gives the following example which exhibits that \underline{${\mathfrak{B}}_{\nogg}$ is not neces-}\\
\underline{sarily self-dual}.
\par{A counter-example}: {\it
Let ${\nogg}=2$. Define a map $f$ as
$$
f(X_1\sigma_1+X_2\sigma_2)=X_1\sigma_2 \forany X_1, X_2\in\mathbb{R}.
$$
Then, remarking that $(b_0+b_1\sigma_1\sigma_2)(X_1\sigma_1+X_2\sigma_2)=b_0(X_1\sigma_1+X_2\sigma_2)$,
we have readily $f\in {\bf{L}}_{{\mathfrak{B}}_{2,{\mathrm{ev}}}}({\mathfrak{B}}_{2,{\mathrm{od}}}:{\mathfrak{B}}_{2})$.
If we assume that there exists a $u_f\in {\mathfrak{B}}_{2}$ such that $f(X)=X\!{\cdot} u_f$, then
$\sigma_1  f(\sigma_1)=\sigma_1\cdot\sigma_1\!{\cdot} u_f=0$ but
$\sigma_1f(\sigma_1)=\sigma_1\cdot\sigma_2\neq0$, contradiction!
Hence, there exists no $u_f\in {\mathfrak{B}}_{2}$ such that $f(X)=X\!{\cdot} u_f$.
}

\subsection{${\fC}$-valued functions and superdomains}
\begin{lem}\label{lem3-1.1}
Let $\phi(t)$ and $\Phi(t)$ be continuous ${\mathfrak{C}}$-valued functions on an interval 
$[a,b]\subset{\euc}$. Then,
\par
(1) $\int_a^b dt\,\phi(t)$ exists,
\par
(2) if $\Phi'(t)=\phi(t)$ on $[a,b]$, then
$\displaystyle{
\int_a^b dt\,\phi(t)=\Phi(b)-\Phi(a)},
$
\par
(3) if $\lambda\in{\fC}$ is a constant, then
$$
\int_a^b dt\,(\phi(t)\cdot\lambda)=\bigg(\int_a^b dt\,\phi(t)\bigg)\cdot\lambda
\et
\int_a^b dt\,(\lambda\cdot\phi(t))=\lambda\cdot\int_a^b dt\,\phi(t).
$$
\end{lem}

Moreover, we may generalize above lemma for
a ${\fC}$-valued function $\phi(q)$ on an open set 
$\Omega\subset{\euc}^m$.

\begin{defn}\label{def3-1.1}
We put $\pi_{\mathrm{B}}^{-1}(U)=\{X\in \supermo\;|\; \pi_{\mathrm{B}}(X)\in U\}$ for a set $U\subset\mathbb{R}^m$.
A set ${\mathfrak{U}}_{\mathrm{ev}}\subset\supermo$ 
is called an \underline{even superdomain} 
if $\pi_{\mathrm{B}}({\mathfrak{U}}_{\mathrm{ev}})={\mathfrak{U}}_{\mathrm{ev},\mathrm{B}}\subset {\mathbb R}^m$ is open and connected and
$\pi_{\mathrm{B}}^{-1}({\mathfrak{U}}_{\mathrm{ev},\mathrm{B}})={\mathfrak{U}}_{\mathrm{ev}}$.
When $ {\mathfrak{U}} \subset \mathfrak{R}^{m|n}$ is represented by
${\mathfrak{U}}={\mathfrak{U}}_{\mathrm{ev}} \times {\mathfrak{R}}_{\mathrm{od}}^n$ with a even superdomain 
${\mathfrak{U}}_{\mathrm{ev}}\subset\supermo$,
${\mathfrak{U}}$ is called a \underline{superdomain} in ${\fR}^{m|n}$.
A set ${\mathfrak{U}}$ is called \underline{coarse-open} in $\mathfrak{R}^{m|n}$ if there exists an open set $U\subset{\euc}^m$ such that ${\mathfrak{U}}=\pi_{\mathrm{B}}^{-1}(U)$.
\end{defn}

\subsection{Superdifferentiability of functions on ${\fR}^{m|n}$}
Since ${\fR}$, ${\fC}$ and ${\fR}^{m|n}$ are Fr\'echet spaces, we define $G$- or $F$-differentiability of functions between them as in subsection \ref{GFonF}.

\begin{defn}\label{yagi247} 
Let $f$ be a ${\fC}$-valued function on a superdomain ${\mathfrak{U}}\subset {\fR}^{m|n}$.
Then, a function $f$ is said to be super $C_G^{1-}$-differentiable, denoted by  $f\in {\G}_{S\!D}^{1-}({\mathfrak{U}}:\fC)$ or simply $f\in {\G}_{S\!D}^{1-}$ {if} there exist  ${\fC}$-valued functions $F_A(X)$ $(1\le A\le m+n)$ on ${\mathfrak{U}}$
such that
\begin{equation}
\dt f(X+tH)\bigg|_{t=0}=f_G'(X;H)=\sum_{A=1}^{m+n}H_A F_A(X)\;\;\mbox{for each $X\in {\mathfrak{U}}$ and $H\in {\fR}^{m|n}$}
\label{sg1}
\end{equation}
where $ f(X+tH)$ is considered as a ${\fC}$-valued differentiable function w.r.t. $t\in{\euc}$.
Moreover, if $F_A(X)$ is continuous, we say $f$ is super $C_G^{1}$-differentiable and is denoted by 
$f\in {\G}_{S\!D}^{1}$.
We denote $F_A(X)$ by $f_{X_A}(X)$. 
Moreover, for $r\ge 2$, $f$ is said to be in ${\G}_{S\!D}^{r-}$ if  $F_A$ are ${\G}_{S\!D}^{r-1}$.
$f$ is said to be ${\G}_{S\!D}^{\infty}={\G}_{S\!D}^{\infty}({\mathfrak{U}}:\fC)$ or super G-differentiable if $f$ is ${\G}_{S\!D}^r$ for all $r\ge 1$.
\end{defn}

\begin{defn}
Let $f$ be a ${\fC}$-valued function on a superdomain ${\mathfrak{U}}\subset  {\fR}^{m|n}$.
 A function $f$ is said to be  super $C_F^1$-differentiable, denoted by  $f\in {\F}_{S\!D}^1({\mathfrak{U}}:\fC)$ 
 or simply $f\in {\F}_{S\!D}^1$ {if} there exist  ${\fC}$-valued continuous functions $F_A$ $(1\le A\le m+n)$ on ${\mathfrak{U}}$
and functions $\rho_A:{\mathfrak{U}}\times{\fR}^{m|n}\to {\fC}$ 
such that 
\begin{equation}
\begin{aligned}
& (a)\; f(X+H)-f(X)=\sum_{{j}=1}^{m+n}H_AF_A(X)+\sum_{A=1}^{m+n}H_A \rho_A(X;H)
\for X\in {\fR}^{m|n},\\
& (b)\; \rho_A(X,H)\to 0 \;\;\mbox{in}\;\; {\mathfrak{C}} \when H\to 0 \;\;\mbox{in}\;\; {\mathfrak{R}}^{m|n},
\end{aligned}
\label{sf1}
\end{equation}
for each $X\in {\mathfrak{U}}$ and $X+H\in {\mathfrak{U}}$.
$f$ is said to be super  $C_F^2$-differentiable,
when $F_A \in {\F}_{S\!D}^1(\mathfrak{R}^{m|n}:\fC)$ $(1\le A\le m+n)$. We define analogously the
super $C_F^{r}$-differentiablity. A function $f$ is called super $C_F^{\infty}$-differentiable, or simply
{super F-differentiable} if it is in
${\F}_{S\!D}^{\infty}=\cap_{r=0}^{\infty}{\F}_{S\!D}^r(\mathfrak{R}^{m|n}:\fC)$.
\end{defn}

\begin{remark}
(i) If $f$ is super Fr\'echet-differentiable at $x$, we have
$$
\begin{gathered}
{f}'_G(x;h)={f}'_F(x)h=\sum_{j=1}^m f'_{x_j}(x;h_j)=\sum_{j=1}^m{h_j}F_j(x)=\langle h|(F_j(x))\rangle\\ \with
F_j(x)=\frac{\partial f}{\partial x_j}(x)={f}'_G(x;{\bf{e}}_j),\; 
{\bf{e}}_j=(\underbrace{\overbrace{0,{\cdots},0,1}^{j},0,{\cdots},0}_{m}).
\end{gathered}
$$
(ii) Let ${\mathfrak{V}}$ be an open set ${\fC}^{m|n}$.
When $f:{\mathfrak{V}}\to{\fC}$ is in ${\F}_{S\!D}^{\infty}$, 
$f$ is also said to be {superanalytic}.
It is clear that if $f:{\fC}^{m|0}\to\fC$ is superanalytic, then $f(z_{\mathrm{B}})$ for $z_{\mathrm{B}}\in{\mathbb{C}}^m$ is analytic in ordinary sense. 
Therefore, superanalyticity is another name of super Fr\'echet-differentiability for functions from
${\fC}^{m|0}$ to ${\fC}$.
\end{remark}

\subsection{Definition and properties of the Grassmann continuation}\label{sub4.5}
\paragraph{\bf{Existence of the Grassmann continuation: Proof of Lemma \ref{GC-lemma}}}
Denoting by 
$$
\begin{gathered}
x_1=x_{1,{\mathrm{B}}}+x_{1,{\mathrm{S}}}\with
x_{1,{\mathrm{B}}}=X_{1,\emptyset}\in{\mathbb{R}},\;\;
x_{1,{\mathrm{S}}}=\sum_{|{\bf{I}}|=k_1=\mathrm{ev}\ge2}X_{1,{\bf{I}}}\sigma^{\bf{I}} \where X_{1,{\bf{I}}}\in{\mathbb{C}},\\
x_{1,{\mathrm{S}}}^{[{k}_1]}=\sum_{|{\bf{I}}|=k_1}X_{1,{\bf{I}}}\sigma^{\bf{I}}
=\mbox{ 
the ${k}_1$-th degree component of $x_{1,{\mathrm{S}}}$},
\end{gathered}
$$
we put
$$
(x_{1,{\mathrm{S}}}^{\alpha_1 })^{[{k}_1]}
=\sum_{\scriptstyle{\alpha_1=p_{1,1}+{\cdots}+p_{1,\ell_1}}\atop\scriptstyle{\sum_{i=1}^{\ell_1} r_{1,i} p_{1,i} ={k}_1,\, r_{1,i}\ge2 }}
(x_{1,{\mathrm{S}}}^{[r_1]})^{p_{1,1}}
{\cdots} \,
(x_{1,{\mathrm{S}}}^{[{r_{\ell_1}}]})^{p_{1,\ell_1}}\et
(x_{\mathrm{S}}^\alpha)^{[k]}=\sum_{k=k_1+{\cdots}+k_m}(x_{1,{\mathrm{S}}}^{\alpha_1 })^{[{k}_1]}{\cdots}
(x_{m,{\mathrm{S}}}^{\alpha_m })^{[{k}_m]}.
$$
Using these notations, we define
\begin{equation}
{\tilde f}^{[k]}(x) = \sum_{2|\alpha|\le k}\frac{\partial_q^\alpha f(q)} {\alpha ! }\Bigg|_{q=x_{\mathrm{B}}}
\sum_{k=k_1+{\cdots}+k_m}(x_{1,{\mathrm{S}}}^{\alpha_1})^{[{k}_1]} {\cdots} 
(x_{m,{\mathrm{S}}}^{\alpha_m})^{[{k}_m]}.
\label{II-3-3-1}
\end{equation}
For example, we have
{\allowdisplaybreaks
\begin{align*}
{\tilde f}^{[0]}(x)& = f(x_{\mathrm{B}} ),\\
{\tilde f}^{[1]}(x) & =0,\\
{\tilde f}^{[2]}(x) &= \sum_{{j}=1}^m\partial_{q_j} f (x_{\mathrm{B}})(x_{j,{\mathrm{S}}})^{[2]},\\
{\tilde f}^{[3]}(x) & =0,\\
{\tilde f}^{[4]}(x)  &= \sum_{{j}=1}^m\partial_{q_j} f (x_{\mathrm{B}})(x_{j,{\mathrm{S}}})^{[4]}
+\frac{1}{2}\sum_{{j}=1}^m\partial_{q_j}^2 f(x_{\mathrm{B}})(x_{j,{\mathrm{S}}}^2)^{[4]}
+\sum_{j\neq k}\partial_{q_j\,q_k}^2 f(x_{\mathrm{B}})
(x_{j,{\mathrm{S}}})^{[2]}(x_{k,{\mathrm{S}}})^{[2]},\quad etc.\\
\end{align*}
}
Since ${\tilde f}^{[j]}(x)\ne {\tilde f}^{[k]}(x)\,\,(j\ne k)$ in ${\fC}$,
we may take the sum 
$$
{\tilde f}(x)=\sum_{k=0}^\infty {\tilde f}^{[k]}(x)=\sum_{k=0}^\infty \sum_{2|\alpha|\le k}\frac{\partial_q^\alpha f(x_{\mathrm{B}})} {\alpha ! }(x_{\mathrm{S}}^\alpha)^{[k]}
=\sum_{|\alpha|=0}^{\infty}\frac{\partial_q^{\alpha}f(x_{\mathrm{B}})}{\alpha ! }x_{\mathrm{S}}^\alpha
\in {\fC}=\oplus_{k=0}^\infty{\fC}^{[k]}.\qed
$$

\begin{remark}
In \eqref{II-3-3-1}, the sum w.r.t. $\alpha$ is finite but $(x_{1,{\mathrm{S}}}^{\alpha_1 })^{[{k}_1]}$ is not finite sum w.r.t. ${\bf{J}}\in{\mathcal{I}}$ for $|{\bf{J}}|=k_1$.
\end{remark}

\begin{cor} 
Let $U\subset{\euc}^m$ be an open set and let $f\in C^{\infty}(U:{\fC})$ be represented by
\begin{equation}
f(q)=\sum_{{\bf{J}}\in{\mathcal{I}}} f_{\bf{J}}(q)\,\sigma^{\bf{J}} \with 
f_{\bf{J}}(q)\in C^\infty({\mathfrak{U}}_{\mathrm{ev,B}}:{\mathbb{C}})
\quad \text{for each ${\bf{J}}\in{\mathcal{I}}$}.
\label{II-3-1}
\end{equation}
Then, we may define a mapping
$\tilde f$ from
${\mathfrak{U}}_{\mathrm{ev}}$ into $ {\fC} $, 
called the Grassmann continuation of $f$, by
\begin{equation}
{\tilde f}(x) = \sum_{ |\alpha|\ge 0 } \frac{1}{\alpha ! }
\partial_q^\alpha f(x_{\mathrm{B}} )\,{x_{\mathrm{S}}^\alpha}
\where
\partial_q^\alpha f(x_{\mathrm{B}})
=\sum_{\bf{J}} \partial_q^\alpha f_{\bf{J}}(x_{\mathrm{B}})\,\sigma^{\bf{J}} .
\label{II-3-2}
\end{equation}
\end{cor}
\par
Since this is obtained analogously as  above, proof is omitted here.

\begin{cor}[Corollary 2.3 of \cite{IM91}]\label{cor3-1.3}
If $f$ and $\tilde f$ be given as above, 
then (i) $\tilde f$ is continuous, (ii) $\tilde f(x)=0$ in ${\mathfrak{U}}_{\mathrm{ev}}$ 
implies $f(x_{\mathrm{B}})=0$ in ${\mathfrak{U}}_{\mathrm{ev},\mathrm{B}}$ and (iii)
 if we define the partial derivatives of $\tilde f$ by
\begin{equation}
\partial_{x_j} \tilde f (x)
=\left.{\frac{d}{dt}}\tilde f(x+t\,{\bf{e}}_{j})\right|_{t=0} 
\where
{\bf{e}}_{j}=({\overbrace{0,{\cdots},0,1}^{j},0,{\cdots},0})\in{\supermo},
\label{II-3-4}
\end{equation}
then we get
\begin{equation}
\partial_{x_j} \tilde f (x)={\widetilde{\partial_{q_j} f}}(x)
\for j=1,{\cdots},m.
\label{II-3-5}
\end{equation}
Analogously, we have
\begin{equation}
\partial_x^{\alpha} \tilde{f}(x)=\widetilde{\partial_q^{\alpha}f({\cdot})}(x).
\label{6.6}
\end{equation}
(iv) Moreover, for $y=(y_1,{\cdots},y_m)\in\supermo$,
\begin{equation}
\left.{\frac{d}{dt}}\tilde f(x+ty)\right|_{t=0}=
\sum_{{j}=1}^m y_j 
\sum_\alpha {\frac{1}{\alpha!}} \partial_q^\alpha \partial_{q_j}f(x_{\mathrm{B}})
\,x_{\mathrm{S}}^\alpha=\sum_{{j}=1}^m y_j\partial_{x_j} \tilde f (x).
\label{II-3-6}
\end{equation}
\end{cor}
\par
{\it Proof}.
Let $y_j=y_{j,\mathrm{B}}+y_{j,\mathrm{S}} \in\rev$. 
For $y_{(j)}=y_j{\bf{e}}_{j}=y_{j,\mathrm{B}}{\bf{e}}_{j}+y_{j,\mathrm{S}}{\bf{e}}_{j}
=y_{(j),\mathrm{B}}+y_{(j),\mathrm{S}}\in{\supermo}$,
as
$$
{\frac{d}{dt}}\tilde f(x+ty_{(j)})={\frac{d}{dt}}
\left\{
\sum_{|\alpha|=0}^{\infty} {\frac{1}{\alpha!}}
\left(
\sum_{j} \partial_q^\alpha f_{j}(x_{\mathrm{B}}+ty_{(j),\mathrm{B}})\sigma^{j}
\right)
(x_{\mathrm{S}}+ty_{(j),\mathrm{S}})^\alpha \right\},
$$
we get easily, 
$$
\begin{aligned}
{\frac{d}{dt}}\tilde f(x+ty_{(j)})\Big|_{t=0}
&=y_{(j),\mathrm{B}}\sum_{|\alpha|=0}^{\infty}{\frac{1}{\alpha!}}
\left(
\sum_{j} \partial_{q_j}\partial_q^\alpha f_{j}(x_{\mathrm{B}})\sigma^{j}
\right)
x_{\mathrm{S}}^\alpha
+y_{(j),\mathrm{S}}\sum_{|\check{\alpha}|=0}^{\infty} {\frac 1{{\check{\alpha}}!}}
\left(
\sum_{j} \partial_q^{\alpha} f_{j}(x_{\mathrm{B}})\sigma^{j}
\right)x_{\mathrm{S}}^{\check{\alpha}}\\
&=y_j\sum_\alpha {\frac{1}{\alpha!}} 
\partial_q^\alpha \partial_{q_j}f(x_{\mathrm{B}}) \,x_{\mathrm{S}}^\alpha 
=y_j{\widetilde{\partial_{q_j}f}}(x).
\end{aligned}
$$
Here ${\check{\alpha}}=(\alpha_1,{\cdots},\alpha_j-1,{\cdots},\alpha_m)$.
Putting $y_j=1$ in the above and by \eqref{II-3-4}, we have \eqref{II-3-5}.
Last equality is proved by induction with the length $|\alpha|$. \qed

More generally,
\begin{lem}\label{XXX}
Let $f(q)\in C^\infty({\euc}^m)$, 
we have the Taylor formula for ${\tilde f}$:
For any $N$, there exists $\tilde\tau_N(f;x,y)\in{\fC}$ such that
\begin{equation}
{\tilde f}(x+y)
=\sum_{|\alpha|=0}^N\frac{1}{\alpha!}
\partial_x^{\alpha}{\tilde f}(x)y^\alpha+\tilde\tau_N(f;x,y),
\label{taylor-exp}
\end{equation}
with
$$
\tilde\tau_N(f;x,y)=\sum_{|\alpha|=N+1} y^\alpha\int_0^1 dt\frac{1}{N!}(1-t)^N
\partial_x^\alpha {\tilde f}(x+ty).
$$
\end{lem}
\par{\it Proof. } Substituting $q=x_{\mathrm{B}}$ and $q'=y_{\mathrm{B}}$ in 
$$
{f}(q+q')
=\sum_{|\alpha|=0}^N\frac{1}{\alpha!}
\partial_q^{\alpha}{f}(q)q^{\prime\alpha}
+\sum_{|\alpha|=N+1} q^{\prime\alpha}\int_0^1 dt\frac{1}{N!}(1-t)^N
\partial_q^\alpha {f}(q+tq'),
$$
and extending both sides, we have the desired result by \eqref{II-3-5} because
$\widetilde{\partial_q^{\alpha}f}(x)=\partial_x^{\alpha} f(x),\; \widetilde{q^{\prime\alpha}}=y^{\alpha}$.  \qed

\begin{cor}\label{XXX-cor}
For $f(q)\in C^\infty({\euc}^m)$,
$\tilde{f}(x)\in {\F}_{S\!D}^1({\fR}^{m|0})$.
\end{cor}
\par
{\it Proof}. For $N=1$ in \eqref{taylor-exp}, we have
$$
\tilde{f}(x+y)=\tilde{f}(x)+\sum_{{j}=1}^my_j\partial_{x_j}\tilde{f}(x)+\sum_{{j}=1}^my_j\rho_j(x,y)
$$
with
\begin{equation}
\begin{gathered}
\partial_{x_j} \tilde{f}(x)=\widetilde{\partial_{q_j}f}(x)\in{\fC},\\
\rho_j(x,y)=
\sum_{k=1}^m y_k g_{j,k}(x,y),\;\;
g_{j,k}(x,y)=\int_0^1(1-t)\partial^2_{x_jx_{k}}\tilde{f}(x+ty)dt.
\end{gathered}
\label{f-smooth}
\end{equation}
We want to show
\begin{claim}\label{1025}
If $y\to0$ in ${\fR}_{\mathrm{ev}}^m$, then for each $x\in{\fR}_{\mathrm{ev}}^m$ and $j=1,{\cdots},m$, $\rho_j(x,y)\to0$, i.e,
for any $\epsilon>0$, $j$ and $x\in{\fR}_{\mathrm{ev}}^m$, there exists $\delta>0$ such that if $\dist_{m|0}(y)<\delta$ then
$\dist_{1|0}\rho_j(x,y)<\epsilon$.
\end{claim}
Take ${\bf{I}}\in{\mathcal{I}}_{\mathrm{ev}}$ arbitrarily and decompose it as ${\bf{I}}={\bf{J}}+{\bf{K}}$.
Remarking Corollary \ref{cor3-1.3}, we have
$$
\partial^2_{x_jx_{k}}\tilde{f}(x+ty)=\sum_{|\alpha|=0}^{\infty}\frac{\partial^2_{q_jq_{k}}\partial_q^{\alpha}f(x_{\mathrm{B}}+ty_{\mathrm{B}})}{\alpha!}
(x_{\mathrm{S}}+ty_{\mathrm{S}})^{\alpha}.
$$
If ${\bf{I}}=\tilde0$, we have
$$
|\proj_{\tilde0}\rho_j(x,y)|\le \sum_{k=1}^m |y_{k,\mathrm{B}}||\pi_{\mathrm{B}}g_{j,k}(x,y)|
\to0\when y_{\mathrm{B}}\to0.
$$
For each fixed ${\bf{I}}\neq\tilde0$, there exist a family of finite index sets $\{{\bf{K}}\;|\;{\bf{K}}\subset {\bf{I}}\}$,  
$\{x_{\mathrm{B}}+ty_{\mathrm{B}}\;|\; t\in[0,1],\;|y_{\mathrm{B}}|\le1\}$ is compact 
and ${\proj}_{\bf{K}}(x_{\mathrm{S}}+ty_{\mathrm{S}})^{\alpha}=0$ if $2|\alpha|> |{\bf{K}}|$.
Therefore, 
we may find constant $C_{\bf{K}}=C_{\bf{K}}(x_{\mathrm{S}},y_{\mathrm{S}})$ such that
$$
\bigg|{\proj}_{\bf{K}}\big(\sum_{|\alpha|=0}^{\infty}\frac{\partial^2_{q_jq_{k}}\partial_q^{\alpha}f(x_{\mathrm{B}}+ty_{\mathrm{B}})}{\alpha!}
(x_{\mathrm{S}}+ty_{\mathrm{S}})^{\alpha}\big)\bigg|
\le
\sum_{\alpha}\frac{\big|\partial^2_{q_jq_{k}}\partial_q^{\alpha}f(x_{\mathrm{B}}+ty_{\mathrm{B}})\big|}{\alpha!}\big|{\proj}_{\bf{K}}(x_{\mathrm{S}}+ty_{\mathrm{S}})^{\alpha}\big|\le C_{\bf{K}}.
$$
In fact, as
$$
\begin{aligned}
\big|{\proj}_{\bf{K}}(g_{j,k}(x,y))\big|&\le\int_0^1 dt\, (1-t)\big|{\proj}_{\bf{K}}(\partial^2_{x_jx_{k}}\tilde{f}(x+ty))\big|\\
&\le\sum_{2|\alpha|\le|{\bf{I}}|}\frac{\max_{t}\big|\partial^2_{q_jq_{k}}\partial_q^{\alpha}f(x_{\mathrm{B}}+ty_{\mathrm{B}})\big|}{\alpha!}
\int_0^1dt\,\big|{\proj}_{\bf{K}}(x_{\mathrm{S}}+ty_{\mathrm{S}})^{\alpha}\big|,
\end{aligned}
$$
with ${\proj}_{\bf{K}}(x_{\mathrm{S}}+ty_{\mathrm{S}})^{\alpha}=0$ if $2|\alpha|> |{\bf{K}}|$,
$$
C_{\bf{K}}=C_{\bf{K}}(x_{\mathrm{S}},y_{\mathrm{S}})=\int_0^1dt\,\big|{\proj}_{\bf{K}}(x_{\mathrm{S}}+ty_{\mathrm{S}})^{\alpha}\big|\to0\;\;\mbox{in ${\mathbb{C}}$ when $y_{\mathrm{S}}\to 0$ in ${\mathfrak{C}}$},
$$
we have
$$
\big|{\proj}_{\bf{I}}(\rho_j(x,y))\big|\le\sum_{k=1}^m\sum_{{\bf{I}}={\bf{J}}+{\bf{K}}}\big|{\proj}_{\bf{J}}(y_k)\big|\big|{\proj}_{\bf{K}}(g_{j,k}(x,y))\big|
\le \sum_{k=1}^m\sum_{{\bf{I}}={\bf{J}}+{\bf{K}}}\big|{\proj}_{\bf{J}}(y_k)\big|C_{\bf{K}},
$$
and this finite sum tends to $0$ when $y\to0$, which implies $\tilde{f}(x)\in {\F}_{S\!D}^1({\fR}^{m|0})$. \qed

\begin{remark}
 Concerning the meaning of the Grassmann continuation $\tilde{f}$ of $f$, 
in p.7 of \cite{DW2}, de Witt claimed as follows:
{\begin{quotation}
``The presence of a soul in the independent variable evidently has little practical effect on the variety of functions with which one may work in applications of the theory. In this respect $\rev$ is a harmless generalization of its own subspace ${\euc}$, the real line."
\end{quotation}}
\end{remark}

To guarantee his intuitional claim in a certain sense, we have
\begin{prop}\label{dew-cl}
Let $F\in C_G^{\infty}({\fR}^{m|0}:{\fC})$.  
Putting  $f(q)=F(q)$ for $q\in{\euc}^m$, we have $\tilde{f}=F$.
\end{prop}
\par
{\it Proof.} 
We claim that
if $H=H(x)=\sum_{{\bf{J}}\in{\mathcal{I}}}\sigma^{\bf{J}}H_{\bf{J}}(x)$ is $C_G^{\infty}$-differentiable from ${\fR}^{m|0}$ to ${\fC}$ which is $0$ on ${\euc}^m$, i.e. $\partial_q^{\alpha}H_{\bf{J}}(q)=0$ for any $\alpha$ and $q\in\euc^m$,
then  $H$  is $0$ on ${\fR}^{m|0}$, that is, for any ${\bf{I}}\in{\mathcal{I}}$, 
we should have ${\proj}_{\bf{I}}(H(x))=0$. 

To show this, 
we apply the Taylor formula \eqref{tay-f} on general Fr\'echet space to have, for any $N$ and ${\bf{J}}$,
$$
H_{\bf{J}}(x_{\mathrm{B}}+x_{\mathrm{S}})=\tau_N(H_{\bf{J}};x_{\mathrm{B}},x_{\mathrm{S}})
=\sum_{|\alpha|=N+1} x_{\mathrm{S}}^\alpha\int_0^1 dt\frac{1}{N!}(1-t)^N\partial_x^\alpha H_{\bf{J}}(x_{\mathrm{B}}+tx_{\mathrm{S}}).
$$
Now, we need to show ${\proj}_{\bf{I}}(\tau_N(H_{\bf{J}};x_{\mathrm{B}},x_{\mathrm{S}}))=0$.
Since all terms in $x_{\mathrm{S}}^{\alpha}$ have degree at least $2|\alpha|$,
if $2|\alpha|>|{\bf{I}}|\ge0$, then ${\proj}_{\bf{I}}(x_{\mathrm{S}}^{\alpha})=0$.
Taking $N$ sufficiently large such that for all $\alpha$ with $|\alpha|=N+1$
with ${\proj}_{\bf{I}}(x_{\mathrm{S}}^{\alpha})=0$, we have ${\proj}_{\bf{I}}(\tau_N(H_{\bf{J}};x_{\mathrm{B}},x_{\mathrm{S}}))=0$ for any ${\bf{J}}$.
Therefore, we have
$$
{\proj}_{\bf{I}}(\sum_{{\bf{J}}\in{\mathcal{I}}}\sigma^{\bf{J}}\tau_N(H_{\bf{J}};x_{\mathrm{B}},x_{\mathrm{S}}))=0.
$$
Putting $H(x)=F(x)-\tilde{f}(x)$, we have the desired result.
$\qquad\square$

{\it Notation}: Hereafter, for the sake of notational simplicity, 
\underline{$\tilde f$ is denoted simply by $f$} unless there occurs confusion.

Following is claimed in (1.1.17) of \cite{DW2} without proof and cited as Theorem 1 in \cite{MK}
with ``proof'':
\begin{claim}
Let $f$ be an analytic function on an open set $V\subset{\mathbb{C}}$ to $\mathbb{C}$. 
Then, we may extend $f$ uniquely to a function $\tilde{f}:{\fC}^{1|0}\to{\fC}$ as
\begin{equation}
\tilde{f}(z)=\sum_{n=0}^{\infty}\frac{1}{n!}f^{(n)}(z_{\mathrm{B}})z_{\mathrm{S}}^n
\for z=z_{\mathrm{B}}+z_{\mathrm{S}} \with z_{\mathrm{B}}\in V,
\label{AGC}
\end{equation}
which is superanalytic. That is,
$$
{\tilde f}(z+w)={\tilde f}(z)+wF(z)
+w\rho_j(z,w)
\with
\rho_j(z,w)\to 0\quad\mbox{in ${\fC}$ when $w\to0$ in ${\fC}^{1|0}$}.
$$
\end{claim}
\begin{remark}
We know that above claim is derived from Corollary \ref{XXX-cor} because F-differentiability in $\mathbb{C}$ 
to $\mathbb{C}$ implies $\tilde{f}\in {\F}_{S\!D}^1({\fC}^{1|0}:{\fC})$. Moreover, by Theorem \ref{yag-thm3} bellow, we have $\tilde{f}\in {\F}_{S\!D}^{\infty}({\fC}^{1|0}:{\fC})$, that is, $\tilde{f}$ is superanalytic.

But I feel some insufficiency in their proof itself.  They first claim the convergence of the right-hand side of \eqref{AGC} as is re-guaranteed in subsection \ref{sub4.5}. And using this, they have
$$
\begin{aligned}
{\tilde f}(z+w)&=\sum_{n=0}^\infty \frac{1}{n!}
f^{(n)}(z_{\mathrm{B}}+w_{\mathrm{B}})(z_{\mathrm{S}}+w_{\mathrm{S}})^n\\
&=\sum_{n=0}^\infty \frac{1}{n!}\bigg(\sum_{\ell=0}^\infty \frac{1}{\ell!}
f^{(\ell+n)}(z_{\mathrm{B}})w_{\mathrm{B}}^\ell\bigg)
\bigg(\sum_{k=0}^n
\frac{n!}{k!(n-k)!}z_{\mathrm{S}}^{n-k}w_{\mathrm{S}}^k\bigg)
\quad\text{(analyticity of $f$ on $\mathbb{C}$)}\\
&=\sum_{n=0}^\infty\bigg[\sum_{\ell=0}^\infty \frac{1}{\ell!}
f^{(\ell+j+k)}(z_{\mathrm{B}})w_{\mathrm{B}}^\ell 
\bigg(\sum_{k+j=n}
\frac{1}{k!j!}z_{\mathrm{S}}^{j}w_{\mathrm{S}}^k\bigg)\bigg]\quad\text{(renumbering)}\\
&=\sum_{n=0}^\infty\frac{1}{n!}\bigg[\sum_{{j}=0}^\infty \frac{1}{j!}
f^{(n+j)}(z_{\mathrm{B}})z_{\mathrm{S}}^{j}
\bigg(\sum_{\ell+k=n}
\frac{n!}{\ell!k!}w_{\mathrm{B}}^\ell w_{\mathrm{S}}^k\bigg)\bigg]\quad\text{(rearranging)}\\
&=\sum_{n=0}^\infty\frac{1}{n!}\bigg(\sum_{{j}=0}^\infty \frac{1}{j!}
f^{(n+j)}(z_{\mathrm{B}})z_{\mathrm{S}}^{j}\bigg)
(w_{\mathrm{B}}+w_{\mathrm{S}})^n
=\sum_{n=0}^\infty\frac{1}{n!}{\tilde f}^{(n)}(z)w^n.
\end{aligned}
$$
From this, they claim that $\tilde{f}$ is superanalytic. From their argument, we should have
$$
{\tilde f}(x+y)-{\tilde f}(x)
=F(x)y+y\rho(x,y),
$$
with
$$
F(z)={\tilde f}^{(1)}(z),\;\;
\rho(z,w)=\sum_{n=2}^\infty\frac{1}{n!}{\tilde f}^{(n)}(z)w^{n-1}.
$$
Therefore $F(x)$ should be continuous w.r.t. $z$ and $\rho(z,w)$ should be horizontal w.r.t. $w$.
But this last part of horizontality is not so transparent at least for me.
\end{remark}
\begin{remark}
If $f$ is real analytic on ${\euc}^m$, there exists a function $\delta(q)>0$ such that 
when $|q'|\le\delta(q)$, $f(q+q')$ has the Taylor series expansion at $q$.
From above proof, $\tilde{f}(x+y)$ is Pringsheim regular for $|y_{\mathrm{B}}|\le\delta(x_{\mathrm{B}})$.
\end{remark}

\subsection{Supersmooth functions and their properties}
\begin{defn}\label{def3-1.2}
(1) For a given even superdomain ${\mathfrak{U}}_{\mathrm{ev}}\subset\supermo$,
a mapping $\tilde f $ from ${\mathfrak{U}}_{\mathrm{ev}}$ into ${\fC}$
is called a \underline{supersmooth} function
if $\tilde f$ is the Grassmann continuation of a smooth mapping $f$
from ${\mathfrak{U}}_{\mathrm{ev},\mathrm{B}}=\pi_{\mathrm{B}}({\mathfrak{U}}_{\mathrm{ev}})$ into $ {\mathfrak{C}}$.
We denote by
${\CSS}({\mathfrak{U}}_{\mathrm{ev}}:{\fC} )$, the set of supersmooth functions on ${\mathfrak{U}}_{\mathrm{ev}}$.
\par
(2) A mapping $f$ from a superdomain ${\mathfrak{U}}\subset{\fR}^{m|n}$ to ${\fC}$ 
is called \underline{supersmooth}, 
if it has the following form:
\begin{equation}
f(X)=f(x,\theta) = \sum_{ |a|\le n } \theta^a  f_a (x)\with f_a (x) \in {\CSS} ({\mathfrak{U}}_{\mathrm{ev}}:{\fC})
\label{II-3-7}
\end{equation}
where $ a=(a_1, {\cdots} a_n) \in \{0,1\} ^n $ and
$\theta^a = \theta_1^{a_1} {\cdots} \, \theta_n^{a_n} $.
In the following, {\it supersmooth functions are assumed to be homogeneous}
(i.e., for each $a$, $f_a(x)$ is homogeneous, $f_a(x)\in\cev$ or $\cod$),
unless otherwise mentioned and
we denote the set of them by $ {\CSS}({\mathfrak{U}}:{\fC})$.
More especially, we put
$$
\ccsl_{\mathrm{SS}}({\mathfrak{U}})=\{f(x,\theta)\in{\CSS}({\mathfrak{U}}:{\fC})\,|\, f_a(x_{\mathrm{B}})\in{\mathbb{C}}\}.
$$
\par
(3) For $ f(X)= \sum_{ |a|\le n } \theta^a  f_a (x) \in  {\CSS} ({\mathfrak{U}}:{\fC})$,
$ j=1,2, {\cdots} ,m $ and $ s = 1,2, {\cdots} ,n$,
we put
\begin{equation}
\left \{
\begin{aligned}
&F_{j}(X)= \sum_{|a|\le n}\theta^a{\cdot} \partial_{x_j}  f_a(x) ,\\
&F_{s+m} (X)=
\sum_{|a| \le n} (-1)^{l(a)} 
\theta_1^{a_1}{\cdots}
\theta_{s}^{{a_s}-1}
{\cdots}
\theta_n^{a_n}{\cdot}f_a(x)
\end{aligned}
\right .
\label{II-3-8}
\end{equation}
where $l(a)= \sum_{{j}=1}^{s-1} a_j $ and
$\theta_{s}^{-1}=0$.
$F_{A}(X)$ are called the partial derivatives of $f$ 
with respect to $ X_{A} $ at
$ X = (x, \theta ) $ and are denoted by
\begin{equation}
\begin{cases}
F_{j} (X) = {\displaystyle{{\partial}\over{\partial x_j}}} f(x, \theta )
= \partial_{x_j} f(x, \theta )
=f_{x_j}(x, \theta ) \for j=1,2, {\cdots} ,m,\\
F_{m+s}(X) = {\displaystyle{{\partial}\over{\partial\theta_s}}}f(x,\theta )
= \partial_{\theta_s} f(x, \theta )
=f_{\theta_s}(x, \theta )  \for s = 1,2, {\cdots} ,n \\
\end{cases}
\label{II-3-9}
\end{equation}
or simply by
\begin{equation}
F_{A}(X)={\partial}_{X_{A}}f(X)=f_{X_{A}}(X) 
\for {A}=1,{\cdots},m+n.
\label{II-3-10}
\end{equation}
\end{defn}

\begin{remark} 
(1) We only use the derivatives defined above which are called 
the {\it left derivatives} with respect to odd variables.
(Some people call these as right derivatives, 
cf. Vladimirov and Volovich~\cite{VV83}, etc.) 
Similarly, we define the {\it right derivatives} 
with respect to odd variables as follows:
Put
$$
 \CSS^{(r)} ({\mathfrak{U}}:{\fC})=\{f(x,\theta)=\sum_{|a|\le n} \tilde{f}_a(x)\theta^a\;|\; f_a(x)\in{\CSS} ({\mathfrak{U}}_{\mathrm{ev}}:{\fC})\}.
$$
For $ f \in  \CSS^{(r)} ({\mathfrak{U}}:{\fC})$,
$ j=1,2, {\cdots} ,m $ and $ s = 1,2, {\cdots} ,n$,
we put
$$ 
\left \{
\begin{aligned} 
&F^{(r)}_j(X) = \sum_{|a|\le n} \partial_{x_j} f_a(x){\cdot} \theta^a,\\
&F^{(r)}_{s+m} (X) =
\sum_{|a| \le n} (-1)^{r(a)} f_a(x){\cdot}
\theta_1^{a_1}{\cdots}
\theta_{s}^{{a_s}-1}
{\cdots}
\theta_n^{a_n}
\end{aligned}
\right.
$$ 
where $r(a)= \sum_{{j}=s+1}^{n} a_j $.
$F^{(r)}_{A}(X)$ are called the (right) partial derivatives of $f$ 
with respect to $ X_{A} $ at
$ X = (x, \theta ) $ and are denoted by
$$
F^{(r)}_j (X) = {{\partial}\over{\partial x_j}} f(x, \theta ) 
= \partial_{x_j} f(x, \theta ),\quad
F^{(r)}_{m+s}(X) = f(x, \theta )
\overset{\leftarrow}{\frac{\partial}{\partial\theta_s}}
= f(x, \theta )
{\overset{\leftarrow}{\partial}}_{\theta_s}
$$
for $ j=1,2,{\cdots},m $ and $ s = 1,2,{\cdots},n.$
\newline
(2) As we use the infinite dimensional Grassmann algebras, 
the expression \eqref{II-3-8} is unique. 
In fact, $ \sum_a\theta^a  f_a(x)\equiv 0 $ on $U$ implies $ f_a(x) \equiv 0$
(compare the underlined part in p. 322 of Vladimirov and Volovich~\cite{VV83}.) 
\newline
(3) The higher derivatives are defined analogously and 
we use the following notations.
$$
\partial_x^\alpha = \partial_{x_1}^{\alpha_1} 
{\cdots} \partial_{x_m}^{\alpha_m} \et
\partial_{\theta}^a = 
\partial_{\theta_n}^{a_n} {\cdots}\partial_{\theta_1}^{a_1} ,
$$
for multi-indeces $\alpha=(\alpha_1,{\cdots},\alpha_m)\in({\mathbb N}\cup\{0\})^m$
and $a=(a_1,{\cdots},a_n)\in\{0,1\}^n$.
\end{remark}

Repeating the argument in proving Corollary \ref{cor3-1.3} if necessary, 
we get 
\begin{equation}
 f\in{\CSS}({\mathfrak{U}}:{\fC})\Longrightarrow 
 d_Gf(X;Y)={\frac{d}{dt}}f(X+tY)\bigg|_{t=0}=
\sum_{{j}=1}^{m}
y_j{\frac{\partial}{\partial x_j}}f(X)
+\sum_{k=1}^{n}
\omega_k{\frac {\partial}{\partial \theta_k}}f(X)
\label{II-3-11-2}
\end{equation}
where $X=(x,\theta),Y=(y,\omega)\in {\fR}^{m|n}$ 
such that $X+tY\in {\mathfrak{U}}$ for any $t\in[0,1]$. Therefore,
\begin{cor}\label{SS2GSD}
${\CSS}({\mathfrak{U}}:{\fC})\Longrightarrow {\G}_{S\!D}^{\infty}({\mathfrak{U}}:{\fC})$.
\end{cor}

\begin{prop}\label{2-1VV-mod} 
Let ${\mathfrak{U}}$ be an open subset of ${\fR}^{m|n}$.
Then, $f\in {\G}_{S\!D}^1({\mathfrak{U}}:{\fC})$ {if and only if} $f$ is $G$-differentiable  
and there exist continuous functions $F_A(X)$ defined on ${\mathfrak{U}}$ such that
\begin{equation}
f'_{X_A}(X;H_A{\bf{e}}_A)=H_A{\,} F_A(X),\quad A=1,{\cdots},m+n \forany H\in{\fR}^{m|n}.
\label{vv2.8-mod}
\end{equation}
\end{prop}
\par{\it Proof}.
$\Longrightarrow$) $f\in {\G}_{S\!D}^1({\mathfrak{U}}:{\fC})$ implies $G$-differentiability and we have readily
\eqref{vv2.8-mod} by taking $H=(H_A)_{A=1}^{m+n}$ suitably.\\
$\Longleftarrow$) $G$-differentiability \eqref{FD11-1} with \eqref{vv2.8-mod} clearly implies superdifferentiabilty. \qed

\begin{prop}\label{2.1VV-infty} 
Let ${\mathfrak{U}}$ be an open subset of ${\fR}^{m|n}$.
Then, $f\in {\G}_{S\!D}^1({\mathfrak{U}}:{\fC})$ {if and only if} $f$ is $G$-differentiable 
and its derivatives satisfy the following equations for $A=1,{\cdots},m+n$,
\begin{equation}
\left\{
\begin{aligned}
&G_A f_{X_A}'(X;H_A{\bf{e}}_A)-(-1)^{p(H_A)p(G_A)}H_Af_{X_A}'(X;G_A{\bf{e}}_A)=0 
\for p(X_A)=p(G_A)=p(H_A),\\
&f_{X_A}'(X;H_AG_A{\bf{e}}_A)=H_A f_{X_A}'(X;G_A{\bf{e}}_A) 
\for p(H_A)=0,\; p(X_A)=p(G_A).
\end{aligned}
\right.
\label{vv2.10-2}
\end{equation}
\end{prop}
\par
{\it Proof}. $\Longrightarrow)$ Multiplying $G_A$ from the left to \eqref{vv2.8-mod} and changing the role of $G_A$ and $H_A$ in the obtained equality, we have the first equation of \eqref{vv2.10-2}.
The second equation in \eqref{vv2.10-2} is derived from \eqref{vv2.8-mod} by
$$
f_{X_A}'(X;H_AG_A{\bf{e}}_A)=H_AG_AF_A(X)=H_Af_{X_A}'(X;G_A{\bf{e}}_A) \for p(H_A)=0,\; p(X_A)=p(G_A).
$$
$\Longleftarrow)$ Putting $G_A=1$ in \eqref{vv2.10-2} and defining $F_A(X)= f_{X_A}'(X;{\bf{e}}_A)$ which yields
\eqref{vv2.8-mod} for $A=1,{\cdots},m$.
The case for $A=m+1,{\cdots}, m+n$  \underline{is given in the Proof of Proposition \ref{propII-3.5-1} in the next subsection}.
 \qed

To relate functions in ${\CSS}$ and ${\G}_{S\!D}^{\infty}$ or ${\F}_{S\!D}^{\infty}$, we need the following notion which are not necessary on the finite dimensional space ${\mathfrak{B}}_{\nogg}$.
\begin{defn}[p.246 of \cite{yag88}]
Let ${\mathfrak{U}}$ be an open set in ${\mathfrak{R}}^{m|n}$ and  $f:{\mathfrak{U}}\to\mathbb{R} $(or $\to\mathbb{C}$). 
$f$ is said to be \underline{admissible} on ${\mathfrak{U}}$ if there exists some $L\ge0$ and a ${\euc}$(or $\mathbb{C}$)-valued function $\phi$ defined on ${\mathfrak{U}}_{\nogg}=p_{\nogg}({\mathfrak{U}})$ such that $f(X)=\phi{\circ}p_{\nogg}(X)=\phi(p_{\nogg}(X))$.
For $r$ $(0\le r\le{\infty})$,
$f$ is said to be admissible $C^r$ (or simply $f\in C_Y^r({\mathfrak{U}}:{\fC})$ where ${Y}$ in ${}_{Y}$ stands for Yagi) 
if $\phi\in C^r({\mathfrak{U}}_{\nogg}:{\mathbb{R}})$ (or $C^r({\mathfrak{U}}_{\nogg}:\mathbb{C})$).
\par
Let $f(X)=\sum_{{\bf{I}}\in{\mathcal{I}}}\sigma^{\bf{I}}{\cdot}f_{\bf{I}}(X)$ with $f_{\bf{I}}$ is $\mathbb{R}$(or $\mathbb{C}$)-valued on ${\mathfrak{U}}$.
For each ${\bf{I}}\in{\mathcal{I}}$, if $f_{\bf{I}}$ is admissible $C^r$ (or simply $f\in C_Y^r$) on ${\mathfrak{U}}$, $f$ is called \underline{admissible} $C_Y^r({\mathfrak{U}}:{\fC})$ on ${\mathfrak{U}}$.
More precisely, there exists some $L_{\bf{I}}\ge0$ and a ${\euc}$(or $\mathbb{C}$)-valued function $\phi_{\bf{I}}$ defined on ${\mathfrak{U}}_{\nogg}=p_{\nogg}({\mathfrak{U}})$ such that $f_{\bf{I}}(X)=\phi_{\bf{I}}{\circ}p_{\nogg}(X)=\phi_{\bf{I}}(p_{\nogg}(X))$.
Moreover, we define its partial derivatives by  
\begin{equation}
\frac{\partial f}{\partial X_{A,{\bf{K}}}}=\sum_{\bf{J}}\sigma^{\bf{J}}{\cdot}\frac{\partial f_{\bf{J}}}{\partial X_{A,{\bf{K}}}}=
\begin{cases}
|{\bf{K}}|={\mathrm{ev}} & \mbox{if $1\le A\le m$},\\
|{\bf{K}}|={\mathrm{od}} & \mbox{if $m+1\le A\le m+n$}
\end{cases}.
\label{adm-der}
\end{equation}
\end{defn}
\begin{remark}
In the above, $\frac{\partial f_{\bf{J}}}{\partial X_{A,{\bf{K}}}}$ is considered as the ordinary partial derivative
on ${\fR}_L^{m|n}$.
\end{remark}

\begin{defn}[p.246 of \cite{yag88}]
A ${\fR}$ (or ${\fC}$)-valued function $f$ on ${\mathfrak{U}}$ is said to be  \underline{projectable} if for each $L\ge0$, 
there exists a ${\fR}$(or ${\fC}$)-valued function $f_{\nogg}$ defined on ${\mathfrak{U}}_{\nogg}\subset {\fR}^{m|n}_{\nogg}$ such that $p_{\nogg}{\circ}f=f_{\nogg}{\circ}p_{\nogg}$ on ${\mathfrak{U}}$.
\end{defn}
\begin{claim}
A projectable function on ${\mathfrak{U}}$ is also admissible on ${\mathfrak{U}}$.
\end{claim}
\par{\it Proof}.
For each ${\bf{I}}\in{\mathcal{I}}$, we
define $L=L_{\bf{I}}=\max_{j}\{j\;|\; {\bf{I}}=(i_1,{\cdots}, i_j,0,{\cdots}) \,\; i_j=1\}$.
Using the map ${\proj}_{\bf{I}}$ introduced in \S 2, we have 
$$
{\begin{CD}
{\mathfrak{U}} @>f>> {\mathfrak{R}}\\
@V{p_{\nogg}}V{}V 
@V{}V{p_{\nogg}}V\\
{\mathfrak{U}}_{\nogg} @>>f_{\nogg}>{\mathfrak{R}}_{\nogg}\\
\end{CD}}\quad
\Longrightarrow\quad
{\begin{CD}
{\mathfrak{U}} @>{\proj}_{\bf{I}}{\circ}f>>{\mathbb{C}}\\
@V{p_{\nogg}}V{}V  
@V{}V{\mbox{Id}}V\\ 
{\mathfrak{U}}_{\nogg} @>>{\proj}_{\bf{I}}{\circ}f_{\nogg}> {\mathbb{C}}
\end{CD}}. 
$$
In fact, $f_{\bf{I}}={\proj}_{\bf{I}}{\circ}f=f_{{\bf{I}},L}{\circ}p_L$ with $f_{{\bf{I}},L}={\proj}_{\bf{I}}{\circ}f_{\nogg}$. \qed

\begin{lem}[Theorem 1 of \cite{yag88}]\label{yagi-thm1}
Let ${\mathfrak{U}}$ be a convex open set in ${\fR}^{m|n}$.
If $f:{\mathfrak{U}}\to{\euc}$ is in ${\G}_{S\!D}^1$, then $f$ is projectable 
and $C_Y^1$ on ${\mathfrak{U}}$.
\end{lem}
\par{\it Proof}.
Since
$\frac{d}{ds} f(X+sH)=\sum_{A=1}^{m+n} H_A F_A(X+sH)$, we have
\begin{equation}
f(X+H)-f(X)=\int_0^1\frac{d}{ds} f(X+sH)ds=\sum_{A=1}^{m+n} H_A \int_0^1F_A(X+sH) ds.
\label{4.20-}
\end{equation}
This means that if $p_{\nogg}(H_A)=0$, then $p_{\nogg}(f(X+H)-f(X))=0$. Therefore if we define $f_{\nogg}:{\mathfrak{U}}_{\nogg}\to {\mathfrak{R}}_{\nogg}$ by $f_{\nogg}(p_{\nogg}(Z))=p_{\nogg}(f(Z))$, then it implies that $f$ is projectable and so admissible.

Putting $E_{{A},{\bf{I}}}=\sigma^{\bf{I}}{\bf{e}}_A=
(\overbrace{0,{\cdots},0,\sigma^{\bf{I}}}^{A},0,{\cdots},0)\in{\mathfrak{R}^{m|n}}$, 
from the definition of partial derivatives in ${\fR}_L^{m|n}$, we define
\begin{equation}
{\frac{\partial}{\partial  X_{{A},{\bf{I}}}}}f(X) =
\left.{\frac{d}{dt}}f(X+tE_{{A},{\bf{I}}})\right|_{t=0}
\label{II-3-13-1}
\end{equation}
and since $f\in {\G}_{S\!D}^1$, there exists 
continuous $F_A(X)$ satisfying
\begin{equation}
\frac{\partial}{\partial X_{A,{\bf{I}}}}f(X)=\sigma^{\bf{I}} F_A(X), \quad |A|=|{\bf{I}}|,
\label{II-4-08-1}
\end{equation}
thus the admissible function $f$ is $C_Y^1$ on ${\mathfrak{U}}$. \qed.

\subsection{Cauchy-Riemann relation}
To understand the meaning of supersmoothness, we consider the dependence with
respect to the ``coordinate" more precisely.

\begin{prop}[Theorem 2 of \cite{yag88}]\label{propII-3.5-1}
Let $ f(X)=\sum_{\bf{I}}f_{\bf{I}}(X)\sigma^{\bf{I}} \in {\mathcal{G}}^1_{S\!D} ({\mathfrak{U}}:{\fC})\cap
C_G^{\infty}({\mathfrak{U}}:{\fC})$ 
where ${\mathfrak{U}}$ is a superdomain in ${\fR}^{m|n}$.
Let $X=(X_{A})$ be represented by 
$X_{A}=\sum_{{\bf{I}}} X_{{A},{\bf{I}}}\sigma^{\bf{I}}$ where ${A}=1,{\cdots}, m+n$,
$ X_{{A},{\bf{I}}}\in {\mathbb{C}}$ for $|{\bf{I}}|\ne 0$ and $ X_{{A},\tilde 0}\in {\euc}$.
Then, $f(X)$, considered as a function 
of countably many variables $\{X_{{A},{\bf{I}}}\}$ with values in ${\fC}$,
satisfies the following (Cauchy-Riemann type) equations.
\begin{equation}
\left\{
{\begin{aligned}
&{\frac{\partial}{\partial  X_{{A},{\bf{I}}}}}f(X) =
\sigma^{\bf{I}} {\frac{\partial}{\partial X_{{A},\tilde 0}}}f(X)
 \for 1\le{A}\le m,\; |{\bf{I}}|={\mathrm{ev}},\\
&\sigma^{\bf{K}}{\frac{\partial}{\partial X_{{A},{\bf{J}}}}}f(X)
+\sigma^{\bf{J}}{\frac{\partial}{\partial X_{{A},{\bf{K}}}}}f(X)=0\for m+1\le{A}\le m+n,\;|{\bf{J}}|={\mathrm{od}}=|{\bf{K}}|.
\end{aligned}}
\right.
\label{II-3-12-1-1}
\end{equation}
\par
Conversely, let a function $f(X)=\sum_{\bf{I}}f_{\bf{I}}(X)\sigma^{\bf{I}}$ be given such that
$f_{\bf{I}}(X+tY)\in C^\infty([0,1]:{\mathbb{C}})$ for each fixed $X,Y\in {\mathfrak{U}}$ and
$f(X)$ satisfies above \eqref{II-3-12-1-1} with \eqref{II-3-13-1}. 
Then,  $ f \in {\mathcal{G}}^1_{S\!D}({\mathfrak{U}}:{\fC})$.
\end{prop}

{\it Proof}. 
Putting ${\bf{J}}=\tilde0$ in \eqref{II-4-08-1}, we have
$$
{\frac{\partial}{\partial  X_{{A},{\tilde0}}}}f(X) =
{\frac{\partial}{\partial X_{A}}}f(X)
 \for 1\le{A}\le m,
$$
Therefore, for $1\le{A}\le m$ and $|{\bf{J}}|=$even in \eqref{II-4-08-1},
we get readily the first equation of \eqref{II-3-12-1-1}. 
Replacing ${\bf{I}}$ with ${\bf{J}}$ or ${\bf{K}}$ in \eqref{II-4-08-1},
for $m+1\le {A} \le m+n$ and $|{\bf{J}}|=\text{odd}=|{\bf{K}}|$ and multiplying $\sigma^{\bf{K}}$ or $\sigma^{\bf{J}}$ from left, respectively,
we have the second equality in \eqref{II-3-12-1-1} readily. 	
\par
To prove the converse statement, we have to construct functions 
$F_A$($1\le A\le m+n$) which satisfies
\begin{equation}
d_Gf(X;H)={\dt}f(X+tH)\big|_{t=0}=\sum_{A=1}^{m+n} H_A F_A(X)\;\;\mbox{for $X\in {\mathfrak{U}}$ and $H=(H_A)_{A=1}^{m+n}\in\mathfrak{R}^{m|n}$
}.
\label{fsd1}
\end{equation}

For $1\le{A}\le m$, we put
$\displaystyle{
F_{A}(X)=\frac{\partial}{\partial X_{{A},\tilde 0}}f(X)
\for \;X\in {\mathfrak{U}}}$.

On the other hand,
from the second equation of \eqref{II-3-12-1-1} and Lemma \ref{yg-vv},
we have an element $F_{A}(X)$($m+1\le {A}\le m+n$) such that
$\displaystyle{
\sigma^{\bf{J}}F_{A}(X)=\frac{\partial}{\partial X_{{A},{\bf{J}}}}f(X)}$.

Using these $\{F_A(X)\}$ defined above, we claim that \eqref{fsd1} holds following Yagi's argument.

Since $f$ is admissible, for any $L\ge0$, $p_{\nogg}{\circ}f$ is so also, therefore there exist some $N\ge0$ and a ${\mathfrak{R}}_{\nogg}$-valued $C^{\infty}$ function $f_N$ such that $p_{\nogg}{\circ}f(X)=f_N{\circ}p_N(X)$ on $X\in {\mathfrak{U}}$.
By natural imbedding from ${\fR}_{\nogg}$ to ${\fR}_N$, we may assume $N\ge L$.
Then, we can show that
$$
\frac{\partial}{\partial X_{A,{\bf{K}}}}f_N(p_N(X))=\begin{cases}
p_{\nogg}\big(\displaystyle{\frac{\partial}{\partial X_{A,{\bf{K}}}}f(X)}\big) &\mbox{ if ${\bf{K}}\in{\mathcal{I}}_N$},\\
\quad 0 &\mbox{ if otherwise.}
\end{cases}
$$ 
Therefore, for any $L\ge 0$, 
{\allowdisplaybreaks
\begin{align*}
p_{\nogg}\big(\dt f(X+tH)\bigg|_{t=0}\big) &=\dt p_{\nogg}(f(X+tH))\bigg|_{t=0}
\big(\because)\, p_{\nogg}(\dt g(t)\big|_{t=0})=\dt(p_{\nogg}(g(t))\big|_{t=0}\big),\\
&=\dt f_N(p_N(X+tH))\bigg|_{t=0}
\big(\because)\, p_{\nogg}(f(X))=f_N(p_N(X)) \big)\\
&=\sum_{A}\sum_{\bf{K}}(p_N(H))_{A,{\bf{K}}}{\cdot}\frac{\partial}{\partial X_{A,{\bf{K}}}}f_N(p_N(X)) \big(\because)\, \mbox{finite dimensional case}\big)\\
&=\sum_{A}\sum_{\bf{K}}(p_N(H))_{A,{\bf{K}}}{\cdot}p_{\nogg}\big(\frac{\partial}{\partial X_{A,{\bf{K}}}}f(X)\big) \big(\because)\, p_{\nogg}(g(X))=g_N(p_N(X)) \big)\\
&=\sum_{A}\sum_{\bf{K}}(p_N(H))_{A,{\bf{K}}}{\cdot}p_{\nogg}(\sigma^{\bf{K}} F_A(X)) \big(\because)\, \mbox{by \eqref{II-3-12-1-1}}\big) \\
&=\sum_{A}\sum_{\bf{K}}(p_N(H))_{A,{\bf{K}}}{\cdot}p_{\nogg}(\sigma^{\bf{K}}){\cdot}p_{\nogg}(F_A(X))\\
&=\sum_{A}\big(\sum_{\bf{K}}(p_N(H))_{A,{\bf{K}}}{\cdot}p_{\nogg}(\sigma^{\bf{K}})\big)p_{\nogg}(F_A(X))\\
&=\sum_{A}p_{\nogg}((p_N(H))_{A,{\bf{K}}}{\cdot}p_{\nogg}(\sigma^{\bf{K}}))p_{\nogg}(F_A(X))\\
&=\sum_{A}(p_{\nogg}(H))_{A}{\cdot}p_{\nogg}(F_A(X))
=p_{\nogg}(\sum_{A}H_AF_A(X)).
\end{align*}}
Thus, we have \eqref{fsd1}. The continuity of $F_A(X)$ is clear. \qed

\begin{remark}
For function with finite number of independent variables, it is well-known how to define its partial derivatives.
But when that number is infinite, it is not so clear whether the change of order of differentiation affects the result, etc.
Therefore, we reduce the calculation to the cases with finite number $L$ of generators and making that $L$ to infinity.
\end{remark}

\begin{thm}[Theorem 3 of \cite{yag88}]\label{yag-thm3}
Let ${\mathfrak{U}}\subset {\fR}^{m|n}$ be a coarse open set.
If $f\in C_G^{\infty}({\mathfrak{U}}:{\fC})\cap{\G}_{S\!D}^1({\mathfrak{U}}:{\fC})$, then $f$ is ${\G}_{S\!D}^{\infty}({\mathfrak{U}}:{\fC})$.
\end{thm}
\par{\it Proof}.
Since $f\in {\G}_{S\!D}^1$, it satisfies Cauchy-Riemann equation \eqref{II-3-12-1-1}.
As $f$ is $C^{\infty}({\mathfrak{U}}:{\fC})$, 
$F_A(X)=\frac{\partial}{\partial X_{A,{\tilde{0}}}}f(X)$ also satisfies the C-R equation, for $1\le A\le m$.
In fact, for $1\le B\le m$, $|{\bf{J}}|=$even,
$$
\begin{aligned}
\frac{\partial}{\partial X_{B,{\bf{J}}}}F_A(X)
&=\frac{\partial}{\partial X_{B,{\bf{J}}}}\frac{\partial}{\partial X_{A,{\tilde{0}}}}f(X)
\overset{admissible}{=}\frac{\partial}{\partial X_{A,{\tilde{0}}}}\frac{\partial}{\partial X_{B,{\bf{J}}}}f(X)\\
&\overset{\eqref{II-4-08-1}}{=}\frac{\partial}{\partial X_{A,{\tilde{0}}}}\sigma^{\bf{J}}\frac{\partial}{\partial X_{B,{\tilde{0}}}}f(X)
=\sigma^{\bf{J}}\frac{\partial}{\partial X_{B,{\tilde{0}}}}\frac{\partial}{\partial X_{A,{\tilde{0}}}}f(X)
=\sigma^{\bf{J}}\frac{\partial}{\partial X_{B,{\tilde{0}}}}F_A(X).
\end{aligned}
$$
And for $m+1\le B\le m+n$, $|{\bf{J}}|=|{\bf{K}}|=$odd,
$$
\begin{aligned}
\sigma^{\bf{K}}{\frac{\partial}{\partial X_{{B},{\bf{J}}}}}F_A(X)
+\sigma^{\bf{J}}{\frac{\partial}{\partial X_{{B},{\bf{K}}}}}F_A(X)
&=\sigma^{\bf{K}}{\frac{\partial}{\partial X_{{B},{\bf{J}}}}}\frac{\partial}{\partial X_{A,{\tilde{0}}}}f(X)
+\sigma^{\bf{J}}{\frac{\partial}{\partial X_{{B},{\bf{K}}}}}\frac{\partial}{\partial X_{A,{\tilde{0}}}}f(X)\\
&\overset{admissible}{=}\frac{\partial}{\partial X_{A,{\tilde{0}}}}\bigg(\sigma^{\bf{K}}{\frac{\partial}{\partial X_{{B},{\bf{J}}}}}f(X)
+\sigma^{\bf{J}}{\frac{\partial}{\partial X_{{B},{\bf{K}}}}}f(X)\bigg)\overset{\eqref{II-3-12-1-1}}{=}0.
\end{aligned}
$$
Hence $F_A(X)$(for $1\le A\le m$) is ${\G}_{S\!D}^1({\mathfrak{U}}:{\fC})$. 

Analogously, for $m+1\le A\le m+n$, $\frac{\partial}{\partial X_{A,{\bf{J}}}}f=\sigma^{\bf{J}}{\cdot}\frac{\partial}{\partial X_A}f$ is also ${\G}_{S\!D}^1$ on ${\mathfrak{U}}$. 
In fact, we have, for $|{\bf{K}}|$=even,
$$
\frac{\partial}{\partial X_{B,{\bf{K}}}}\sigma^{\bf{J}}{\cdot}\frac{\partial}{\partial X_A}f
=\sigma^{\bf{J}}\sigma^{\bf{K}}\frac{\partial}{\partial X_{B,{\tilde{0}}}}\frac{\partial}{\partial X_A}f
=\sigma^{\bf{K}}\frac{\partial}{\partial X_{B,{\tilde{0}}}}\bigg(\sigma^{\bf{J}}\frac{\partial}{\partial X_A}f\bigg).
$$
And for $|{\bf{I}}|, |{\bf{J}}|, |{\bf{K}}|$=odd, $m+1\le B\le m+n$,
$$
=\bigg(\sigma^{\bf{K}}
{\frac{\partial}{\partial X_{{B},{\bf{J}}}}}
+\sigma^{\bf{J}}
{\frac{\partial}{\partial X_{{B},{\bf{K}}}}}\bigg)
\sigma^{\bf{I}}{\cdot}\frac{\partial f(X)}{\partial X_A}
=-\sigma^{\bf{I}}\bigg(\sigma^{\bf{K}}
{\frac{\partial}{\partial X_{{B},{\bf{J}}}}}
+\sigma^{\bf{J}}{\frac{\partial}{\partial X_{{B},{\bf{K}}}}}\bigg)
=0.
$$
Thus for $m+1\le A\le m+n$, $\frac{\partial}{\partial X_{A}}f(X)$ also satisfies the C-R  equations on ${\mathfrak{U}}$ and hence ${\G}_{S\!D}^1$ on ${\mathfrak{U}}$. Therefore $f(X)$ is ${\G}_{S\!D}^2$ on ${\mathfrak{U}}$. 
By induction, $f\in {\G}_{S\!D}^{\infty}$. \qed

\begin{lem}[Lemma 5.1  of \cite{yag88}]\label{yagi5-1}
Let $f\in {\G}_{S\!D}^{\infty}({\fR}^{0|n}:{\fC})$. Then
$$
f(\theta)=f(\theta_1,{\cdots},\theta_n)=\sum_{|a|\le n}\theta^a f_a\with f_a\in{\fC}.
$$
Moreover, if a function $f$ has this structure, then $f\in {\G}_{S\!D}^{\infty}({\fR}^{0|n}:{\fC})$.
\end{lem}
\par{\it Proof}.
For  $n=1$ and $|{\bf{J}}|$=odd, we have,
$$
\frac{\partial}{\partial \theta_{\bf{J}}}f(\theta)=\dt f(\theta+t\sigma^{\bf{J}})\big|_{t=0}=\sigma^{\bf{J}}{\cdot}\frac{d}{d\theta} f(\theta)\with
\theta=\sum_{{\bf{I}}\in{\mathcal{I}}_{\mathrm{od}}}\theta_{\bf{I}}\sigma^{\bf{I}},\; \theta_{\bf{I}}\in{\mathbb{C}}.
$$
Since $f(\theta+t\sigma^{\bf{J}}+s\sigma^{\bf{K}})$ is $C^2$ w.r.t. $t, s\in{\euc}$, we have
$$
\begin{aligned}
\frac{\partial}{\partial \theta_{\bf{K}}}\frac{\partial}{\partial \theta_{\bf{J}}}f(\theta)
&=\frac{\partial}{\partial s}\frac{\partial}{\partial t}f(\theta+t\sigma^{\bf{J}}+s\sigma^{\bf{K}})\bigg|_{t=s=0}
=\sigma^{\bf{K}}{\cdot}\sigma^{\bf{J}}{\cdot}
\frac{d}{d\theta} \frac{d}{d\theta} f(\theta)\\
&=\frac{\partial}{\partial t}\frac{\partial}{\partial s}f(\theta+t\sigma^{\bf{J}}+s\sigma^{\bf{K}})\bigg|_{t=s=0}
=\sigma^{\bf{J}}{\cdot}\sigma^{\bf{K}}{\cdot}
\frac{d}{d\theta} \frac{d}{d\theta} f(\theta).
\end{aligned}
$$
Since $|{\bf{J}}|, |{\bf{K}}|$ are odd, we have $\sigma^{\bf{J}}\sigma^{\bf{K}}=-\sigma^{\bf{K}}\sigma^{\bf{J}}$ and therefore
$$
\frac{\partial}{\partial \theta_{\bf{K}}}\frac{\partial}{\partial \theta_{\bf{J}}}f(\theta)=\frac{1}{2}\sigma^{\bf{K}}{\cdot}\sigma^{\bf{J}}{\cdot}
\frac{d}{d\theta} \frac{d}{d\theta} f(\theta)=0.
$$
This implies that
by representing $f(\theta)=\sum_{\bf{I}} f_{\bf{I}}(\theta)\sigma^{\bf{I}}$, the each component $f_{\bf{I}}(\theta)$ is a polynomial of degree $1$ with variables $\{\theta_{\bf{J}}\;|\; {\bf{J}}\in{\mathcal{I}}_{\mathrm{od}}\}$.
Then $\displaystyle{\sigma^{\bf{I}}{\cdot}\frac{d}{d\theta} f(\theta)=\frac{\partial}{\partial \theta_{\bf{I}}}f(\theta)}$ is constant for any $|{\bf{I}}|=$odd.
Thus $\displaystyle{\frac{d}{d\theta} f(\theta)}$ is constant denoted by $a\in{\fC}$.
Then, $\displaystyle{\frac{d}{d\theta}( f(\theta)-{\theta}a)=0}$. Therefore there exists $b\in{\fC}$ such that $f(\theta)={\theta}a+b$.
\par
We proceed by induction w.r.t. $n$.
Let $f$ be a ${\G}_{S\!D}^{\infty}$ function on an open set $U\subset {\fR}^{0|n}$. Fixing $\theta_1,{\cdots}, \theta_{n-1}$,
$f(\theta_1,{\cdots}, \theta_{n-1},\theta_n)$ is a ${\G}_{S\!D}^{\infty}$ function with one variable $\theta_n$.
Thus, we have
$$
f(\theta_1,{\cdots}, \theta_{n-1},\theta_n)=\theta_n g(\theta_1,{\cdots}, \theta_{n-1})+h(\theta_1,{\cdots}, \theta_{n-1})
\with
\frac{\partial}{\partial \theta_n}f(\theta)=g(\theta_1,{\cdots}, \theta_{n-1}).
$$
Therefore $g$ is ${\G}_{S\!D}^{\infty}$ w.r.t. $(\theta_1,{\cdots}, \theta_{n-1})$, $h$ is also ${\G}_{S\!D}^{\infty}$ w.r.t. $(\theta_1,{\cdots}, \theta_{n-1})$. \qed
\begin{remark}
Though this lemma with a sketch of the proof is announced in \cite{DW2} and is cited in \cite{MK} without proof, but I feel some unclearness of their proof. This point is ameliorated by \cite{yag88} like as above.
\end{remark}

\begin{lem}[Lemma 5.2  of \cite{yag88}]\label{yag-lem5-2}
Let $\mathcal{U}\subset{\fR}^{m|0}$ be a coarse open set.
Let $f\in {\G}_{S\!D}^{\infty}(\mathcal{U}:{\fC})$ which vanishes identically 
on ${\mathcal{U}}_{\mathrm{B}}=\pi_{\mathrm{B}}(\mathcal{U})$. 
Then, $f$ vanishes identically on $\mathcal{U}$.
\end{lem}
\par{\it Proof}.
It is essential to prove the case $m=1$.
Take an arbitrary point $t\in {\mathcal{U}}_{\mathrm{B}}$ and we consider the behavior of $f$ on $\pi_{\mathrm{B}}^{-1}(t)$.
Let $X\in \pi_{\mathrm{B}}^{-1}(t)$ and $X_{\nogg}=p_{\nogg}(X)$.
Then $\{X_{\bf{K}}\;|\; {\bf{K}}\in{\mathcal{I}}_{\nogg},\; |{\bf{K}}|={\mathrm{ev}}\ge 2\}$ is a coordinate for $(\pi_{\mathrm{B}}^{-1}(t))_{\nogg}$ as the ordinary space ${\mathbb{C}}$.
Let $f_{\nogg}$ be the $L$-th projection of $f$. Then,
$$
\frac{\partial}{\partial X_{\bf{K}}}f_N(X_{\nogg})=\sigma^{\bf{K}}{\cdot}\frac{\partial}{\partial X_{{\tilde{0}}}}f_{\nogg}(X_{\nogg})\for {\bf{K}}\in{\mathcal{I}}_{\nogg} \et |{\bf{K}}|={\mathrm{ev}}.
$$
If ${\bf{K}}_1, {\cdots}, {\bf{K}}_h\in {\mathcal{I}}_{\nogg}$, $|{\bf{K}}_j|=$even$>0$ and $2h>L$, then $\sigma^{{\bf{K}}_1}{\cdots}\sigma^{{\bf{K}}_h}=0$
and $\displaystyle{\frac{\partial}{\partial x_{{\bf{K}}_1}}{\cdots}\frac{\partial}{\partial x_{{\bf{K}}_h}}f_{\nogg}(X_{\nogg})=0}$.
This implies that $f_{\nogg}$ is a polynomial on $(\pi_{\mathrm{B}}^{-1}(t))_{\nogg}$.
Moreover, for any $h\ge 0$,
$$
\frac{\partial}{\partial x_{{\bf{K}}_1}}{\cdots}\frac{\partial}{\partial x_{{\bf{K}}_h}}f_{\nogg}(t)
=\sigma^{{\bf{K}}_1}{\cdots}\sigma^{{\bf{K}}_h}\bigg(\frac{\partial}{\partial X_{{\tilde{0}}}}\bigg)^h f_{\nogg}(t).
$$
Since $f$ vanishes on ${\mathcal{U}}_{\mathrm{B}}$, we have
$$
\bigg(\frac{\partial}{\partial X_{{\tilde{0}}}}\bigg)^h f_{\nogg}(t)=0 \quad \mbox{on}\quad {\mathcal{U}}_{\mathrm{B}}
$$
and hence
$$
\frac{\partial}{\partial x_{{\bf{K}}_1}}{\cdots}\frac{\partial}{\partial x_{{\bf{K}}_h}}f_{\nogg}(t)=0 \forany h\ge 0 \et
{\bf{K}}_1, {\cdots}, {\bf{K}}_h\in {\mathcal{I}}_{\nogg} \with |{\bf{K}}_j|={\mathrm{even}}>0.
$$
Thus the polynomial $f_{\nogg}\big|_{\pi_{\mathrm{B}}^{-1}(t)}$ must vanish identically and hence $f_{\nogg}\equiv 0$ on ${\mathcal{U}}_{\nogg}$. This holds for any $L\ge0$. Thus $f\equiv 0$ on $\mathcal{U}$. \qed

\subsection{Proof of Main Theorem \ref{mthm}}
It is clear from outset that $(a)\Rightarrow (b)\Rightarrow  (c)\Rightarrow  (d)$.
From Proposition \ref{propII-3.5-1}, $(d)\Rightarrow (a)$. Lastly, the equivalence of $(d)$ and $(e)$ is given by

\begin{thm}[Thorem 4 of \cite{yag88}]
Let $f$ be a ${\G}_{S\!D}^{\infty}$ function on a convex open set ${\mathcal{U}}\subset{\fR}^{m|n}$. Then, there exist ${\fR}$-valued $C^{\infty}$ functions $u_a$ on ${\mathcal{U}}_{\mathrm{B}}$ such that
$$
f(x,\theta)=\sum_{|a|\le n}\theta^a \tilde{u}_a(x).
$$
Moreover, this expression is unique.
\end{thm}
\par
{\it Proof}.
$\Longrightarrow)$
For fixed $x$, by Lemma \ref{yagi5-1}, $f(x,\theta)$ has  the representation 
$f(x,\theta)=\sum_{|a|\le n}\theta^a\varphi_a(x)$
with $\varphi_a(x)\in {\fC}$.
Since $f\in {\G}_{S\!D}^{\infty}$, it is clear that for each $a$, $\varphi_a(x)\in {\fC}$ is on ${\fR}^{m|0}$ and moreover
$\varphi_a(x_{\mathrm{B}})$ is in $C^{\infty}({\euc}^m)$. Denoting the Grassmann continuation of it by $\tilde{\varphi}_a(x)$, we should have $\tilde{\varphi}_a(x)=f_a(x)$ by Lemma \ref{yag-lem5-2}.

$\Longleftarrow)$ Since the supersmoothness leads the C-R relation, we get the superdifferentiability. \qed

\appendix
\section{Is the Hamilton flow supersmooth w.r.t. initial data?} 
Let an even supersmooth function ${\mathcal{H}}(t,x,\xi,\theta,\pi)$ on $\euc\times {\fR}^{2m|2n}$ be given.
We consider a Hamilton flow
$(x(t),\xi(t),\theta(t),\pi(t))$ defined by
\begin{equation}
\allowdisplaybreaks
\left\{
\begin{aligned}
&\dt x_j
=\frac{\partial{\mathcal{H}}(t,x,\xi,\theta,\pi)}{\partial \xi_j},\qquad
\dt \xi_j
=-\frac{\partial{\mathcal{H}}(t,x,\xi,\theta,\pi)}{\partial x_j},\\
&\dt \theta_k=-\frac{\partial{\mathcal{H}}(t,x,\xi,\theta,\pi)}{\partial \pi_k},
\qquad
\dt \pi_k=-\frac{\partial{\mathcal{H}}(t,x,\xi,\theta,\pi)}{\partial \theta_k},
\end{aligned}
\right.
\label{CM2}
\end{equation}
with initial data at time $t={\unbt}$ given by
\begin{equation}
(x({\unbt}),\xi({\unbt}),\theta({\unbt}),\pi({\unbt}))=({\inidata}).
\label{CM2b}
\end{equation}

\begin{problem}
Is the Hamilton flow supersmooth w.r.t. the initial data ${\inidata}$, for a given supersmooth Hamiltonian function ${\mathcal{H}}(t,x,\xi,\theta,\pi)$?
\end{problem}

Since this problem is rather vague, we explain by taking a simple concrete example;
the time-independent Hamiltonian stemming from the Pauli equation.

Let a system of PDE be given by
\begin{equation}
\begin{aligned}
{\mathbb{H}}\left(q,{\frac{\hbar}{i}}\frac\partial{\partial q}\right)
&=\frac 1{2M}\bigg[\sum_{j=1}^3{\pmb{\sigma}}_j
\big({\frac{\hbar}{i}} \partial_{q_j}-eA_j(q)\big)\bigg]^2+\mu V(q)\\
&=\frac 1{2M}\sum_{j=1}^3 \big({\frac{\hbar}{i}}\partial_{q_j}-eA_j
(q)\big)^2 +\mu V(q)\\
&\qquad\qquad+\frac{e\hbar}{2iM}
\big[F_{12}(q){\pmb\sigma}_1{\pmb\sigma}_2
+F_{23}(q){\pmb\sigma}_2{\pmb\sigma}_3
+F_{31}(q){\pmb\sigma}_3{\pmb\sigma}_1\big].
\end{aligned}
\label{p1-2}
\end{equation}
Here, we put
$$
{\pmb\sigma}_1=\begin{pmatrix}
0&1\\
1&0
\end{pmatrix},\quad
{\pmb\sigma}_2=\begin{pmatrix}
0&-i\\
i&0
\end{pmatrix},\quad
{\pmb\sigma}_3=\begin{pmatrix}
1&0\\
0&-1
\end{pmatrix} \et
F_{jk}(q)=\frac{\partial A_j(q)}{\partial q_k}-\frac{\partial A_k(q)}{\partial q_j}.
$$
Putting $M=1$, $e=1$ and $\mu=1$ above,
we may derive an even super Hamiltonian given by
\begin{equation}
\begin{aligned}
{\mathcal{H}}(x,\xi,\theta,\pi)
=&\frac12\sum_{j=1}^3(\xi_j-A_j(x))^2+V(x)\\
&\quad
+\frac12 F_{12}(x)\langle\theta|\pi\rangle
+\frac{i{\hbar}}{2}\big(F_{23}(x)+iF_{31}(x)\big)\theta_1\theta_2
+\frac{i{\hbar}^{-1}}{2}\big(F_{23}(x)-iF_{31}(x)\big)\pi_1\pi_2\\
\end{aligned}
\label{CM1}
\end{equation}
where $A_j(x)$, $F_{jk}(x)$ are Grassmann continuations of  $A_j(q)$, $F_{jk}(q)$, respectively. 

Then, we restate the above problem as
\begin{problem}
For the initial value problem \eqref{CM2} with the given \eqref{CM1},
do it have the representation
\begin{equation}
\left\{
\begin{aligned}
x(t)&= \sum_{|{\bar{a}}|+|{\bar{b}}|=0,2,4} x_{\bar{a}\bar{b}}(t)\unbtheta^{\bar{a}}\unbpi^{\bar{b}},\quad
\xi(t)= \sum_{|{\bar{a}}|+|{\bar{b}}|=0,2,4} \xi_{\bar{a}\bar{b}}(t)\unbtheta^{\bar{a}}\unbpi^{\bar{b}},\\
\theta(t)&=\sum_{|{\bar{a}}|+|{\bar{b}}|=1,3} \theta_{\bar{a}\bar{b}}(t)\unbtheta^{\bar{a}}\unbpi^{\bar{b}},\quad\;\;
\pi(t)= \sum_{|{\bar{a}}|+|{\bar{b}}|=1,3} \pi_{\bar{a}\bar{b}}(t)\unbtheta^{\bar{a}}\unbpi^{\bar{b}}
\end{aligned}
\right. \label{p2.25}
\end{equation}
where
$$
\left\{
\begin{aligned}
&x_{\bar{a}\bar{b}}(t)=x_{\bar{a}\bar{b}}(t,\unbx,\unbxi)=\partial_{\unbpi_2}^{b_2}\partial_{\unbpi_1}^{b_1}
\partial_{\unbtheta_2}^{a_2}\partial_{\unbtheta_1}^{a_1} 
 x(t,\unbx,\unbxi,0,0),\\
&\xi_{\bar{a}\bar{b}}(t)=\xi_{\bar{a}\bar{b}}(t,\unbx,\unbxi)=\partial_{\unbpi_2}^{b_2}\partial_{\unbpi_1}^{b_1}
\partial_{\unbtheta_2}^{a_2}\partial_{\unbtheta_1}^{a_1} 
 \xi(t,\unbx,\unbxi,0,0),\\
&\theta_{\bar{a}\bar{b}}(t)=\theta_{\bar{a}\bar{b}}(t,\unbx,\unbxi)=\partial_{\unbpi_2}^{b_2}\partial_{\unbpi_1}^{b_1}
\partial_{\unbtheta_2}^{a_2}\partial_{\unbtheta_1}^{a_1} 
\theta(t,\unbx,\unbxi,0,0),\\ 
&\pi_{\bar{a}\bar{b}}(t)=\pi_{\bar{a}\bar{b}}(t,\unbx,\unbxi)=\partial_{\unbpi_2}^{b_2}\partial_{\unbpi_1}^{b_1}
\partial_{\unbtheta_2}^{a_2}\partial_{\unbtheta_1}^{a_1} 
 \pi(t,\unbx,\unbxi,0,0),
\end{aligned}
\right.
$$
with ${\bar{a}}=(a_1,a_2),\;{\bar{b}}=(b_1,b_2)\in\{0,1\}^2$. Here, $x_{\bar{a}\bar{b}}(t,\unbx,\unbxi)$, ${\cdots}$,
$\pi_{\bar{a}\bar{b}}(t,\unbx,\unbxi)$  should be the Grassmann continuations of
$x_{\bar{a}\bar{b}}(t,\unbq,\unbp)$, ${\cdots}$, $\pi_{\bar{a}\bar{b}}(t,\unbq,\unbp)$ with $\unbq=\pi_{\mathrm{B}}(\unbx),\, \unbp=\pi_{\mathrm{B}}(\unbxi)$, respectively.
\end{problem}

We assume the following (analogous to those in Yajima~\cite{Yaj} for Schr\"odinger equation):
\begin{description}
\item[(A)] $A_j(q)\in C^\infty({{\euc}^3})$ is real-valued
and there exists $\epsilon>0$ such that
\begin{equation*}
\begin{aligned}
|\partial_q^\alpha F_{jk}(q)| \le C_\alpha (1+|q|)^{-1-\epsilon}
\for |\alpha|\ge 1,\\
|\partial_q^\alpha A_j(q)| 
\le C_\alpha \for |\alpha|\ge 1.
\end{aligned}
\end{equation*}
\item[(V)] $V(q)\in C^\infty({{\euc}^3})$ is real-valued
and
\begin{equation*}
|\partial_q^\alpha V(q)| \le C_\alpha \for |\alpha|\ge 2.
\end{equation*}
\end{description}

\begin{prop}[Proposition~2.1 of Inoue-Maeda~\cite{IM02}]\label{Prop-CM1}
Let ${\mathcal{H}}(x,\xi,\theta,\pi)$  
be given as in \eqref{CM1}.  
Under Assumptions (A) and (V), for any $T>0$ and any $t\in [-T,T]$,
there exists a unique solution of \eqref{CM2} with \eqref{CM2b}
which is denoted by  
$x(t)$ or $x(t,{\inidata})$, $\cdots $, $\pi(t)$ or $\pi(t,{\inidata})$, 
etc, depending on the context. 
\end{prop} 

Main ingredient of the proof of this proposition is that
we decompose dependent variables, using degree w.r.t. Grassmann generators, as follows:
\begin{equation}
{\begin{aligned}
&x_j(t)=\sum_{\ell=0}^\infty x_j^{[2\ell]}(t)
=x_{j,\mathrm{B}}(t)+x_{j,\mathrm{S}}(t), \\
&\xi_j(t)=\sum_{\ell=0}^\infty \xi_j^{[2\ell]}(t)
=\xi_{j,\mathrm{B}}(t)+\xi_{j,\mathrm{S}}(t),\\
\end{aligned}}
\quad
{\begin{aligned}
&\theta_k(t)=\theta_{k,\mathrm{S}}(t)
=\sum_{\ell=0}^\infty \theta_k^{[2\ell+1]}(t),\\
&\pi_k(t)=\pi_{k,\mathrm{S}}(t)=\sum_{\ell=0}^\infty \pi_k^{[2\ell+1]}(t),
\end{aligned}}
\label{CM-4}
\end{equation}
with
$$
\begin{gathered}
x_{j,\mathrm{B}}(t)=x_j^{[0]}(t), \;
x_{j,\mathrm{S}}(t)=\sum_{\ell=1}^\infty x_j^{[2\ell]}(t),\\
\xi_{j,\mathrm{B}}(t)=\xi_j^{[0]}(t), \;
\xi_{j,\mathrm{S}}(t)=\sum_{\ell=1}^\infty \xi_j^{[2\ell]}(t).
\end{gathered}
$$
\begin{remark}
(i) Here, the weakness of the  topology of $\mathfrak{R}$ is essential.
In fact, to prove that $\sum_{\ell=0}^{\infty}x_j^{[2\ell]}(t)$ exists in $\oplus_{\ell=1}^{\infty}\mathfrak{R}^{[2\ell]}=\mathfrak{R}$, it is sufficient to prove that $x_j^{[2\ell]}(t)$ exists. But it seems  very hard to prove
that $\sum_{\ell=0}^{\infty}x_j^{[2\ell]}(t)$ converges in $\ell^1$ topology.\\
(ii) Moreover, by the uniqueness of the  decomposition \eqref{CM-4}, 
we have the unique solution of \eqref{CM2} with \eqref{CM2b}.
\end{remark}

We investigate the smoothness (in ordinary sense)  of 
$(x(t),\xi(t),\theta(t),\pi(t))$ 
w.r.t. the initial data $({\inidata})$. 
In order to fix the notation, we give the brief proof.

\begin{prop}[Proposition~2.3 of Inoue-Maeda~\cite{IM02}]\label{Prop-CM2}
Let $\{A_j(q)\}_{j=1}^3,\; V(q)\in C^\infty({\euc}^3:{\euc})$
and $T>0$.
The solution $(x(t),\xi(t),\theta(t),\pi(t))$ of \eqref{CM2} and \eqref{CM2b}
on $[-T,T]$ is ``s-smooth'' in $(t,{\inidata})$.
That is, smooth in $t$ for any fixed $({\inidata})$ 
and supersmooth in $({\inidata})$ for any fixed  $t\in [-T,T]$.
\end{prop}
\begin{remark}
We claim there that $(x(t),\xi(t),\theta(t),\pi(t))$ are $C^{\infty}_G$-smooth
w.r.t. the initial data $(\unbt,{\inidata})$. But the super-smoothness is not proved yet there which is pointed out by Y. Inahama.
\end{remark}
\par{\it Proof.}
For notational simplicity, we represent 
$x(t,{\inidata})$ as $x(t)$ or $x$, etc.

(I) In oder to prove the smoothness w.r.t. the initial data, 
we differentiate \eqref{CM2} and \eqref{CM2b} formally w.r.t. 
$({\inidata})$, which
gives us the following differential equation:
\begin{equation}
{d \over dt } {\mathcal{J}}^{\cbra(1)}(t) 
= {\mathcal{H}}^{\cbra(2)}(t){\mathcal{J}}^{\cbra(1)}(t) \with
{\mathcal{J}}^{\cbra(1)}(0)=I.
\label{CM25}
\end{equation}
Here
\begin{equation}
{\mathcal{J}}^{\cbra(1)}(t)= 
\begin{pmatrix}
\partial_{\unbx}x&\partial_{\unbxi}x
&\partial_{\unbtheta}x&\partial_{\unbpi}x\\
\partial_{\unbx}\xi&\partial_{\unbxi}\xi
&\partial_{\unbtheta}\xi&\partial_{\unbpi}\xi\\
 \partial_{\unbx}\theta&  \partial_{\unbxi}\theta
&\partial_{\unbtheta}\theta&\partial_{\unbpi}\theta\\
 \partial_{\unbx}\pi&  \partial_{\unbxi}\pi
&\partial_{\unbtheta}\pi&\partial_{\unbpi}\pi
\end{pmatrix},\;
\partial_{\unbx}x=
\begin{pmatrix}
\partial_{{\unbx}_1}x_1&\partial_{{\unbx}_2}x_1&\partial_{{\unbx}_3}x_1\\
\partial_{{\unbx}_1}x_2&\partial_{{\unbx}_2}x_2&\partial_{{\unbx}_3}x_2\\
\partial_{{\unbx}_1}x_3&\partial_{{\unbx}_2}x_3&\partial_{{\unbx}_3}x_3\\
\end{pmatrix}, 
\quad \text{etc.}
\end{equation}
with arguments $(t,{\inidata})$ and
\begin{equation}
{\mathcal{H}}^{\cbra(2)}(t)=
\begin{pmatrix}
{\mathcal{H}}^{\cbra(2|0)}(t)&{\mathcal{H}}^{\cbra(1|1)}(t)\\
{\mathcal{H}}^{\cbra(1|1)}(t)&{\mathcal{H}}^{\cbra(0|2)}(t)
\end{pmatrix}
=\begin{pmatrix}
  \partial_x\partial_{\xi}{\mathcal{H}} & 
  \partial_{\xi}\partial_{\xi}{\mathcal{H}} &
  -\partial_{\theta}\partial_{\xi}{\mathcal{H}} & 
  -\partial_{\pi}\partial_{\xi}{\mathcal{H}} \\
-\partial_x\partial_x{\mathcal{H}} & -\partial_{\xi}\partial_x{\mathcal{H}} &
\partial_{\theta}\partial_x{\mathcal{H}} &\partial_{\pi}\partial_x{\mathcal{H}} \\
  \partial_x\partial_{\pi}{\mathcal{H}} & 
  \partial_{\xi}\partial_{\pi}{\mathcal{H}} &
-\partial_{\theta}\partial_{\pi}{\mathcal{H}} & 
-\partial_{\pi}\partial_{\pi}{\mathcal{H}} \\
  \partial_x\partial_{\theta}{\mathcal{H}} & 
  \partial_{\xi}\partial_{\theta}{\mathcal{H}} &
-\partial_{\theta}\partial_{\theta}{\mathcal{H}} & 
-\partial_{\pi}\partial_{\theta}{\mathcal{H}}
\end{pmatrix}
\end{equation}
where
$$
{\mathcal{H}}^{\cbra(2|0)}(t)=
\begin{pmatrix}
  \partial_x\partial_{\xi}{\mathcal{H}} & 
  \partial_{\xi}\partial_{\xi}{\mathcal{H}}\\
-\partial_x\partial_x{\mathcal{H}} & -\partial_{\xi}\partial_x{\mathcal{H}}
\end{pmatrix},\;
\partial_x\partial_{\xi}{\mathcal{H}}=
\begin{pmatrix}
\partial_{x_1}\partial_{\xi_1}{\mathcal{H}}&\partial_{x_2}\partial_{\xi_1}{\mathcal{H}}
&\partial_{x_3}\partial_{\xi_1}{\mathcal{H}}\\
\partial_{x_1}\partial_{\xi_2}{\mathcal{H}}&\partial_{x_2}\partial_{\xi_2}{\mathcal{H}}
&\partial_{x_3}\partial_{\xi_2}{\mathcal{H}}\\
\partial_{x_1}\partial_{\xi_3}{\mathcal{H}}&\partial_{x_2}\partial_{\xi_3}{\mathcal{H}}
&\partial_{x_3}\partial_{\xi_3}{\mathcal{H}}\\
\end{pmatrix}, 
\quad \text{etc.}
$$
with arguments $(x(t),\xi(t),\theta(t),\pi(t))$.
We may prove that the solution of \eqref{CM2} and \eqref{CM2b} is 
in fact differentiable w.r.t. $({\inidata})$ and satisfies \eqref{CM25}.

Furthermore, for each positive integer $k\ge 1$ and $\ell\ge 2$, putting 
\begin{equation}
{\mathcal{J}}^{\cbra(k)}(t)=
\left(\partial_{\unbx}^{\alpha }\partial_{\unbxi}^{\beta}\partial_{\unbtheta}^a
\partial_{\unbpi}^b 
\begin{pmatrix}x\\ \xi\\ \theta\\ \pi\end{pmatrix}
\right)_{\scriptstyle{|\alpha+\beta|\;\;\;\;\;\;\;}\atop\scriptstyle{+|a+b|=k}}
\et
{\mathcal{H}}^{\cbra(\ell)}(t)
=\big(\partial_{x}^{\alpha }\partial_{\xi}^{\beta}\partial_{\theta}^a
\partial_{\pi}^b {\mathcal{H}}\big)_{\scriptstyle{|\alpha+\beta|\;\;\;\;\;\;\;}\atop\scriptstyle{+|a+b|=k}}
\end{equation}
respectively, we have the following differential equation for $k\ge 2$:
\begin{equation}
\begin{aligned}
&{d \over dt}{\mathcal{J}}^{\cbra(k)}(t)
={\mathcal{H}}^{\cbra(2)}(t){\mathcal{J}}^{\cbra(k)}(t)+R^{\cbra(k)}(t) 
\with {\mathcal{J}}^{\cbra(k)}(0)=0\\
& \where R^{(k)}(t)=\sum_{p=2}^k\sum_{k=k_1+\cdots k_p}
c_{p,k}{\mathcal{H}}^{\cbra(p+1)}(t){\mathcal{J}}^{\cbra(k_1)}(t)\otimes\cdots\otimes {\mathcal{J}}^{\cbra(k_p)}(t).
\end{aligned}
\label{CM25-1}
\end{equation}
Here, $c_{p,k}$ are suitable constants. 
As above, this equation has the unique solution and therefore
the solution of \eqref{CM2} and \eqref{CM2b} is 
in fact $k$-times differentiable w.r.t. $({\inidata})$.

We need to explain how to make rigorous the formal arguments above:
Denoting the solution of \eqref{CM2} 
with the initial data $({\unbx}+{\epsilon}{\bf{e}}_k{\unby}_k,{\unbxi},{\unbtheta},{\unbpi})$ as
$$
\begin{gathered} 
\tilde{x}_i(t)=x_i(t,{\unbx}+{\epsilon}{\bf{e}}_k{\unby}_k,{\unbxi},{\unbtheta},{\unbpi}),{\cdots},
\tilde{\pi}_b(t)=\pi_b(t,{\unbx}+{\epsilon}{\bf{e}}_k{\unby}_k,{\unbxi},{\unbtheta},{\unbpi}),\;\;
{\bf{e}}_k=(\underbrace{\overbrace{0,{\cdots},0,1}^{k},0,{\cdots},0}_{m}),\\
\tilde{Z}(t)=(\tilde{x}(t),\tilde{\xi}(t),\tilde{\theta}(t),\tilde{\pi}(t))=(\tilde{Z}_A(t))_{A=1}^{2(m+n)},\; \;
Z(t)=(x(t),\xi(t),\theta(t),\pi(t))=(Z_A(t))_{A=1}^{2(m+n)},
\end{gathered}
$$
we have
\begin{equation}
\dt{\tilde{x}}_i(t)={\mathcal{H}}_{\xi_i}(\tilde{x}(t),\tilde{\xi}(t),\tilde{\theta}(t),\tilde{\pi}(t))\;\;\mbox{or}\;\;
{\tilde{x}}_i(t)-{\tilde{x}}_i(0)=\int_0^tds\,{\mathcal{H}}_{\xi_i}(\tilde{x}(s),{\cdots},\tilde{\pi}(s)), etc.
\label{difference-s}
\end{equation}

Considering \eqref{difference-s} as ODE with parameter $\euc\ni{\epsilon}\neq 0$, we have the difference quotients
\begin{equation}
\dt \frac{\tilde{x}_i(t)-x_i(t)}{\epsilon}=\frac{\tilde{Z}_A(t)-Z_A(t)}{\epsilon}g_{A.i}(t,{\epsilon})
\label{diff-quot}
\end{equation}
with
$$
g_{A.i}(t,{\epsilon})=\int_0^1d\tau\,{\mathcal{H}}_{Z_A\xi_i}({\tau}\tilde{Z}(t)+(1-{\tau})Z(t)).
$$
Here, we used below:
$$
f(\tilde{X})-f(X)=\int_0^1d{\tau}\,\frac{d}{d{\tau}}f({\tau}\tilde{X}+(1-{\tau})X)
=(\tilde{X}-X)\int_0^1d{\tau}\,\frac{\partial}{\partial X}f({\tau}\tilde{X}+(1-{\tau})X).
$$
Before making ${\epsilon}\to0$ in \eqref{diff-quot}, we remark
$$
\lim_{{\epsilon}\to0}g_{A.i}(t,{\epsilon})={\mathcal{H}}_{Z_A\xi_i}(Z(t)).
$$
Regarding $\frac{\tilde{x}_i(t)-x_i(t)}{\epsilon},{\cdots}, \frac{\tilde{\pi}_a(t)-{\pi}_a(t)}{\epsilon}$ as unknown functions satisfying \eqref{diff-quot} analogous to the ODE with parameter ${\epsilon}$
and
following the procedure used to prove the continuity and differentiability of solutions w.r.t. ${\epsilon}$, 
we have the limit
$$
\lim_{{\epsilon}\to0}\frac{\tilde{x}_i(t)-x_i(t)}{\epsilon}=\frac{\partial x_i(t)}{\partial{\unbx}_k}
$$
which satisfies
$$
\dt {\unby}_k\frac{\partial x_i(t)}{\partial{\unbx}_k}= {\unby}_k\frac{\partial x_j(t)}{\partial{\unbx}_k}{\mathcal{H}}_{x_j\xi_i}(x(t),\xi(t),\theta(t),\pi(t))
+{\cdots}+{\unby}_k\frac{\partial \pi_{\ell}(t)}{\partial{\unbx}_k}{\mathcal{H}}_{\pi_{\ell}\xi_i}(x(t),\xi(t),\theta(t),\pi(t)).
$$
Taking ${\unby}_k=1$ above, we have
$$
\dt \frac{\partial x_i(t)}{\partial{\unbx}_k}=\frac{\partial x_j(t)}{\partial{\unbx}_k}{\mathcal{H}}_{x_j\xi_i}(x(t),\xi(t),\theta(t),\pi(t))
+{\cdots}+\frac{\partial \pi_{\ell}(t)}{\partial{\unbx}_k}{\mathcal{H}}_{\pi_{\ell}\xi_i}(x(t),\xi(t),\theta(t),\pi(t)).
$$
Re-taking the initial data as $({\unbx},{\unbxi},{\unbtheta},{\unbpi}+{\epsilon}{\bf{e}}_{m+b}{\unbrho}_b)$ and proceeding as above, we have
$$
\dt {\unbrho}_b\frac{\partial \pi_a(t)}{\partial{\unbpi}_b}=-{\unbrho}_b\frac{\partial x_j(t)}{\partial{\unbpi}_b}{\mathcal{H}}_{x_j\xi_i}(x(t),\xi(t),\theta(t),\pi(t))-{\cdots}-{\unbrho}_b\frac{\partial \pi_c(t)}{\partial{\unbpi}_b}{\mathcal{H}}_{\pi_c\xi_i}(x(t),\xi(t),\theta(t),\pi(t)).
$$
Since we may take ${\unbrho}_b\in\rod$ arbitrary,  we have
$$
\dt \frac{\partial \pi_a(t)}{\partial{\unbpi}_b}=-\frac{\partial x_j(t)}{\partial{\unbpi}_b}{\mathcal{H}}_{x_j\xi_i}(x(t),\xi(t),\theta(t),\pi(t))-{\cdots}-\frac{\partial \pi_c(t)}{\partial{\unbpi}_b}{\mathcal{H}}_{\pi_c\xi_i}(x(t),\xi(t),\theta(t),\pi(t)).
$$
Combining these, we show that \eqref{CM25} holds rigorously. Analogously \eqref{CM25-1} holds for any $k$.
Therefore, $(x(t),\xi(t),\theta(t), \pi(t))$ belongs to $C_G^{\infty}$ w.r.t $(\inidata)$.

(II) For notational simplicity, we put
$$
\begin{aligned}
\tilde{x}_i(t;{\epsilon})&=\tilde{x}_i(t,{\inidatamodpara}),\quad
\tilde{x}_i(t;{0})=x_i(t,{\inidata})=x_i(t),\\
\tilde{\xi}_i(t;{\epsilon})&=\tilde{\xi}_i(t,{\inidatamodpara}),\quad
\tilde{\xi}_i(t;{0})=\xi_i(t,{\inidata})=\xi_i(t),\\
\tilde{\theta}_a(t;{\epsilon})&=\tilde{\theta}_a(t,{\inidatamodpara}),\quad
\tilde{\theta}_a(t;{0})={\theta}_a(t,{\inidata})={\theta}_a(t),\\
\tilde{\pi}_a(t;{\epsilon})&=\tilde{\pi}_a(t,{\inidatamodpara}),\quad
\tilde{\pi}_a(t;{0})={\pi}_a(t,{\inidata})={\pi}_a(t).
\end{aligned}
$$
Then, since these are proved to be $C_G^{\infty}$-differentiable, 
we have readily that
$$
\frac{d \tilde{x}_i(t;{\epsilon})}{d{\epsilon}}\bigg|_{{\epsilon}=0}
={\unby}_j\frac{\partial{x}_i(t)}{\partial {\unbx}_j}
+{\unbeta}_j\frac{\partial{x}_i(t)}{\partial {\unbxi}_j}
+{\unbomega}_a\frac{\partial{x}_i(t)}{\partial {\unbtheta}_a}
+{\unbrho}_a\frac{\partial{x}_i(t)}{\partial {\unbpi}_a},
$$
with
$$
\frac{\partial{x}_i(t)}{\partial {\unbx}_j}, \;\; \frac{\partial{x}_i(t)}{\partial {\unbxi}_j}\in{\rev},\quad
\frac{\partial{x}_i(t)}{\partial {\unbtheta}_a}, \;\; \frac{\partial{x}_i(t)}{\partial {\unbpi}_a}\in{\rod}.
$$
Therefore  we have
$$
d_G x_i(t,\underline{Z};H)=\sum_{A=1}^{2m+2n}H_A\frac{\partial{x_i(t,\underline{Z})}}{\partial H_A}
\with
\underline{Z}=({\unbx},{\unbxi},{\unbtheta},{\unbpi}),\; 
H=({\unby},{\unbeta},{\unbomega},{\unbrho}),
$$
this implies $x_i(t,\underline{Z})$ is ${\G}_{S\!D}^1$ differentiable by Proposition \ref{2-1VV-mod}. Same hold for
$d_G\xi(t,\underline{Z}),\; d_G\theta_k(t,\underline{Z})$ and $d_G\pi_k(t,\underline{Z})$. \qed

\section{Inverse and implicit function theorems}
Following is the slight modification of the arguments in Inoue and Maeda~\cite{IM91}.
\subsection{Composition of supersmooth functions}
\begin{defn}\label{3.14b}
Let ${\mathfrak{U}}\subset\supermn$ and ${\mathfrak{V}}\subset {\fR}^{p|q}$
be superdomains 
and let $\varphi$ be a continuous mapping from ${\mathfrak{U}}$ to ${\mathfrak{V}}$, denoted by
$\varphi(X)=(\varphi_1(X),{\cdots}, \varphi_{p}(X),
\varphi_{p+1}(X),{\cdots}, \varphi_{p+q}(X))\in{\fR}^{p|q}$.
$\varphi$ is called a supersmooth mapping from ${\mathfrak{U}}$ to ${\mathfrak{V}}$
if each $\varphi_A(X)\in {\CSS}({\mathfrak{U}}:{\fR})$ for $A=1,{\cdots}, p+q$ and
$\varphi ({\mathfrak{U}})\subset {\mathfrak{V}}$.
\end{defn}

\begin{prop}[Composition of supersmooth mappings]\label{prop-geofcomposition}
Let ${\mathfrak{U}}\subset{\fR}^{m|n}$ and ${\mathfrak{V}}\subset {\fR}^{p|q}$
be superdomains and 
let $F: {\mathfrak{U}}\to  {\fR}^{p|q}$ and $G:{\mathfrak{V}}\to{\fR}^{r|s}$
be supersmooth mappings such that $F(\mathfrak{U})\subset {\mathfrak{V}}$.
Then, the composition $G\circ F:{\mathfrak{U}}\to{\fR}^{r|s}$ 
gives a supersmooth mapping and
\begin{equation}
d_X G(F(X))
=[d_X F(X)][d_Y G(Y)]\big|_{Y=F(X)}.
\label{3-23}
\end{equation}
Or more precisely, 
\begin{equation}
d_X G(F(X))=(\partial_{X_A}(G\circ F)_B(X))
=(\sum_{C=1}^{r+s}\partial_{X_A}G_C(X))(\partial_{Y_C}F_B(Y))\big|_{Y=F(X)}).
\label{3-23-1}
\end{equation}
\end{prop}

\par{\it Proof}. 
Put $F(Y)=(G_B(Y))_{B=1}^{r+s}$, $F(X)=(F_C)_{C=1}^{p+q}$, $X=(X_A)_{A=1}^{m+n}$, and $Y=(Y_C)_{C=1}^{r+s}$.
By smooth G-differentiability of the composition of mappings between Fr\'echet spaces, we have the smoothness of
$\Phi(X+tH)$ w.r.t $t$.
Moreover, we have \eqref{3-23-1}.

By the characterization of supersmoothness, we need to say ${\rev}$-linearity of $d_F\Phi$, 
i.e. $d_F\Phi(X)(\lambda H)=\lambda d_F\Phi(X)(H)$ for $\lambda\in{\rev}$ which is obvious from
$$
\begin{aligned}
d(G\circ F)(X)(\lambda H)&=\dt G(F(X+t\lambda H))\bigg|_{t=0}\\
&=(\sum_{C=1}^{r+s}\lambda H_A\partial_{X_A}G_C(X))(\partial_{Y_C}F_B(Y))\big|_{Y=F(X)})\\
&=\lambda \dt G(F(X+t H))\bigg|_{t=0}=\lambda d(G\circ F)(X)(H). \qed
\end{aligned}
$$

\begin{defn}\label{3.16b}
Let ${\mathfrak{U}}\subset\supermn$ and
${\mathfrak{V}}\subset{\fR}^{p|q}$ be superdomains and
let $\varphi:{\mathfrak{U}}\to {\mathfrak{V}}$ be a supersmooth mapping represented by
$\varphi(X)=(\varphi_1(X),{\cdots},\varphi_{p+q}(X))$ with
$\varphi_A(X)\in {\CSS}({\mathfrak{U}}:{\fR})$.
\newline
(1) $\varphi$ is called a supersmooth diffeomorphism if 
\par(i) $\varphi$ is a homeomorphism between
${\mathfrak{U}}$ and ${\mathfrak{V}}$ and
\par(ii) $\varphi$ and $\varphi^{-1}$ are supersmooth mappings. 
\newline
(2) For any $f\in {\CSS}({\mathfrak{V}}:{\fR})$, 
$(\varphi^* f)(X)=(f\circ \varphi)(X)=f(\varphi(X))$,
called the {\it pull back} of $f$,
is well-defined and belongs to ${\CSS}({\mathfrak{U}}:{\fR})$.
\end{defn}

\begin{remark}
 It is easy to see that if $\varphi$ is a supersmooth diffeomorphism, then
$\varphi_{\mathrm{B}}=\pi_{\mathrm{B}} \circ \varphi$ is 
an (ordinary) $C^\infty$ diffeomorphism from ${\mathfrak{U}}_{\mathrm{B}}$ to ${\mathfrak{V}}_{\mathrm{B}}$.
\end{remark}
\begin{remark}
If we introduce the topologies in
${\CSS}({\mathfrak{V}}:{\fC})$ and ${\CSS}({\mathfrak{U}}:{\fC})$ properly,
$\varphi^*$ gives a continuous linear mapping from 
${\CSS}({\mathfrak{V}}:{\fC})$ to ${\CSS}({\mathfrak{U}}:{\fC})$.
Moreover, if $\varphi:{\mathfrak{U}}\to {\mathfrak{V}}$ is a supersmooth diffeomorphism,
then $\varphi^*$ defines an automorphism from
${\CSS}({\mathfrak{V}}:{\fR})$ to ${\CSS}({\mathfrak{U}}:{\fR})$.
\end{remark}

\subsection{Inverse and implicit function theorems}
We recall
\begin{prop}[Inverse function theorem on ${\euc}^m$]
Let $U$ be an open set in ${\euc}^m$.
Let $f: U\ni x\to y=f(x)\in{\euc}^m$ be a $C^k$ $(k\ge1)$ mapping such that
$f'(x_0)\neq0$ for some $x_0\in U$.
\par
Then, there exist a neighbourhood $W$ of $y_0=f(x_0)$ and a neighbourhood $U_0\subset U$ of $x_0$, such that
$f$ maps $U_0$ injectively onto $W$. Therefore, $f|_{U_0}$ has its inverse $\phi=(f|_{U_0})^{-1}:W\to U_0\in C^k(W)$.
Moreover, for any $y=f(x)\in W$ with $x\in U_0$, we have
$\phi'(y)=f'(x)^{-1}$.
\end{prop}

Applying this, we have
\begin{thm}[Inverse function theorem on $ {\supermn}$]\label{3.17b} 
Let $F=(f,g): {\supermn}\ni X \rightarrow \,Y=F(X)\in {\supermn}$ 
be a supersmooth mapping on some superdomain containing $\tilde{X}$.
That is,
$$
f(X)=(f_i(X))_{i=1}^m\in{\fR}_{\mathrm{ev}}^m,\;\;
g(X)=(g_k(X))_{k=1}^n\in{\fR}_{\mathrm{od}}^n,
$$
More precisely, we put
$$
f(X)=(f_i(X))_{i=1}^m\in {\fR}_{\mathrm{ev}}^m,\;\;
g(X)=(g_k(X))_{k=1}^n\in {\fR}_{\mathrm{od}}^n,
$$
such that
\begin{equation}
\begin{gathered}
f_i(x,\theta)=\sum_{|a|={\mathrm{ev}}\le n}f_{ia}(x)\theta^a,\;\;
g_k(x,\theta)=\sum_{|b|={\mathrm{od}}\le n}g_{kb}(x)\theta^b \in \ccsl_{S\!S}({\supermn})\\
\with
\begin{cases}
f_{ia}(x_{\mathrm{B}}), \;\; g_{kb}(x_{\mathrm{B}})\in {\mathbb{C}}&\mbox{if $|a|, |b|\neq0$},\\
f_{ia}(x_{\mathrm{B}})\in{\euc}&\mbox{if $|a|=0$}.
\end{cases}
\end{gathered}
\label{ccsl-assume}
\end{equation}
\par
We assume the super matrix $[d_X F(X)]$
is invertible at ${\tilde X}$, i.e. $\pi_{\mathrm{B}}(\sdet [d_X \Phi(X)]|_{X={\tilde X}})\neq0$. 
Then, there exist a superdomain ${\mathfrak{U}}$, a neighbourhood of ${\tilde X}$ and another  superdomain ${\mathfrak{V}}$, a neighbourhood of 
${\tilde Y} = F({\tilde X}) $ such that $F: {\mathfrak{U}}\to {\mathfrak{V}}$ has a unique supersmooth inverse $\Phi=F^{-1}=(\phi,\psi):{\mathfrak{V}}\to {\mathfrak{U}}$
satisfying 
$$
\phi(Y)=(\phi_{i'}(Y))_{i'=1}^m\in{\fR}_{\mathrm{ev}}^m,\;\;
\psi(Y)=(\psi_{k'}(Y))_{k'=1}^n\in{\fR}_{\mathrm{od}}^n
$$
and
\begin{equation} 
\Phi(F(X)) = X \for X\in {\mathfrak{U}}\et
F(\Phi(Y))=Y\for Y\in {\mathfrak{V}}.
\label{02-13-1}
\end{equation}
Moreover, we have
\begin{equation}
\left.{d_Y}\Phi(Y)= ({d_X}F(X))^{-1} \,\right|_{X=\Phi(Y)} 
\quad \text{in}\quad {\mathfrak{V}}.
\label{3-31}
\end{equation}
\end{thm}
\par{\it Proof}.
(I) To make clear the point, we consider the case $m=1$, $n=2$, that is, ${\mathfrak{U}}, {\mathfrak{V}}\subset{\fR}^{1|2}$.
Let 
$$
F(X)=F(x,\theta)=({f}(x,\theta), {g}_1(x,\theta),{g}_2(x,\theta)):{\mathfrak{U}}\to {\mathfrak{V}}
$$ 
with
\begin{equation}
\left\{
\begin{aligned}
&{f}(x,\theta)=f_{(0)}(x)+f_{(12)}(x)\theta_1\theta_2, \\
&{g_1}(x,\theta)=g_{1(1)}(x)\theta_1+ g_{1(2)}(x)\theta_2,\\
&{g_2}(x,\theta)=g_{2(1)}(x)\theta_1+ g_{2(2)}(x)\theta_2,
\end{aligned}
\right.\et
f_{(0)}(x_{\mathrm{B}})\in\euc,\; f_{(12)}(x_{\mathrm{B}}),\, g_{k(l)}(x_{\mathrm{B}})\in{\mathbb{C}}.
\label{02-20-1}
\end{equation}
In this case, we have
$$
d_XF(X)=\begin{pmatrix}
f'_{(0)}(x)+f'_{(12)}(x)\theta_1\theta_2&g_{1(1)}'(x)\theta_1+ g_{1(2)}'(x)\theta_2&g_{2(1)}'(x)\theta_1+ g_{2(2)}'(x)\theta_2\\
f_{(12)}(x)\theta_2&g_{1(1)}(x)&g_{2(1)}(x)\\
-f_{(12)}(x)\theta_1&g_{1(2)}(x)&g_{2(2)}(x)
\end{pmatrix}
=\begin{pmatrix}
A&C\\
D&B
\end{pmatrix}
$$
with
$$
\sdet(d_XF(X))=\det[A-CB^{-1}D](\det B)^{-1}
\et
\det B=\beta(x)=g_{1(1)}(x)g_{2(2)}(x)-g_{1(2)}(x)g_{2(1)}(x).
$$
Therefore, 
\begin{equation}
\pi_{\mathrm{B}}(\sdet(d_XF(X)))=f'_{(0)}(x_{\mathrm{B}})\beta(x_{\mathrm{B}})^{-1}.
\label{12-sdet}
\end{equation}
We need to find $\Phi=(\phi,\psi_1,\psi_2)$ such that
\begin{equation}
{\left\{
\begin{aligned}
&\phi(f(x_{\mathrm{B}},\theta),g_{1}(x_{\mathrm{B}},\theta),g_{2}(x_{\mathrm{B}},\theta))=x_{\mathrm{B}},\\
&\psi_{1}(f(x_{\mathrm{B}},\theta),g_{1}(x_{\mathrm{B}},\theta),g_{2}(x_{\mathrm{B}},\theta))=\theta_1,\\
&\psi_{2}(f(x_{\mathrm{B}},\theta),g_{1}(x_{\mathrm{B}},\theta),g_{2}(x_{\mathrm{B}},\theta))=\theta_2
\end{aligned}
\right.}
\with
{\left\{
\begin{aligned}
&\phi(y,\omega)=\phi_{(0)}(y)+\phi_{(12)}(y)\omega_1\omega_2,\\
&\psi_1(y,\omega)=\psi_{1(1)}(y)\omega_1+\psi_{1(2)}(y)\omega_2,\\
&\psi_2(y,\omega)=\psi_{2(1)}(y)\omega_1+\psi_{2(2)}(y)\omega_2.
\end{aligned}
\right.}
\label{12-s-diff11}
\end{equation}
To state more precisely, we have
$$
\left\{
\begin{aligned}
&y_{\mathrm{B}}=f_{(0)}(x_{\mathrm{B}}),\quad
y_{\mathrm{S}}=f_{(12)}(x_{\mathrm{B}})\theta_1\theta_2,\\
&\phi_{(0)}(y_{\mathrm{B}}+y_{\mathrm{S}})=\phi_{(0)}(y_{\mathrm{B}})+\phi'_{(0)}(y_{\mathrm{B}})y_{\mathrm{S}},\\
&\phi_{(12)}(y_{\mathrm{B}}+y_{\mathrm{S}})=\phi_{(12)}(y_{\mathrm{B}})+\phi'_{(12)}(y_{\mathrm{B}})y_{\mathrm{S}},
\end{aligned}
\right.
$$
and from the first equation of \eqref{12-s-diff11}, 
\begin{equation}
\left\{
\begin{aligned}
&\phi_{(0)}(y_{\mathrm{B}})=\phi_{(0)}(f_{(0)}(x_{\mathrm{B}}))=x_{\mathrm{B}},\\
&\phi'_{(0)}(y_{\mathrm{B}})y_{\mathrm{S}}+\phi_{(12)}(y_{\mathrm{B}})\beta(x_{\mathrm{B}})\theta_1\theta_2=[\phi'_{(0)}(y_{\mathrm{B}})f_{(12)}(x_{\mathrm{B}})+\phi_{(12)}(y_{\mathrm{B}})\beta(x_{\mathrm{B}})]\theta_1\theta_2=0.
\end{aligned}
\right.
\label{03-14}
\end{equation}

Since $f'_{(0)}(\tilde{x}_{\mathrm{B}})\neq0$, there exists a neighborhood $U_0\subset\euc$ of $\tilde{x}_{\mathrm{B}}$ and $V_0\subset\euc$ of $f_{(0)}(\tilde{x}_{\mathrm{B}})$ where
we find a function $\phi_{(0)}(y_{\mathrm{B}})$ satisfying the first equation \eqref{03-14}.
Moreover, since $\beta(x_{\mathrm{B}})\neq0$, taking the smaller neighborhood if necessary,  we define
$$
\phi_{(12)}(y_{\mathrm{B}})=-\phi'_{(0)}(y_{\mathrm{B}})f_{(12)}(x_{\mathrm{B}})\beta(x_{\mathrm{B}})^{-1}\big|_{x_{\mathrm{B}}=\phi_{(0)}(y_{\mathrm{B}})}.
$$

On the other hand, putting
$$
\begin{aligned}
&\omega_1=g_{1(1)}(x_{\mathrm{B}})\theta_1+g_{1(2)}(x_{\mathrm{B}})\theta_2,\\
&\omega_2=g_{2(1)}(x_{\mathrm{B}})\theta_1+g_{2(2)}(x_{\mathrm{B}})\theta_2,
&\omega_1\omega_2=\beta(x_{\mathrm{B}})\theta_1\theta_2
\end{aligned}
$$
and remarking  $y_{\mathrm{S}}\omega_j=0$ for $j=1,2$, from the last two equations of \eqref{12-s-diff11}, we should have
$$
\begin{aligned}
&\psi_{1(1)}(y_{\mathrm{B}})
(g_{1(1)}(x_{\mathrm{B}})\theta_1+g_{1(2)}(x_{\mathrm{B}})\theta_2)
+\psi_{1(2)}(y_{\mathrm{B}})
(g_{2(1)}(x_{\mathrm{B}})\theta_1+g_{2(2)}(x_{\mathrm{B}})\theta_2)=\theta_1,\\
&\psi_{2(1)}(y_{\mathrm{B}})
(g_{1(1)}(x_{\mathrm{B}})\theta_1+g_{1(2)}(x_{\mathrm{B}})\theta_2)
+\psi_{2(2)}(y_{\mathrm{B}})
(g_{2(1)}(x_{\mathrm{B}})\theta_1+g_{2(2)}(x_{\mathrm{B}})\theta_2)=\theta_2,
\end{aligned}
$$
that is,
$$
\begin{pmatrix}
\psi_{1(1)}(y_{\mathrm{B}})&\psi_{1(2)}(y_{\mathrm{B}})\\
\psi_{2(1)}(y_{\mathrm{B}})&\psi_{2(2)}(y_{\mathrm{B}})
\end{pmatrix}
\begin{pmatrix}
g_{1(1)}(x_{\mathrm{B}})&g_{1(2)}(x_{\mathrm{B}})\\
g_{2(1)}(x_{\mathrm{B}})&g_{2(2)}(x_{\mathrm{B}})
\end{pmatrix}
=\begin{pmatrix}
1&0\\
0&1
\end{pmatrix}.
$$
Therefore, we have $\psi_*(y_{\mathrm{B}})$, which satisfy the desired property.\\
(II) Do analogously as above for general $m,n$ by putting
$$
f_i(x,\theta)=\sum_{|a|={\mathrm{ev}}\le n}f_{i,a}(x)\theta^a,\;\;
g_k(x,\theta)=\sum_{|b|={\mathrm{od}}\le n}g_{k,b}(x)\theta^b,
$$
and
$$
\phi_{i'}(y,\omega)=\sum_{|a'|={\mathrm{ev}}\le n}\phi_{i',a'}(y)\omega^{a'},\;\;
\psi_{k'}(y,\omega)=\sum_{|b'|={\mathrm{od}}\le n}\psi_{k',b'}(y)\omega^{b'},
$$
but with more patience.  \qed

\begin{remark}
Above theorem holds for functions $f_i\in {\mathcal{C}}_{S\!S}({\fR}^{m|n}:{\rev})$ and
$g_k\in {\mathcal{C}}_{S\!S}({\fR}^{m|n}:{\rod})$.
\end{remark}

Moreover, we have
\begin{prop}[Implicit function theorem]\label{3.18b}
Let $\Phi (X,Y): {\mathfrak{U}} \times {\mathfrak{V}} \to {\fC}^{p|q}$ 
be a supersmooth mapping and $( {\tilde X }, {\tilde Y}) \in {\mathfrak{U}} \times {\mathfrak{V}}$,
where ${\mathfrak{U}}$ and ${\mathfrak{V}}$ are superdomains of ${\supermn}$ and ${\fR}^{p|q}$,
respectively.
Suppose $\Phi ( {\tilde X }, {\tilde Y} )=0$ and
$ {\partial}_Y \Phi=[\partial_{y_j} \Phi, \partial_{\omega_r} \Phi ]$
is a continuous and invertible supermatrix 
at $({\tilde X}_{\mathrm{B}},{\tilde Y}_{\mathrm{B}})
\in \pi_{\mathrm{B}}({\mathfrak{U}}) \times \pi_{\mathrm{B}}({\mathfrak{V}})$.
Then, there exist a superdomain ${\mathfrak{V}} \subset {\mathfrak{U}} $ 
satisfying ${\tilde X}_{\mathrm{B}}\in \pi_{\mathrm{B}}({\mathfrak{V}})$ 
and a unique supersmooth mapping $Y=f(X)$ on ${\mathfrak{V}}$
such that $ {\tilde Y} = f( {\tilde X} )$ and $ \Phi (X,f(X))=0 $ in ${\mathfrak{V}}$.
Moreover, we have
\begin{equation}
{\partial_X} f(X) 
= -\left.[\partial_X \Phi(X,Y)]{[ \partial_Y \Phi(X,Y)]^{-1}}\right|_{Y=f(X)}.
\label{3-46}
\end{equation}
\end{prop}
\par
{\it Proof}.
\eqref{3-46} 
is easily obtained by
$$
0=\partial_X \Phi(X,f(X))=\left.
(\partial_X \Phi(X,Y)+\partial_Xf(X)\partial_Y\Phi(X,Y))
\right|_{Y=f(X)}.
$$
The existence proof is omitted here because the arguments in proving 
Proposition \ref{3.17b}  
work well in this situation.	$\qed$

\subsection{Global inverse function theorem}
We have the following theorem of Hadamard type:
\begin{prop}[Global inverse function theorem on ${\euc}^m$]\label{3.17b-1} 
Let $f: {\euc}^m\ni x \rightarrow \,y=f(x)\in {\euc}^m$ 
be a smooth mapping on ${\euc}^m$.
We assume the Jacobian matrix $[d_x f(x)]$
is invertible on ${\euc}^m$, 
and $\Vert (\det [d_x f(x)])\Vert \ge \delta>0$ for any $x$.
Then, $f$ gives a smooth diffeomorphism from ${\euc}^m$ onto ${\euc}^m$.
\end{prop}

\begin{prop}[Global inverse function theorem on $\supermn$]\label{3.17b-2} 
Let $F=(f,g): {\supermn}\ni X \rightarrow \,Y=F(X)\in {\supermn}$ 
be a supersmooth mapping on ${\supermn}$.
We assume the super matrix 
$$
d_X F(X)=\begin{pmatrix}
\frac{\partial f_i}{\partial x_j}&\frac{\partial g_k}{\partial x_j}\\
\frac{\partial f_i}{\partial \theta_l} &\frac{\partial g_k}{\partial \theta_l} 
\end{pmatrix}
$$
is invertible at any ${X}\in \supermn$, 
and there exists $\delta>0$ such that  for any $x$
$$
\Vert \pi_{\mathrm{B}}(\sdet \frac{\partial f_i}{\partial x_j})\Vert \ge \delta>0,
\et
\{x\;|\; \pi_{\mathrm{B}}\det(\frac{\partial g_k}{\partial \theta_l} )=0\}=\emptyset.
$$
Then, $F$ gives a supersmooth diffeomorphism from ${\supermn}$ onto ${\supermn}$.
\end{prop}
\par
{\it Proof}. From the proof  of above Theorem \ref{3.17b}, it is obvious. \qed

\end{document}